\documentclass{aa}  

\usepackage{graphicx}
\usepackage{hyperref}

\usepackage{txfonts}
\usepackage{xcolor}
\usepackage{comment}
\usepackage[normalem]{ulem}
\newcommand{\teff}{\ensuremath{\mathrm{T_{eff}}}}
\newcommand{\kms}{\ensuremath{\mathrm{km\,s^{-1}}}}
\newcommand{\logg}{\ensuremath{\mathrm{\log g}}}
\newcommand{\vt}{\ensuremath{\mathrm{v_{turb}}}}
\newcommand{\vbroad}{\ensuremath{\mathrm{v_{broad}}}}
\newcommand{\mygi}{MyGIsFOS}

\newcommand{\Msol}{$\rm M_\odot$}
\newcommand{\av}{\ensuremath{A_{V}}}

\usepackage{lscape}
\usepackage{longtable}
\usepackage{adjustbox}
\usepackage{pdflscape}

\begin{document}

   \title{Chemical Evolution of R-process Elements in Stars (CERES)
   }

   \subtitle{I. Stellar parameters and chemical abundances from Na to Zr
  \thanks{Chemical abundances (Table 3) are only available in electronic form at the CDS via anonymous ftp to \url{cdsarc.u-strasbg.fr} (130.79.128.5) or via \url{http://cdsweb.u-strasbg.fr/cgi-bin/qcat?J/A+A/} }
   \thanks{Based on observations collected at the European Southern Observatory under ESO programme 0104.D-0059 and on data obtained from the ESO Science Archive Facility. }
}

   \author{Linda Lombardo
          \inst{1}
 \and Piercarlo Bonifacio \inst{1}
 \and Patrick  Fran\c{c}ois \inst{1}
 \and Camilla J. Hansen \inst{2}
 \and Elisabetta Caffau \inst{1}
 \and Michael Hanke \inst{3}
 \and \'Asa Sk\'ulad\'ottir \inst{4,5}
 \and Almudena Arcones \inst{6,7}
 \and Marius Eichler \inst{6}
 \and Moritz Reichert \inst{6,8}
 \and Athanasios Psaltis \inst{6} 
 \and Andreas J. Koch Hansen \inst{3}
 \and Luca Sbordone \inst{9}
          }

   \institute{GEPI, Observatoire de Paris, Universit\'e PSL, CNRS, 5 place Jules Janssen 92195 Meudon, France \\
              \email{Linda.Lombardo@observatoiredeparis.psl.eu}
    \and
    Goethe University Frankfurt, Institute for Applied Physics, Max-von-Laue Str. 11, 60438 Frankfurt am Main
    \and
    Zentrum f\"ur Astronomie der Universit\"at Heidelberg, Astronomisches Rechen-Institut, M\"onchhofstr. 12, 69120 Heidelberg, Germany
    \and
    Dipartimento di Fisica e Astronomia, Università degli Studi di Firenze, Via G. Sansone 1, I-50019 Sesto Fiorentino, Italy
    \and
    INAF/Osservatorio Astrofisico di Arcetri, Largo E. Fermi 5, I-50125 Firenze, Italy
    \and
    Institut f\"ur Kernphysik, Technische Universit\"at Darmstadt, Schlossgartenstr. 2, Darmstadt 64289, Germany
    \and
    GSI Helmholtzzentrum f\"ur Schwerionenforschung GmbH, Planckstr. 1, Darmstadt 64291, Germany
    \and
    Departament d'Astronomia i Astrof\'isica, Universitat de Val\`encia, Edifici d'Investigaci\'o Jeroni Munyoz, C/ Dr. Moliner, 50, E-46100 Burjassot (Val\'encia), Spain
    \and
    ESO – European Southern Observatory, Alonso de Cordova 3107, Vitacura, Santiago, Chile
             }

   \date{Received September 15, 1996; accepted March 16, 1997}

 
  \abstract
   {}
   {
   The Chemical Evolution of R-process Elements in Stars (CERES) project aims to provide a homogeneous analysis of a sample of metal-poor stars ([Fe/H]<--1.5). We present the stellar parameters and the chemical abundances of elements up to Zr for a sample of 52 giant stars.}
   {We relied on a sample of high signal-to-noise UVES spectra. We determined stellar parameters from {\it Gaia} photometry and parallaxes. Chemical abundances were derived using spectrum synthesis and model atmospheres.}
   {We determined chemical abundances of 26 species of 18 elements: Na, Mg, Al, Si, Ca, Sc, Ti, V, Cr, Mn, Fe, Co, Ni, Cu, Zn, Sr, Y, and Zr. For several stars, we were able to measure both neutral and ionised species, including Si, Sc, Mn, and Zr. We have roughly doubled the number of measurements of Cu for stars at [Fe/H] $\leq -2.5$.
   The homogeneity of the sample made it possible to highlight the presence of two Zn-rich stars ([Zn/Fe]$\sim$+0.7), one r-rich and the other r-poor. We report the existence of two branches in the [Zn/Fe] versus [Ni/Fe] plane and suggest that the high [Zn/Fe] branch is the result of hypernova nucleosynthesis. We  discovered two stars with peculiar light neutron-capture abundance patterns: CES1237+1922 (also known as BS 16085-0050), which is $\sim$1 dex underabundant in Sr, Y, and Zr with respect to the other stars in the sample, and CES2250-4057 (also known as HE 2247-4113), which shows a $\sim$1 dex overabundance of Sr with respect to Y and Zr. }
  {The high quality of our dataset allowed us to measure hardly detectable ions. This can provide guidance in the development of line formation computations that take deviations from local thermodynamic equilibrium and hydrodynamical effects into account.}


   \maketitle
%

\section{Introduction}
The elements beyond the Fe-peak, that is, all elements with atomic number $Z > 30$, 
are commonly referred to as `heavy elements' or `neutron-capture' (n-capture) elements, as the most efficient way to form these elements is through neutron captures.
A neutron capture tends to create a nucleus that is away from the nuclear stability valley, and the system tries to fall back to the valley through $\beta$ decay. 

From a theoretical point of view, three neutron-capture processes can be distinguished, depending on the neutron flux. 
When the seed nucleus has time to $\beta$ decay after each neutron capture, one refers to this as the slow neutron-capture process ($s$ process), which occurs at neutron densities of less than about 
$10^{8}$\,cm$^{-3}$ \citep[e.g.][and references therein]{busso}.
The rapid neutron-capture process ($r$ process) occurs when the neutron flux is so high that a seed nucleus can capture several neutrons before decaying. 
The $r$ process requires neutron densities in excess of $10^{24}$ cm$^{-3}$ \citep[e.g.][and references therein]{Kratz}.
In the intermediate neutron density regime ($ 10^{14}\rm\,cm^{-3}\le N_n \le 10^{16}\, cm^{-3}$) 
one talks about the intermediate neutron-capture process \citep[$i$ process;][]{CR1977}.
Sometimes in the literature, Zn ($Z=30$) is not considered
part of this set of elements since it may also be formed
through different nucleosynthetic channels, but neutron captures are
definitely a possibility for synthesising Zn \citep{bisterzo04}.
The same is true for Cu, but we consider Cu to be part of the iron-peak elements.

Observationally, it has been known for many years that, in metal-poor stars,  the abundance ratios [X/Fe], where X is an n-capture element, show a large scatter as a function of [Fe/H] \citep{mcwilliam95,burris00,Francois2007,Hansen2012}, at variance with the lighter elements, for example Mg, which show a very tight relation with [Fe/H] \citep{mcwilliam95,Cayrel2004,Bonifacio2009}.
This has generally been interpreted as evidence that the production of n-capture elements occurs in different sites and under different physical conditions, in contrast to the lighter elements that are formed in either hydrostatic or explosive burning in Type II or Type Ia supernovae (SNe)
\citep[see e.g.][]{Arnett}.

Based on then-existing observations and theoretical considerations, \citet{Truran1981} argued that in metal-poor stars the n-capture elements can only be formed through the $r$ process, essentially because the only then-recognised source of $s-$process nucleosynthesis, asymptotic giant branch stars, would not have had enough time to enrich the interstellar medium before the metallicity rose above $\rm [Fe/H] =-1.5$.
This is sometimes referred to as the $r-$only paradigm.
It is currently believed that fast rotating massive stars can produce $s-$process elements and deliver them through their winds, prior to the SN explosion \citep{Prantzos1990,Pignatari2010,Choplin2018,Banerjee2018,Sk2020}.
Although the site of the $i$ process has not yet been robustly identified, it is considered for the production of some isotopes \citep{Hampel2016,Cote2018,Denissenkov2019,Koch2019A&A...622A.159K,Sk2020,Choplin2021}.
More exotic phenomena, such as proton ingestion phenomena, are also believed to be responsible for the production of some n-capture elements \citep{Hollowell1990,Fujimoto2000,Cristallo2009,Caffau2019}.
Even if Truran's intuition is probably correct and the majority of n-capture elements at low metallicities are formed by the $r$ process, other processes cannot be ignored and may, in fact, contribute to the large scatter observed in the abundance ratios of these elements and to the variety of abundance patterns.
Furthermore, it is now accepted that the $r$ process is not universal, but can occur in different astrophysical sites under different physical conditions, of which at least two are necessary to explain the observations \citep{Qian2001,Qian2007,Hansen2014b,Spite2018,SkSal2020A&A...634L...2S}.

The Chemical Evolution of R-process Elements in Stars (CERES) project has the objective of characterising the abundances of as many n-capture elements as possible in a sample of giant stars of low metallicity ([Fe/H]$< -1.5$). 
The aim of CERES is to provide a high quality set of abundances that can be used to test different theoretical scenarios. {To achieve this, we rely on a set of high resolution and high signal-to-noise ratio (S/N) spectra,} and on the photometry and parallaxes provided by the {\it Gaia} satellite \citep{Gaia} to analyse all the stars in a homogeneous way.
In this first paper of the series, we provide the atmospheric parameters and abundances of 18 elements, Na, Mg, Al, Si, Ca, Sc, Ti, V, Cr, Mn, Fe, Co, Ni, Cu, Zn, Sr, Y, and Zr, the last three of which are n-capture elements.
Abundances of other n-capture elements shall be provided in subsequent papers of the series, based on the same atmospheric parameters.

The current paper is organised as follows: Sect. 2 describes the sample and observations, Sect. 3 the analysis, including stellar parameter and abundance determination, Sect. 4 details the results, and Sect. 5 provides a discussion. Finally, Sect. 6 presents our conclusions.

\section{Sample selection and observational data}
\subsection{Sample selection}
With the goal to derive as complete abundance patterns as possible -- in particular with regard to the heavy elements -- we targeted metal-poor stars ([Fe/H]$<-1.5$) with fewer than five measured heavy elements (Z>30). 
The initial target sample was based on stars from \citet{Frebel2006,Francois2007,Hansen2012,Hansen2020}, the metal-poor tail of GALactic Archaeology with HERMES \citep[GALAH;][]{GALAH2018}, with further metal-poor candidates from \citet{Roederer2014, Yong2013,Barklem2005}. 
We avoided overlap with the R-Process Alliance survey \citep[e.g.][]{TTHansen2018, Sakari2018} and the Hamburg/ESO R-process Enhanced Star (HERES) survey \citep{Christlieb2004} as these stars typically have already been observed to target n-capture elements.
The candidates were then checked against the literature, for example the Stellar Abundances for Galactic Archaeology (SAGA) database \citep{Suda2008}, to ensure that there were only a few available heavy element abundance measurements. 
Finally, we removed binaries and mainly avoid stars classified in the literature as carbon-enhanced metal-poor (CEMP) stars, as the strong CH and CN molecular bands affect the accuracy of heavy element abundance measurements.

As most of the heavy element absorption transitions fall in the blue part of the spectrum, a high S/N is needed in the blue wavelength range (S/N=50 in more metal-rich stars, and S/N=120 in the most metal-poor ones at 390 nm), resulting in considerable exposure times. 
To keep the observations feasible we thus limited the target sample to V<12.2.
Finally, from the stars where the Eu abundance is known, we included a mixture of high and low values to probe r-poor as well as r-rich stars. In case no Eu abundance is known, we keep the star in the sample for follow-up observations. 
Details about the final sample and observations can be found in Table \ref{tab:obs_log} in the appendix.

\subsection{Observations}
The target stars were observed with {the Ultraviolet and Visual Echelle Spectrograph (UVES) of the Very Large Telescope (VLT) at the European Southern Observatory (ESO)} 
\citep{UVES} during two runs (November 2019 and March 2020) with differing exposures to reach a S/N of 50 to 120 per pixel at 390\,nm for most stars (see Table \ref{tab:obs_log}).
For stars with $\rm[Fe/H] < -2.7$, a S/N of 200 per pixel was requested. The observations were carried out using a 1" slit, 1x1 binning, and a dichroic (Dic 1) where the blue and red arms were centred on 390 and 564\,nm, respectively.
This resulted in high-resolution ($R>40\,000$) spectra that owing to the excellent observing conditions (low airmass, $<1.5$, and seeing, $<$1\farcs{0}), reached a median resolution of 49800 in the blue arm and 47500 in the red arm. 
In a few cases these requests were violated (airmass 1.6 and/or seeing as large as 1\farcs{5}) and the reduced spectra attained, nevertheless the required S/N. 
Further details related to the observations can be found in the observing log (Table \ref{tab:obs_log}). 

Our own observations were complemented with archival data of comparable quality. 
All the archival data used were acquired prior to 2019. 
In Table \ref{tab:obs_log}, the wavelength ranges covered by different UVES arms are the following: $303<\lambda<388$ nm for BLUE346, $326<\lambda<454$ nm for BLUE390, $458<\lambda<668$ nm for RED564, and $476<\lambda<684$ nm for RED580.

\section{Analysis}
\subsection{Chromospheric activity}
As a preliminary step we inspected the \ion{Ca}{ii} H and K
lines of all our targets to find signs of chromospheric activity.
We found four stars that are clearly active and present
emission in the core of the \ion{Ca}{ii} H and K lines:
CES0547-1739\footnote{The name ID of the star is defined as the string CES followed by RA\,J2000 (hm, four digits), the sign of the declination, and DEC\,J2000 (dm, four digits).  },  
CES0747-0405, 
CES0900-6222, 
and CES1116-7250.
Three more stars, show minor signs of activity, and should be
further investigated:
CES0919-6958,
CES0413+0636, and
CES0424-1501.
Of the lines we used for abundance analysis only the \ion{Na}{i} D resonance
lines are sensitive to chromospheric effects, yet we do not notice any systematic
effect on the abundances derived from these lines with respect to those
derived from other lines. Thus, our analysis is likely immune to the effects
of the chromosphere. It would be yet interesting to further investigate
the chromospheres of these stars and their variations with time.
Observations of the \ion{He}{I} 1083 \,nm line, with a proper 
modelling of their chromosphere could provide He abundances for these stars
\citep[see e.g.][and references therein]{2011A&A...531A..35P}.

\subsection{Stellar parameters\label{sp}}

\begin{table}
\centering
 \caption[]{\label{tab:stelpar}Stellar parameters and {\it Gaia} de-reddened G magnitude for stars in our sample.}
\begin{tabular}{lrcccc}
 \hline \hline
 Star            & G$_0$ &  \teff &       \logg &       \vt      & [Fe$/$H]    \\
                 & mag   &  K     &       dex   &       \kms     & dex         \\
 \hline
  CES0031--1647 & 8.20 & 4960 & 1.83 & 1.91 & -2.49\\
  CES0045--0932 & 8.70 & 5023 & 2.29 & 1.76 & -2.95\\
  CES0048--1041 & 10.48 & 4856 & 1.68 & 1.93 & -2.48\\
  CES0055--3345 & 9.36 & 5056 & 2.45 & 1.66 & -2.36\\
  CES0059--4524 & 14.66 & 5129 & 2.72 & 1.56 & -2.39\\
  CES0102--6143 & 13.45 & 5083 & 2.37 & 1.75 & -2.86\\
  CES0107--6125 & 13.36 & 5286 & 2.97 & 1.54 & -2.59\\
  CES0109--0443 & 13.30 & 5206 & 2.74 & 1.69 & -3.23\\
  CES0215--2554 & 8.91 & 5077 & 2.00 & 1.91 & -2.73\\
  CES0221--2130 & 10.21 & 4908 & 1.84 & 1.84 & -1.99\\
  CES0242--0754 & 14.72 & 4713 & 1.36 & 2.03 & -2.90\\
  CES0301+0616 & 12.65 & 5224 & 3.01 & 1.51 & -2.93\\
  CES0338--2402 & 9.67 & 5244 & 2.78 & 1.62 & -2.81\\
  CES0413+0636 & 8.06 & 4512 & 1.10 & 2.01 & -2.24\\
  CES0419--3651 & 12.64 & 5092 & 2.29 & 1.78 & -2.81\\
  CES0422--3715 & 9.26 & 5104 & 2.46 & 1.68 & -2.45\\
  CES0424--1501 & 9.68 & 4646 & 1.74 & 1.74 & -1.79\\
  CES0430--1334 & 9.71 & 5636 & 3.07 & 1.63 & -2.09\\
  CES0444--1228 & 12.21 & 4575 & 1.40 & 1.92 & -2.54\\
  CES0518--3817 & 14.12 & 5291 & 3.06 & 1.49 & -2.49\\
  CES0527--2052 & 13.72 & 4772 & 1.81 & 1.84 & -2.75\\
  CES0547--1739 & 11.53 & 4345 & 0.90 & 2.01 & -2.05\\
  CES0747--0405 & 10.32 & 4111 & 0.54 & 2.08 & -2.25\\
  CES0900--6222 & 10.47 & 4329 & 0.94 & 1.98 & -2.11\\
  CES0908--6607 & 10.85 & 4489 & 0.90 & 2.12 & -2.62\\
  CES0919--6958 & 10.76 & 4430 & 0.70 & 2.17 & -2.46\\
  CES1116--7250 & 10.08 & 4106 & 0.48 & 2.14 & -2.74\\
  CES1221--0328 & 15.72 & 5145 & 2.76 & 1.6 & -2.96\\
  CES1222+1136 & 9.64 & 4832 & 1.72 & 1.93 & -2.91\\
  CES1226+0518 & 7.79 & 5341 & 2.84 & 1.60 & -2.38\\
  CES1228+1220 & 9.29 & 5089 & 2.04 & 1.87 & -2.32\\
  CES1237+1922 & 11.85 & 4960 & 1.86 & 1.95 & -3.19\\
  CES1245--2425 & 10.26 & 5023 & 2.35 & 1.72 & -2.85\\
  CES1322--1355 & 10.26 & 4960 & 1.81 & 1.96 & -2.93\\
  CES1402+0941 & 5.83 & 4682 & 1.35 & 2.01 & -2.79\\
  CES1405--1451 & 6.73 & 4642 & 1.58 & 1.81 & -1.87\\
  CES1413--7609 & 10.04 & 4782 & 1.72 & 1.87 & -2.52\\
  CES1427--2214 & 8.61 & 4913 & 1.99 & 1.85 & -3.05\\
  CES1436--2906 & 7.75 & 5280 & 3.15 & 1.42 & -2.15\\
  CES1543+0201 & 12.6 & 5157 & 2.77 & 1.57 & -2.65\\
  CES1552+0517 & 10.12 & 5013 & 2.30 & 1.72 & -2.60\\
  CES1732+2344 & 8.58 & 5370 & 2.82 & 1.65 & -2.57\\
  CES1804+0346 & 6.45 & 4390 & 0.80 & 2.12 & -2.48\\
  CES1942--6103 & 11.68 & 4748 & 1.53 & 2.01 & -3.34\\
  CES2019--6130 & 11.38 & 4590 & 1.13 & 2.09 & -2.97\\
  CES2103--6505 & 12.95 & 4916 & 2.05 & 1.85 & -3.58\\
  CES2231--3238 & 12.84 & 5222 & 2.67 & 1.67 & -2.77\\
  CES2232--4138 & 13.18 & 5194 & 2.76 & 1.59 & -2.58\\
  CES2250--4057 & 9.96 & 5634 & 2.51 & 1.88 & -2.14\\
  CES2254--4209 & 14.68 & 4805 & 1.98 & 1.79 & -2.88\\
  CES2330--5626 & 13.71 & 5028 & 2.31 & 1.75 & -3.10\\
  CES2334--2642 & 13.30 & 4640 & 1.42 & 2.02 & -3.48\\
\hline
\end{tabular}
\end{table}

The stellar parameters for our sample of stars were derived using {\it Gaia} Early Data Release 3 (EDR3) photometry ($G$, $G_{BP}-G_{RP}$) and parallaxes \citep{Gaia,GaiaEDR3}. 
We defined a grid in the parameter space using ATLAS\,9 
model atmospheres by \citet{CastelliKurucz}.
The sub-grid we used has effective temperatures (\teff), surface gravities (\logg) \ and metallicity ([M/H]) in the range of $3500 \leq \teff \leq 6000$, $0\leq\logg\leq4$, and $-4\leq$ [M/H] $\leq+0.5$,  
The $\alpha$-elements are enhanced by $+$0.4\,dex for all models with [M/H] $\le -1$, and
they are solar-scaled for higher metallicity models. The microturbulent velocity is 2 \kms\ for all models.
Theoretical values of $G_{BP}-G_{RP}$, bolometric correction ($BC_G$), and extinction coefficients $A_{G}$, E($G_{BP}-G_{RP}$), using the reddening law of \citet{Fitzpatrick}, were computed for the entire grid. $G$ and $G_{BP}-G_{RP}$ were de-reddened using the reddening maps by \citet[][\av=0.81]{Schlafly2011}.
Effective temperatures and surface gravities were derived iteratively using the procedure described in \citet{Koch-Hansen}. 
The errors on the effective temperature can be conservatively estimated by changing the $G_{BP}-G_{RP}$ by 0.02 mag. 
This is larger than the purely photometric errors, but we include also the uncertainty on the reddening. 
The new effective temperatures are offset by 88\,K that we round to an error of $\pm 100$\,K. 
According to \citet{Bonifacio2018A&A...611A..68B}, the mean difference between the 3D corrected and 1D bolometric corrections computed from ATLAS 9 models is around 0.02 mag for stars with stellar parameters similar to those of our sample stars. We consider this value as the typical uncertainty on the bolometric correction. 
Surface gravities are offset by approximately 0.035 dex with a 100 K change in \teff, while a 0.02 mag change in the bolometric correction implies a 0.01 dex change in \logg. 
Taking into account the $1\sigma $ errors on parallaxes, the surface gravities are offset by about 0.02\,dex.
Microturbulent velocities (\vt) were estimated using the calibration derived by \citet{Mashonkina2017aug}. 
The uncertainty on \vt\ is 0.5 \kms, according to the maximum discrepancy between microturbulences derived from spectroscopy and from the formula in \citet{Mashonkina2017aug} .
The derived stellar parameters are shown in Table \ref{tab:stelpar}, coordinates and other names for the targets can be found in Table \ref{tab:obs_log}. 
[Fe/H] indicates the iron abundance derived from \ion{Fe}{I}. The mean uncertainty on [Fe/H] is 0.13 dex, which corresponds to the mean line-to-line scatter.

\subsection{Line broadening}

In all the spectra  analysed with a resolving power $R \ga 60\,000$, the line width is often not dominated by the instrumental resolution
but by the macroturbulence. In fact, the observed
width is a convolution of instrumental resolution
and macroturbulence and needs to be determined
for each star and each instrumental resolution.
To derive chemical abundances, we developed a procedure to estimate the line broadening in \kms (\vbroad) for each observed spectrum.
We first measured the full width half maximum (FWHM) for  a set  of isolated and non-saturated lines in the observed spectra.
We then measured the FWHM for the same lines for a set of synthetic spectra, broadened
assuming a Gaussian macroturbulence, for several values of \vbroad.
The stellar parameters of the synthetic spectra were chosen to be close to the parameters of the star analysed.
For this purpose we used the parameters determined as described in Sect.\,\ref{sp} and first guess metallicities derived from our first run of {the code My God It’s Full Of Stars} 
(\mygi; see Sect.\,\ref{mygi}; \citealt{mygisfos}), 
assuming a broadening of 7\,\kms\ for all stars.
The mean FWHM over the set of synthetic lines was determined for each input \vbroad . 
This provided a relation between the input macroturbulence and the mean measured FWHM. Interpolation in this relation to the value of the FWHM measured in the observed spectrum provided the adopted \vbroad.
This value was used to broaden the synthetic grid input to \mygi\ (see Sect.\ref{mygi}). 
The values of \vbroad\ we obtained for each spectrum are listed in Table \ref{tab:obs_log}.
The stellar parameters of the synthetic spectra used for the broadening estimate are listed in Table \ref{table:grids}.
The list of lines used to determine the broadening was different for the blue arm  and the red arm of the UVES spectra, since the two arms often have different slit widths and therefore instrumental resolution. 
To perform this procedure we needed the lines to be on the linear part of the curve of growth. Since some lines that are non-saturated at low metallicities may become saturated at higher metallicities, for each setup we employed two line lists as a function of the stellar metallicity: one at $\rm [Fe/H] <-2.5$ and another at $\rm [Fe/H] \geq -2.5$.
The typical uncertainty on \vbroad\ is 0.3 \kms for stars with \vbroad\ $\sim 7$  \kms, and 2.0 \kms\ for stars with \vbroad\ $\sim 10$  \kms.

\begin{table}[ht]
\caption{Range of atmospheric parameters of the synthetic spectra used for the broadening estimates.  
}             
\label{table:grids}      
\centering                         
\begin{tabular}{r r r r r}        
\hline\hline                 
 \teff\ start & \teff\ end & \teff\ step & \logg & \vt \\ 
 K & K & K & dex & \kms \\ 
\hline                        
 4000 & 5200 & 200 & 1.5 & 2.0 \\       
 5200 & 5600 & 200 & 3.0 & 2.0 \\       
\hline                                   
\end{tabular}
\end{table}

\subsection{Chemical abundances}\label{mygi}
We derived chemical abundances of 
Na, Mg, Al, Si, Ca, Sc, Ti, V, Cr, Mn, Fe, Co, Ni, Cu, Zn, 
Y, and Zr,
for our sample stars using the code MyGIsFOS \citep{mygisfos}.
MyGIsFOS is an automatic pipeline that performs a $\chi^2$ minimisation fit on the profile of the selected lines using a grid of synthetic spectra. The grid has been computed with the code SYNTHE 
(see \citealt{Kurucz2005,Sbordone2004}) based on 1D plane-parallel model atmospheres in local thermodynamic equilibrium (LTE), computed with the code ATLAS\,12 \citep{Kurucz2005}.
Sr abundances were derived by matching the observed spectrum around Sr lines with a synthetic one computed using the LTE spectral line analysis code Turbospectrum \citep{Alvarez1998,Plez2012}.
The atomic data used in this study are provided by the {\it Gaia}-ESO Survey (GES) line list \citep[][and references therein] {GES2021} complemented with atomic data from two lists from Castelli's website\footnote{\url{https://wwwuser.oats.inaf.it/castelli/linelists.html}} that cover the wavelength range not covered by the GES list from 300 to 420 nm. 
We were also able to detect the \ion{Si}{II} line at 385.6\,nm in 22 stars, the \ion{Sc}{I} line at 391.1\,nm in 19 stars, the \ion{Mn}{II} line at 412.8\,nm in 38 stars, and the \ion{Zr}{I} line at 473.9 nm in seven stars. 

The derived chemical abundances with uncertainties are provided in machine readable format at the Centre de Donn\'ees astronomiques de Strasbourg (CDS). 
An example of the table provided at the CDS is shown in Table \ref{tab:abu}. 
Chemical abundances are expressed in the form A(X) and [X/H], where $A(X) = \log(X/H) + 12 $, and $[X/H] = \log_{10}(X/H) - \log_{10}(X/H)_\odot$.
The abundance ratios [X/Fe] are expressed as $\mathrm{[\ion{X}{I}/\ion{Fe}{I}] = [\ion{X}{I}/H] - [\ion{Fe}{I}/H]}$ for neutral species and as $\mathrm{[\ion{X}{II}/\ion{Fe}{II}] = [\ion{X}{II}/H] - [\ion{Fe}{II}/H]}$ for singly ionised species. 
The uncertainties s(X) represent the line-to-line scatter when the number of lines is $\geq 2$. When the abundance was derived from only one line, we adopted as s(X) the mean line-to-line scatter over the stars with $\geq 2$ lines of the same element X. 
For \ion{Si}{II}, \ion{Sc}{I}, \ion{Mn}{II}, and \ion{Zr}{I} abundances, we adopted as s(X) the mean line-to-line scatter of the other ionisation state. 
We adopted the solar abundances provided by \citet{CaffauZr,Caffau2011} and \citet{Lodders2009} (see Table \ref{solar}).  

\begin{table*}[ht]
\centering
 \caption[]{\label{tab:abu}Derived chemical abundances with errors for our sample stars.}
\begin{tabular}{lrcccrcccc}
 \hline \hline
 Star            & Nlines &  A(FeI) & s(FeI) & [FeI/H] & Nlines & A(FeII) & s(FeII) & [FeII/H] & \ldots \\
                 &             &  dex    & dex    &         &             & dex    & dex    &         &     \\
\hline
  CES0031-1647 & 304 & 5.03 & 0.12 & -2.49 & 26 & 5.22 & 0.09 & -2.31 & \ldots  \\
  CES0045-0932 & 176 & 4.57 & 0.14 & -2.95 & 12 & 4.72 & 0.19 & -2.80 & \ldots  \\
  CES0048-1041 & 314 & 5.04 & 0.13 & -2.48 & 26 & 5.19 & 0.13 & -2.33 & \ldots  \\
  CES0055-3345 & 340 & 5.16 & 0.11 & -2.36 & 28 & 5.28 & 0.12 & -2.24 & \ldots  \\
  CES0059-4524 & 146 & 5.13 & 0.09 & -2.39 &  8 & 5.26 & 0.09 & -2.26 & \ldots  \\
  ... &  &  &  &  &  &  &  &  &  \\
\hline
\end{tabular}
\tablefoot{The complete table is available in machine readable format at the CDS. }
\end{table*}

\begin{table}[ht]
\caption{Solar abundance values adopted in this work. } 
\label{solar}
\centering
\begin{tabular}{l c c }
\hline\hline
Element & A(X) & References \\     
\hline
Na & 6.30 & \citet{Lodders2009} \\ 
Mg & 7.54 & \citet{Lodders2009} \\ 
Al & 6.47 & \citet{Lodders2009} \\ 
Si & 7.52 & \citet{Lodders2009} \\ 
Ca & 6.33 & \citet{Lodders2009} \\ 
Sc & 3.10 & \citet{Lodders2009} \\ 
Ti & 4.90 & \citet{Lodders2009} \\ 
V  & 4.00 & \citet{Lodders2009} \\ 
Cr & 5.64 & \citet{Lodders2009} \\ 
Mn & 5.37 & \citet{Lodders2009} \\ 
Fe & 7.52 & \citet{Caffau2011} \\ 
Co & 4.92 & \citet{Lodders2009} \\ 
Ni & 6.23 & \citet{Lodders2009} \\ 
Cu & 4.21 & \citet{Lodders2009} \\ 
Zn & 4.62 & \citet{Lodders2009} \\ 
Sr & 2.92 & \citet{Lodders2009} \\
Y  & 2.21 & \citet{Lodders2009} \\
Zr & 2.62 & \citet{CaffauZr} \\
\hline
\end{tabular}
\end{table}

\section{Results}
\subsection{Comparison with literature}
   \begin{figure*}
   \centering
   \includegraphics[width=9.1 cm]{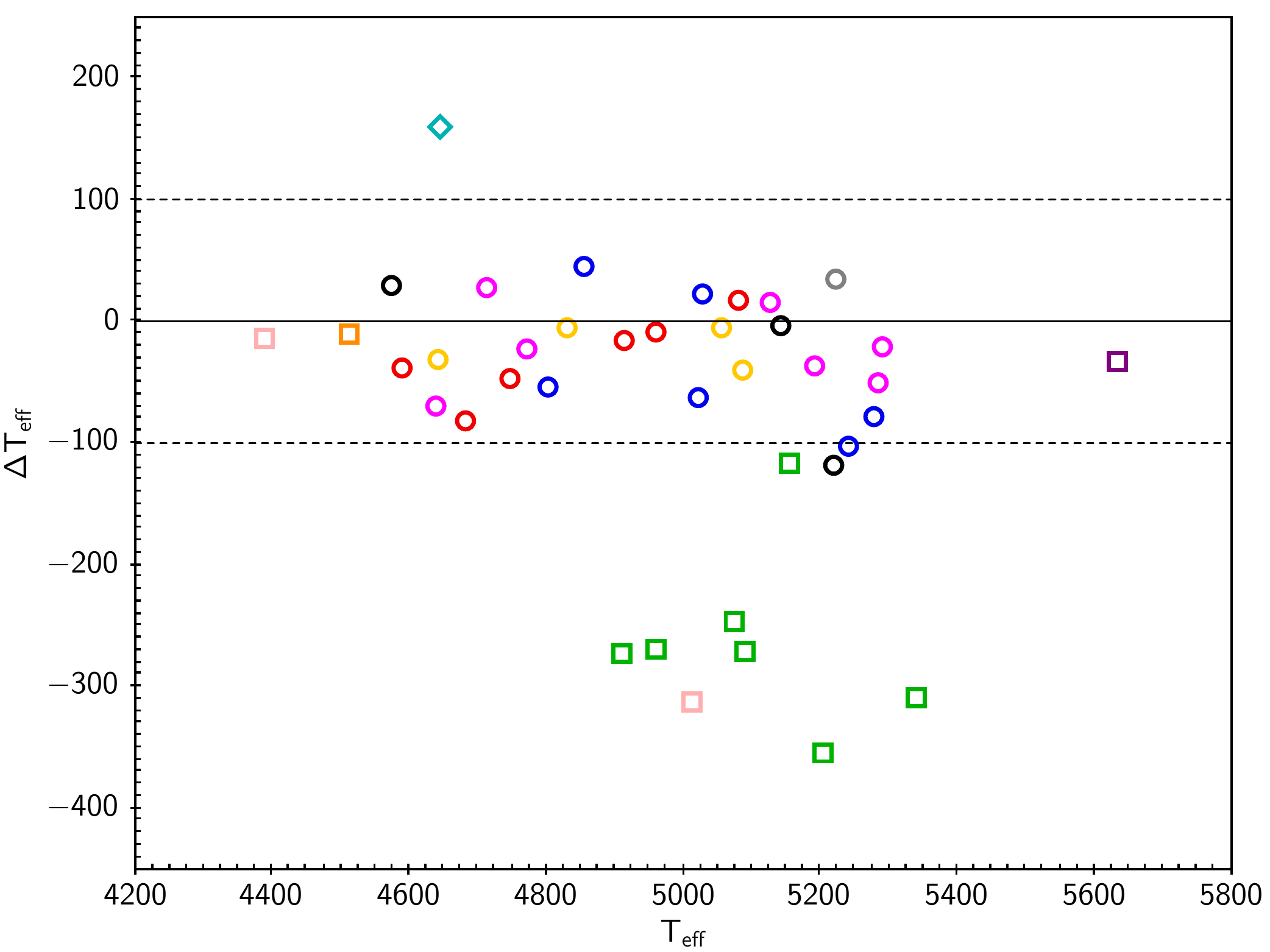}
   \includegraphics[width=9.1 cm]{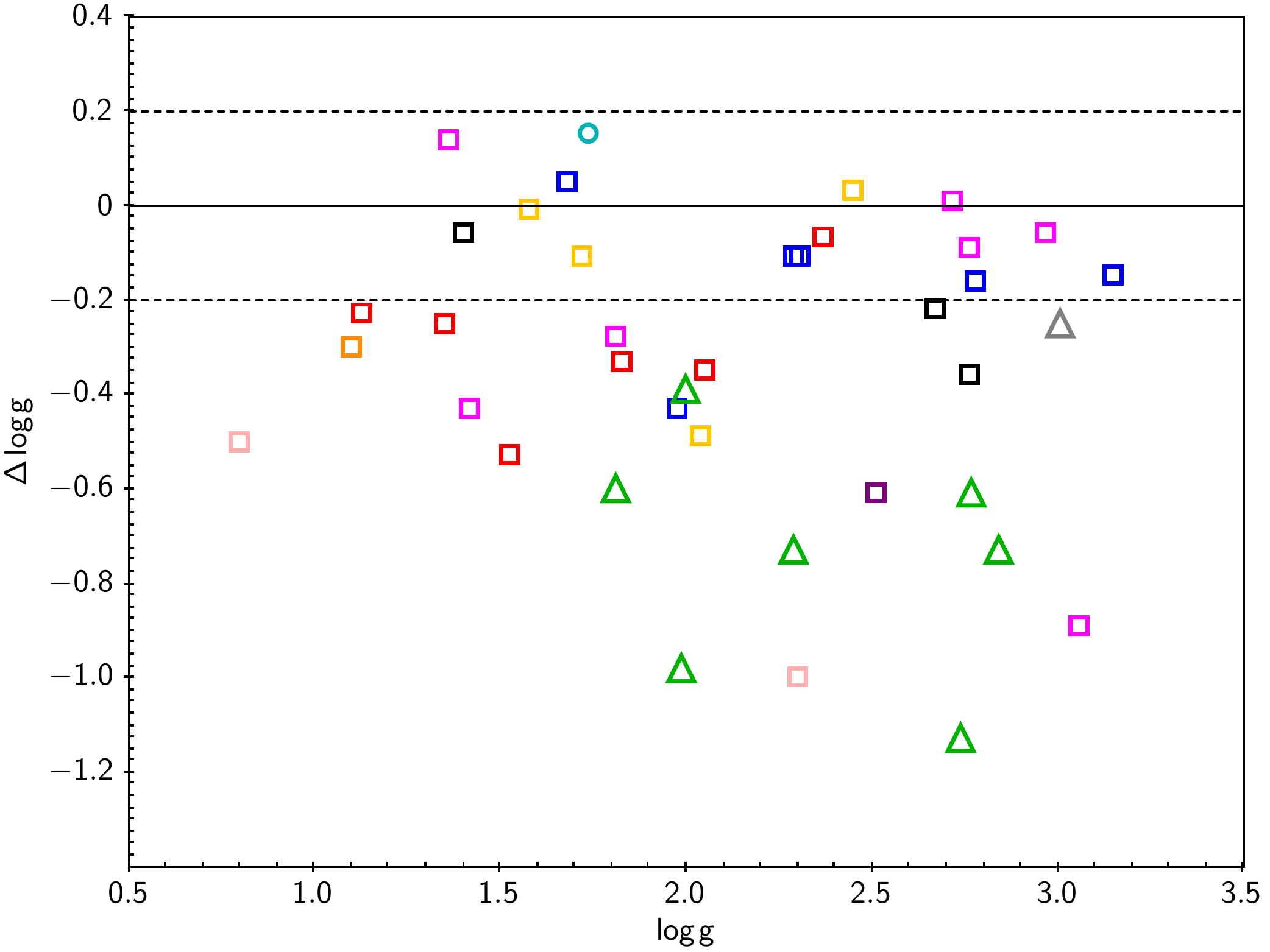}
   \includegraphics[width=9.1 cm]{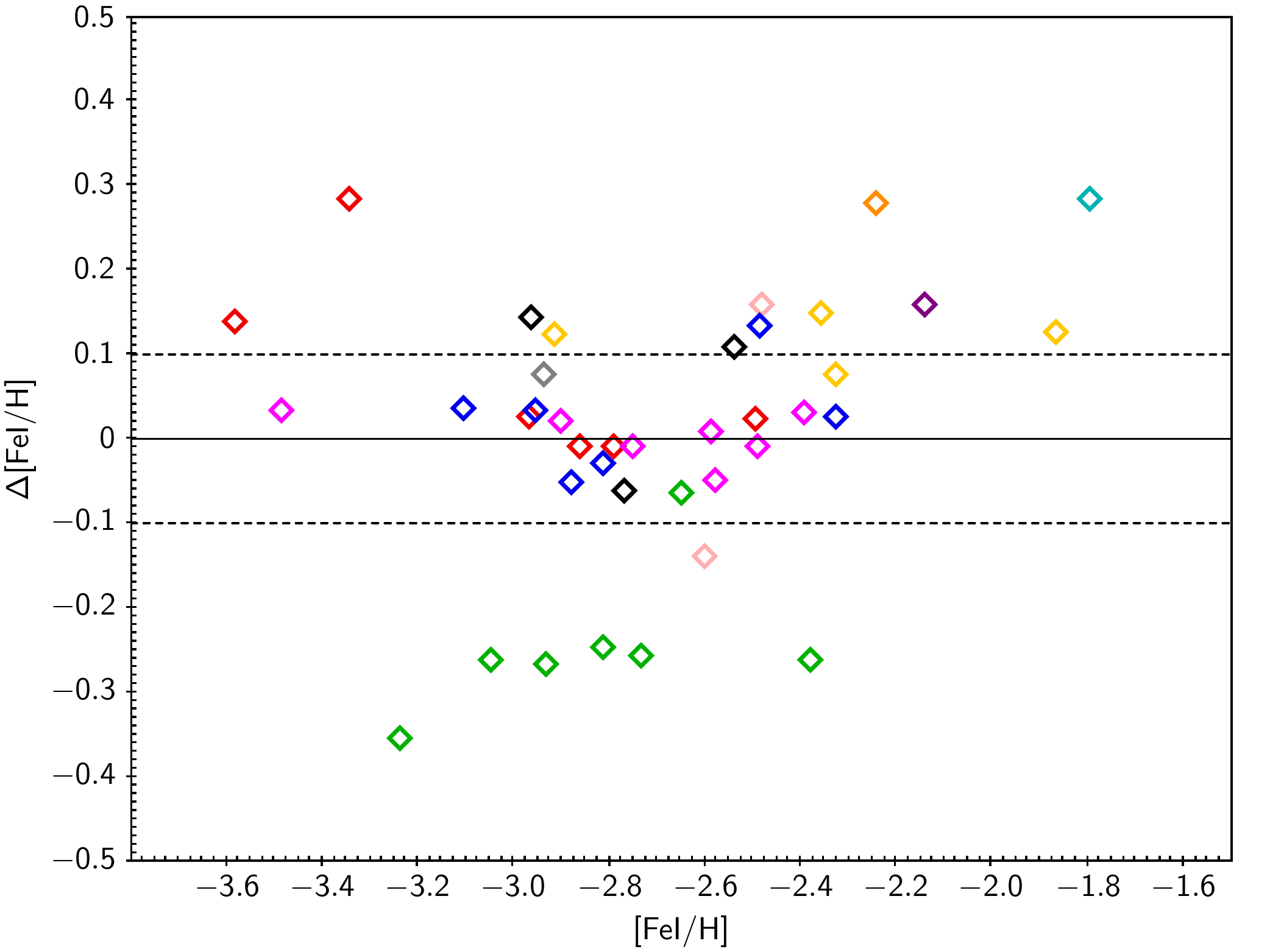}
   \includegraphics[width=9.1 cm]{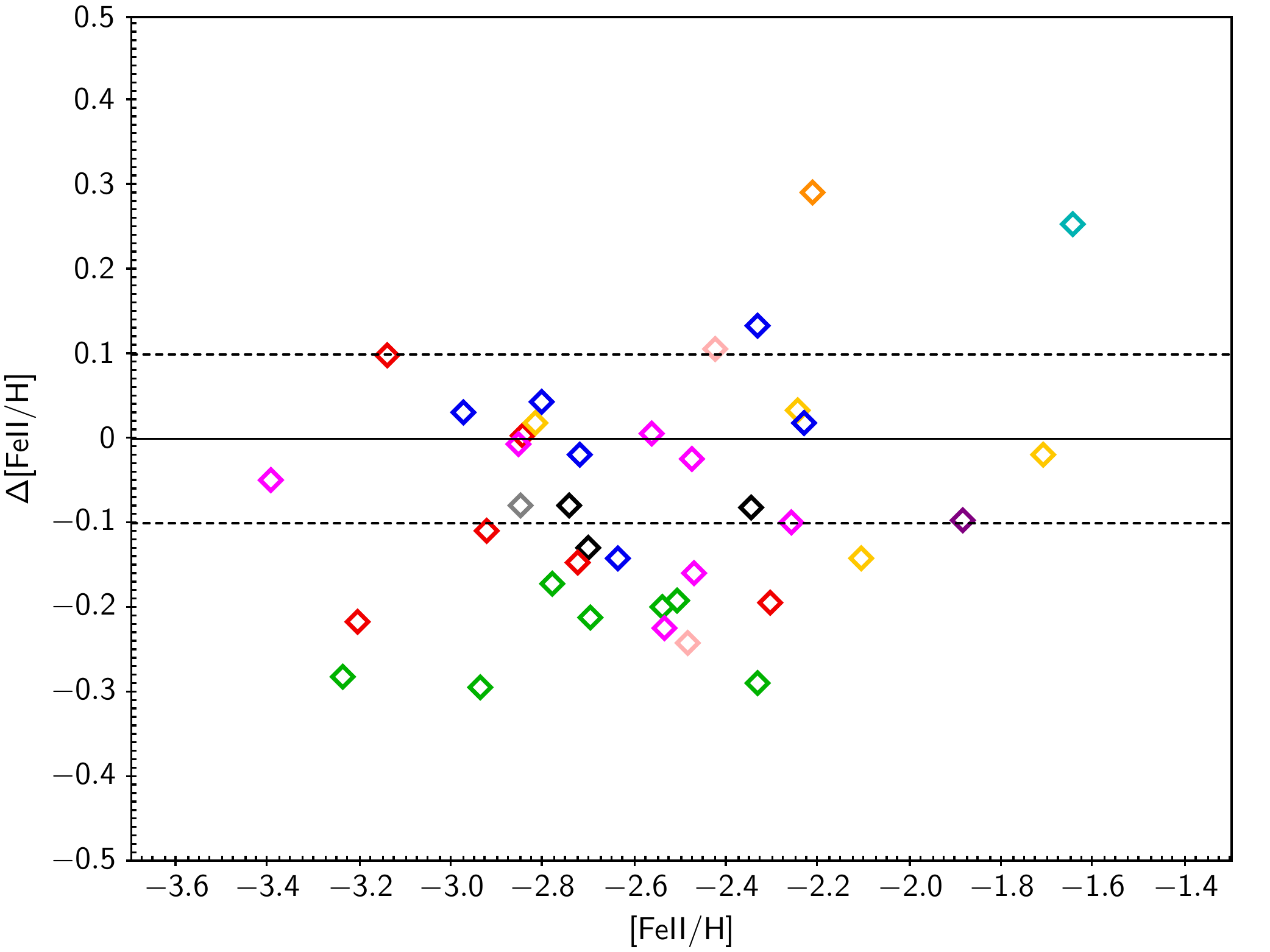}
      \caption{Comparison between this work and the literature: \teff\ (upper-left panel), \logg\ (upper-right panel), \ion{Fe}{i} (lower-left panel), 
and \ion{Fe}{ii} (lower-right panel). Open circles denote \teff\ and \logg\ derived from photometry, open squares denote the same derived from spectroscopy, and open triangles indicate \logg\ derived from theoretical isochrones. Data are from \citet{Barklem2005} (black), \citet{Cayrel2004} (red), \citet{Hansen2020} (cyan), \citet{Ishigaki2012} (yellow), \citet{Johnson2002} (pink), \citet{Lai2008} (grey), \citet{LuckBond1985} (orange), \citet{Mashonkina2017dec} (blue), \citet{Pereira2013} (purple), \citet{Roederer2014} (green), and \citet{Siqueira2014} (magenta). }
        \label{teff_logg}
   \end{figure*}
In Fig. \ref{teff_logg} we compare our derived stellar parameters with those determined by other studies in the literature. 
Some stars were observed in the framework of the Large Program `First Stars'  (PI: R. Cayrel), and for these stars we use the results obtained in \citet{Cayrel2004} for the comparison, even if more recent values are available in literature. 

In this study, we rely on {\it Gaia} EDR3 photometry and parallaxes to derive the stellar parameters (Sect.\,\ref{sp}).
However, in the traditional spectroscopic method, the effective temperature is obtained by requiring that there is no trend between the abundance and excitation potential of \ion{Fe}{i} lines (i.e. excitation equilibrium). The surface gravity, instead, is obtained by requiring that \ion{Fe}{i} and \ion{Fe}{ii} lines provide the same abundance (i.e. ionisation equilibrium). Another way of deriving surface gravity is to use theoretical isochrones, assuming the age, metallicity, and effective temperature of the star.

As shown in Fig. \ref{teff_logg} (upper-left panel), \teff\ values in the literature appear in line with our derived temperatures, with an average $\Delta$\teff=--72\,K ($\sigma$=113\,K), where $\Delta$\teff=\teff(literature)-\teff(this study) is compatible with our $\pm$100 K uncertainties on \teff. 
One exception to this good agreement is \citet{Roederer2014}, where six out of seven stars have effective temperatures $\sim$300\,K lower than our derived values. For the star CES1552+0517, \citet{Johnson2002} also determined a \teff\ $\sim$300\,K lower than our derived value. 
As discussed in \citet{MucciarelliBonifacio2020}, for metal-poor stars with [Fe/H]$\sim-2.5$, the spectroscopic \teff\ and \logg\ appear to be lower than the photometric ones by $\sim$350 K and $\sim$1.0 dex, respectively. This is in agreement with the observed discrepancy. 

Surface gravities in the literature appear lower than our derived values (see Fig. \ref{teff_logg}, upper-right panel), with average $\Delta$\logg=$-0.35$ ($\sigma$=0.32). The largest discrepancy in \logg\ is found for the \citet{Roederer2014} values, which are systematically $\sim-0.75$ dex lower than our derived \logg. This result is not surprising, given their discrepancy with our \teff. 
Since \citet{Roederer2014} determined their \logg\ using theoretical isochrones, we expect that along the red giant branch (RGB) the cooler the star, the lower is the surface gravity. 

We also determined spectroscopic surface gravities for our sample stars and found that those are in agreement with literature values within 0.1 dex. The observed discrepancy between spectroscopic and photometric \logg\ seems to arise from the different iron abundance obtained from \ion{Fe}{i} and \ion{Fe}{ii} lines. At low metallicities, the neutral species are more affected by non-local thermodynamic equilibrium (NLTE) effects than ionised ones \citep[e.g.][]{Amarsi16}, 
which implies that the total abundance derived from each of the two species are different. Hence, when using ionisation equilibrium to derive \logg, these over-ionisation effects can lead to an underestimation of the star's surface gravity. A possible way to avoid this is to apply NLTE corrections before imposing ionisation equilibrium, as  done in \citet{Mashonkina2017aug}, 
whose results are in good agreement with this work (blue squares in Fig. \ref{teff_logg}, upper-right panel). 

In Fig. \ref{teff_logg} (lower panels) we compare our derived chemical abundances for \ion{Fe}{i} and \ion{Fe}{ii} with those in literature. Our results are in general agreement with the literature abundances, with average values $\Delta$[\ion{Fe}{I}/H]=+0.01 ($\sigma=0.15$ dex) and $\Delta$[\ion{Fe}{II}/H]=$-0.08$ ($\sigma=0.14$ dex). We note that the [Fe/H] derived by \citet{Roederer2014} are systematically lower than our values, which is a direct consequence of adopting lower \teff\ and \logg.

\subsection{Alpha elements: Mg, Si, Ca, and Ti}

   \begin{figure*}
   \centering
   \includegraphics[width=9.1 cm]{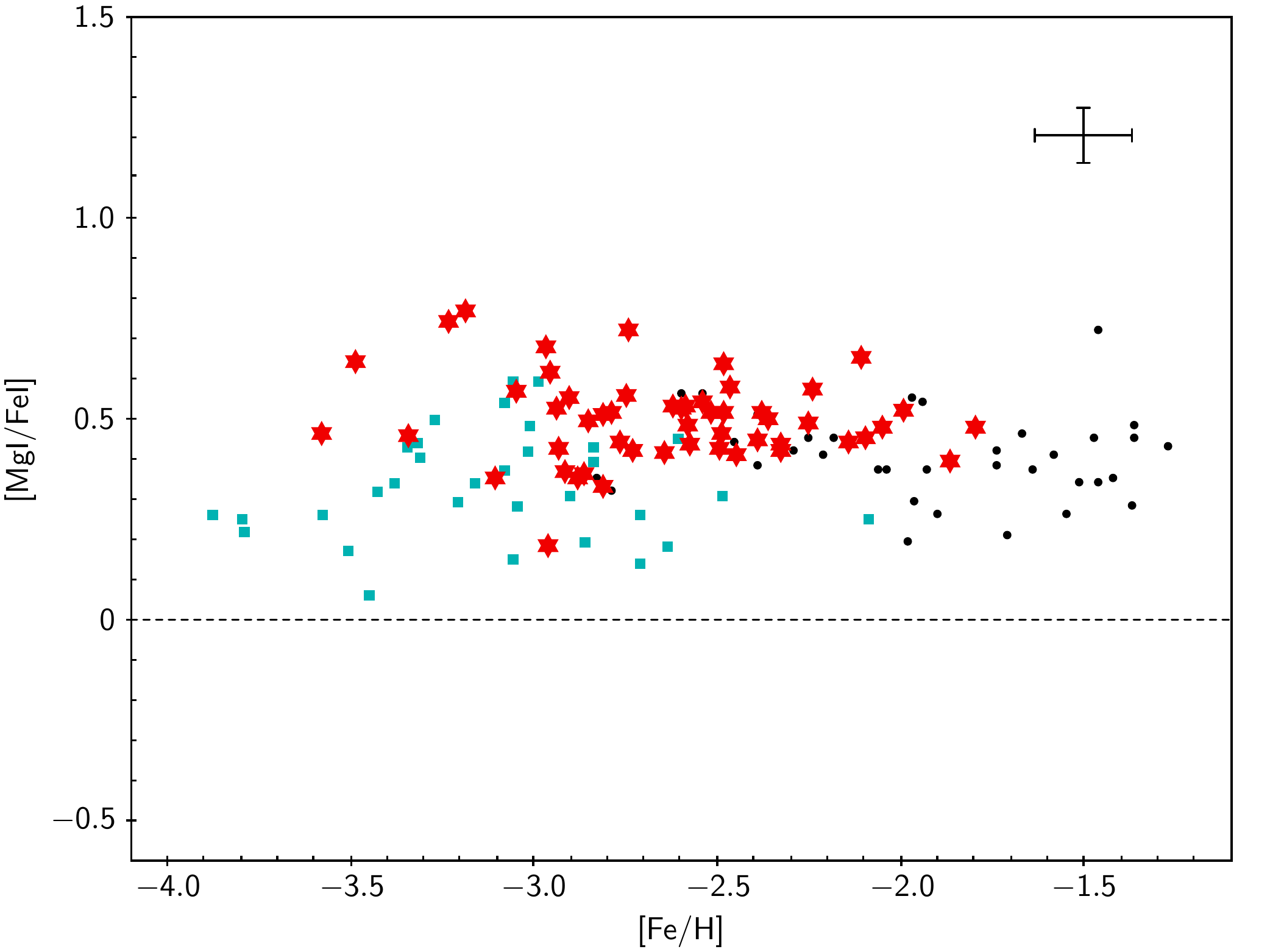}
   \includegraphics[width=9.1 cm]{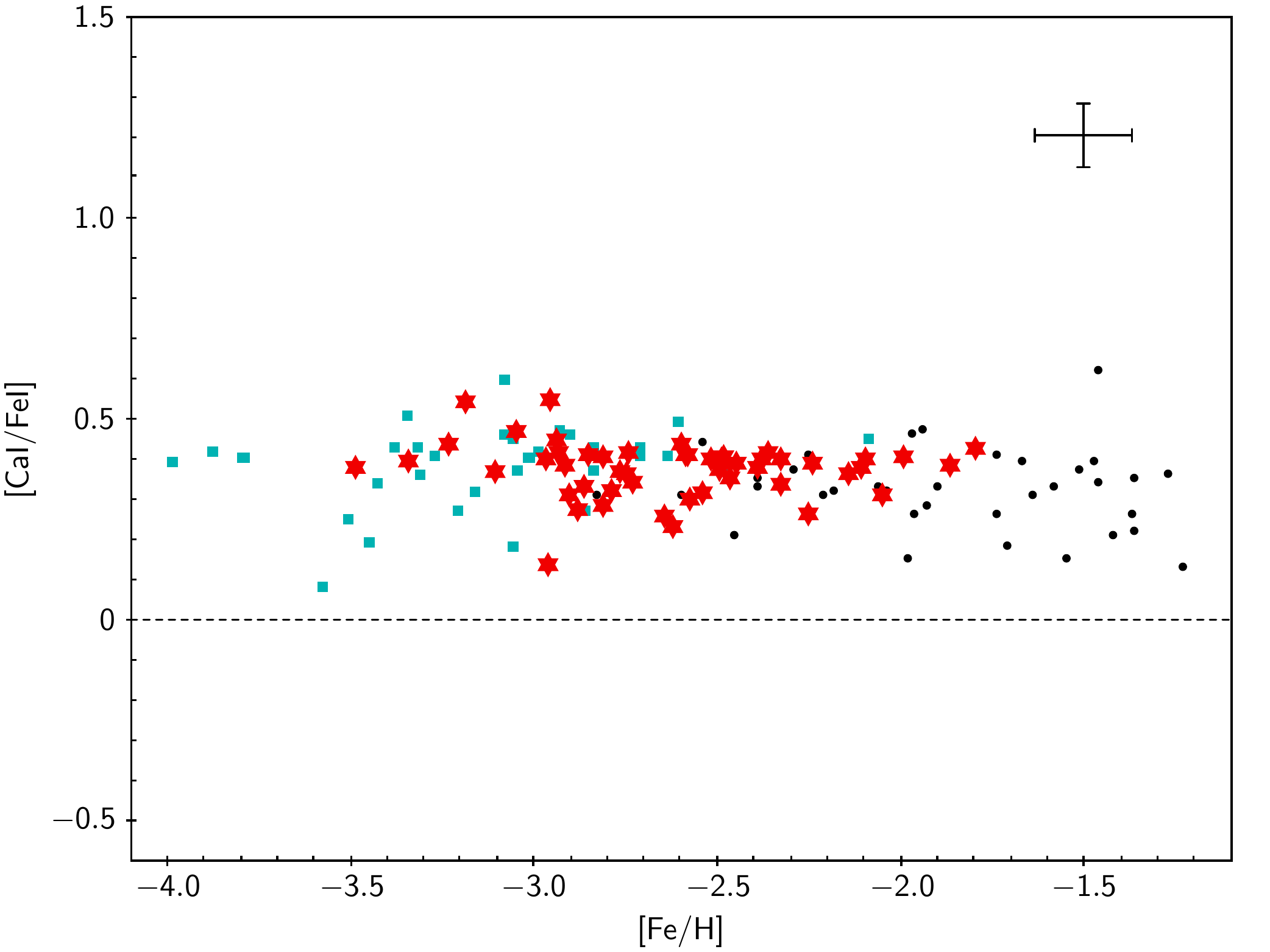}
   \includegraphics[width=9.1 cm]{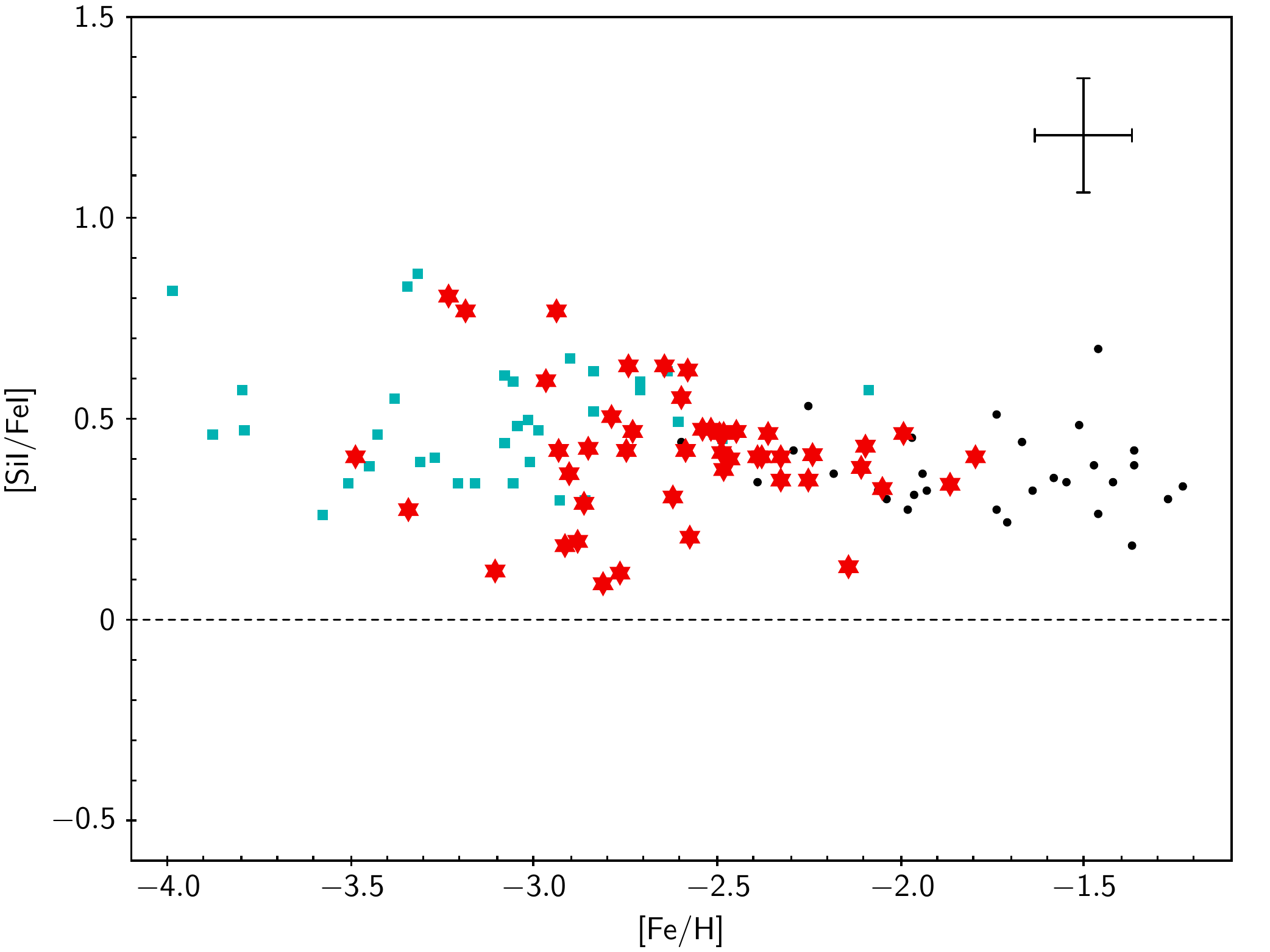}
   \includegraphics[width=9.1 cm]{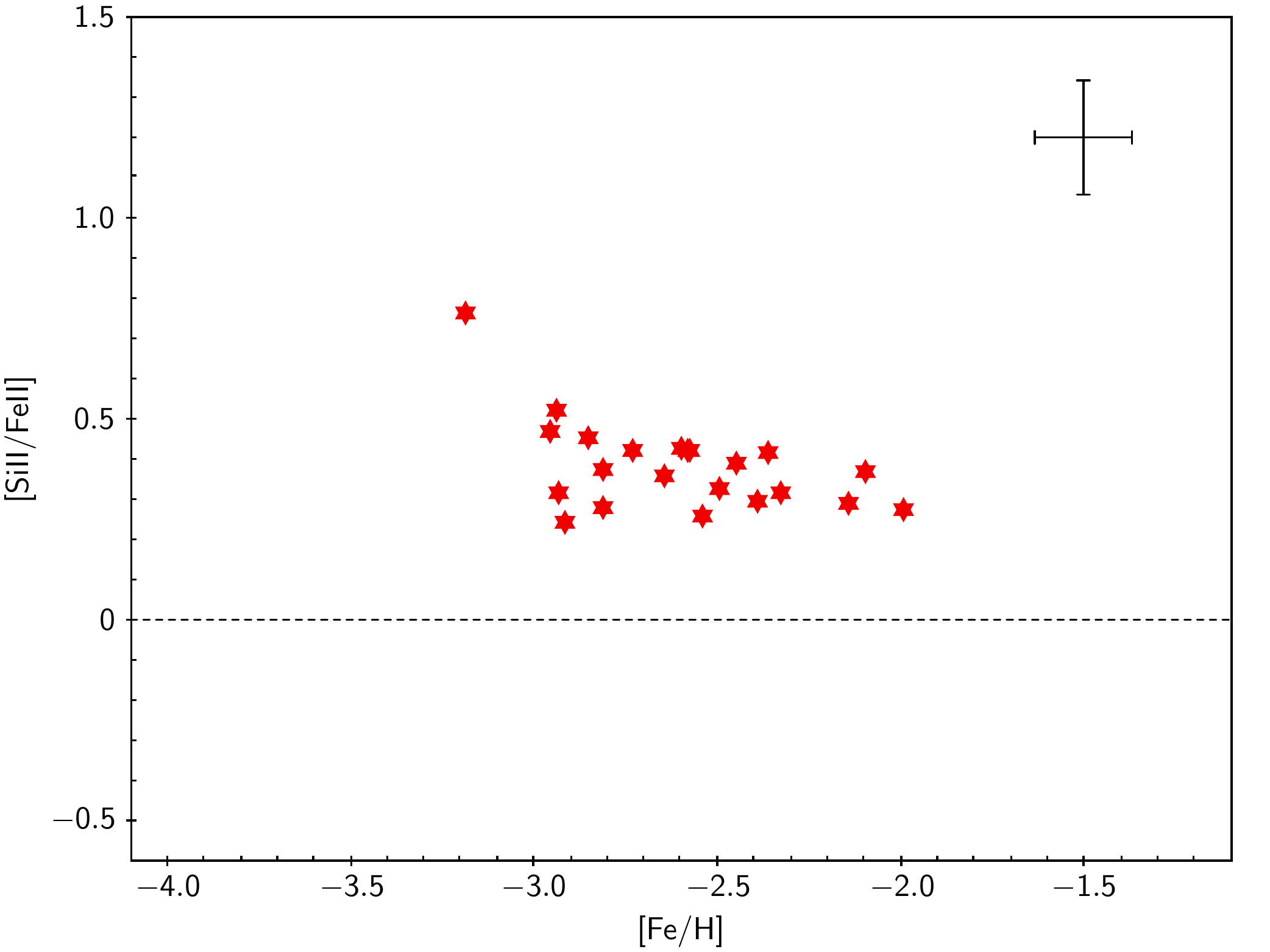}
   \includegraphics[width=9.1 cm]{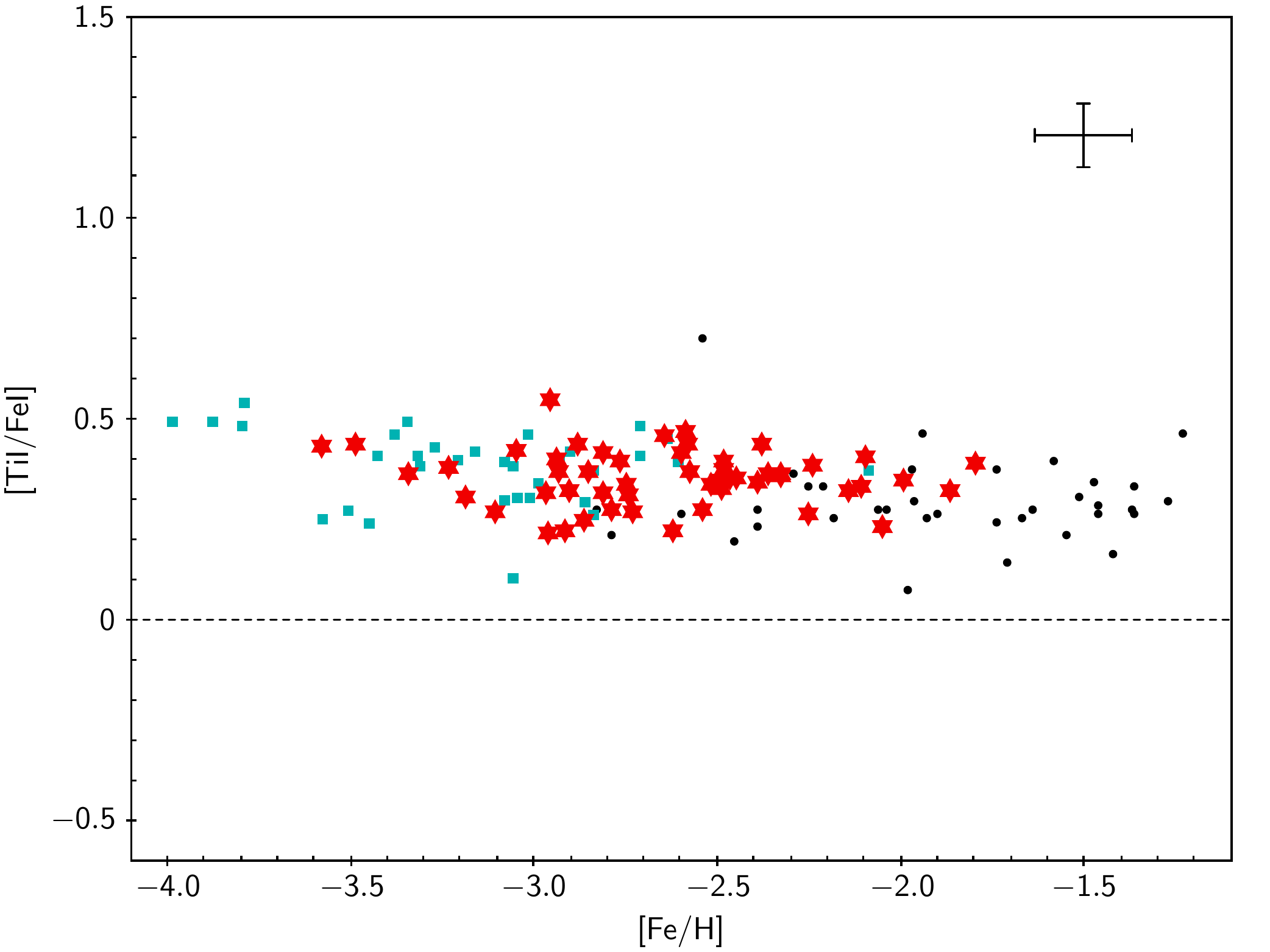}
   \includegraphics[width=9.1 cm]{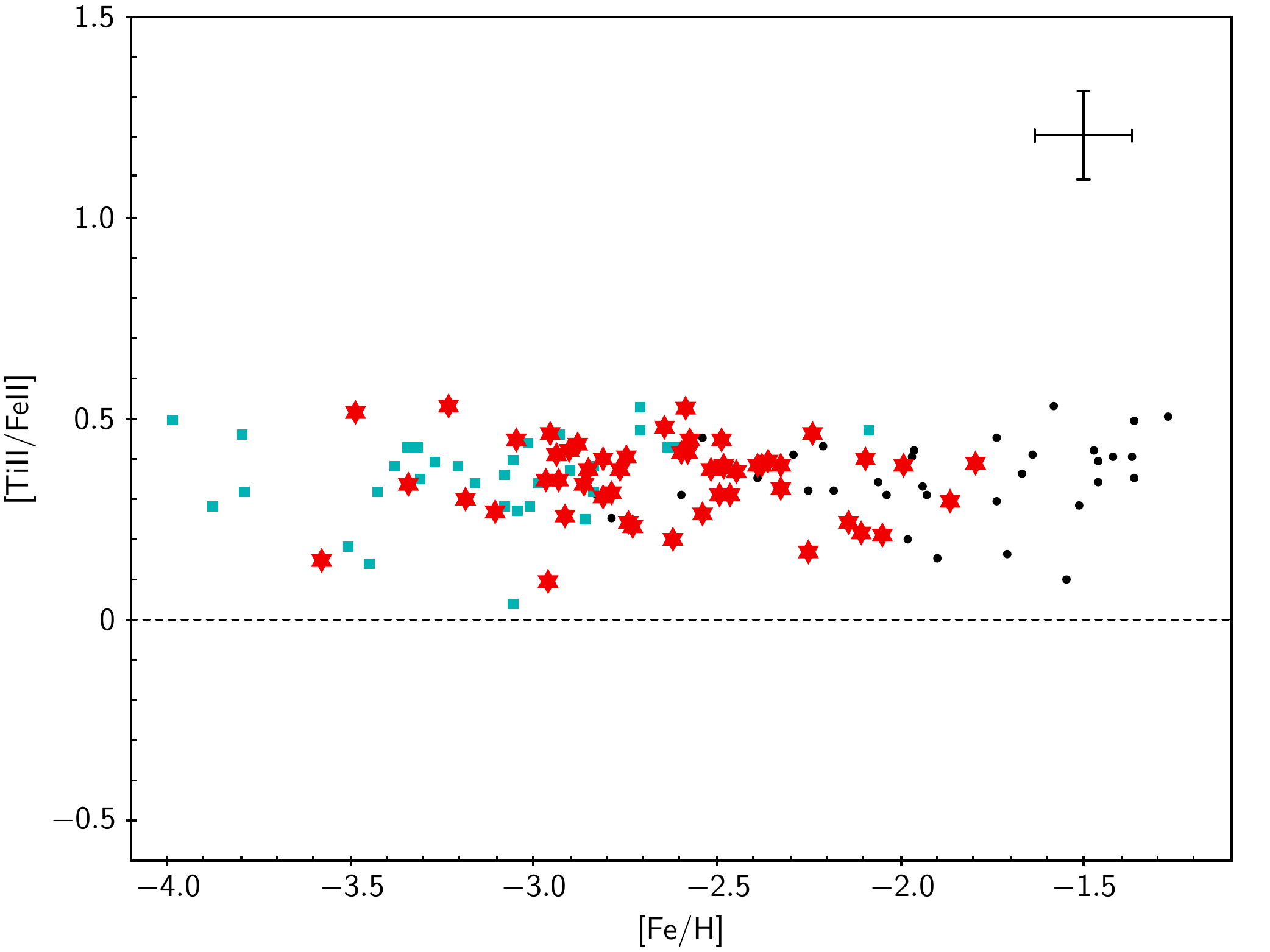}
    \caption{Elemental abundance ratios of \ion{Mg}{I}, \ion{Si}{I}, \ion{Si}{II}, \ion{Ca}{I}, \ion{Ti}{I}, and \ion{Ti}{II} 
      as a function of [Fe/H] for stars in our sample (red star symbols). 
           Cyan squares and black dots represent the same quantities for stars in \citet{Cayrel2004} and \citet{Ishigaki2012}, respectively. A representative error is plotted in the upper-right corner of each panel. }
         \label{alpha_fe}
   \end{figure*}

Figure \ref{alpha_fe} shows the derived \ion{Mg}{I}, \ion{Si}{I}, \ion{Si}{II}, \ion{Ca}{I}, \ion{Ti}{I}, and \ion{Ti}{II} over Fe abundance 
ratios as a function of [Fe/H] for our sample of stars. Our results are compared to the values obtained by \citet{Cayrel2004} and \citet{Ishigaki2012} for giant stars in the same metallicity range. The abundance ratios of all elements are in good agreement with the results found in previous studies. All elements are enhanced respect to the Fe abundance (with sample averages and standard deviations of [\ion{Mg}{I}/\ion{Fe}{I}]\,= +0.50\,$\pm$\,0.11,
[\ion{Si}{I}/\ion{Fe}{I}]\,= +0.41\,$\pm$\,0.16,
[\ion{Si}{II}/\ion{Fe}{II}]\,= +0.38\,$\pm$\,0.11,
[\ion{Ca}{I}/\ion{Fe}{I}]\,= +0.37\,$\pm$\,0.07,
[\ion{Ti}{I}/\ion{Fe}{I}]\,= +0.35\,$\pm$\,0.07,
[\ion{Ti}{II}/\ion{Fe}{II}]\,= +0.35\,$\pm$\,0.10), and the abundance ratios remain constant at different metallicities. 
The dispersion around the mean abundance ratio is equal or smaller than the mean uncertainty ($\sigma$) for Si, Ca and Ti ($\sigma_{\ion{Si}{I}}=0.16$ dex, $\sigma_{\ion{Si}{II}}=0.20$ dex, $\sigma_{\ion{Ca}{I}}=0.08$ dex, $\sigma_{\ion{Ti}{I}}=0.09$ dex, $\sigma_{\ion{Ti}{II}}=0.11$ dex) but slightly larger for Mg ($\sigma_{\ion{Mg}{I}}=0.07$ dex). The scatter appears to become larger at lower metallicities, as it was already observed by \citet{Cayrel2004} in their sample of stars, this is expected since at lower metallicities the lines become weaker. 
The error estimates are in line with what expected from the S/Ns.

We find a mean difference between [\ion{Si}{II}/H] and [\ion{Si}{I}/H] of $0.06\pm0.14$ dex, and a mean difference between [\ion{Ti}{II}/H] and [\ion{Ti}{I}/H] of $0.13\pm0.06$ dex. 
These differences are likely due to NLTE effects.
\citet{2016AstL...42..366M} found minimal departures from LTE for the lower levels of \ion{Si}{I} lines typically used as abundance indicators in F-G-K stars.
On the other hand, NLTE effects are particularly strong for \ion{Ti}{I} lines \citep[see][]{Mashonkina2016}. According to \citet{Mashonkina2016}, for stars with stellar parameters similar to those of our targets, the NLTE corrections for \ion{Ti}{I} are all positive, up to 0.4 dex, while the corrections for \ion{Ti}{II} are positive and $<0.1$ dex.
These values are compatible with the difference in Ti abundances observed in our stars. 

\subsection{Light odd-Z elements: Na and Al}

   \begin{figure*}
   \centering
   \includegraphics[width=9.1 cm]{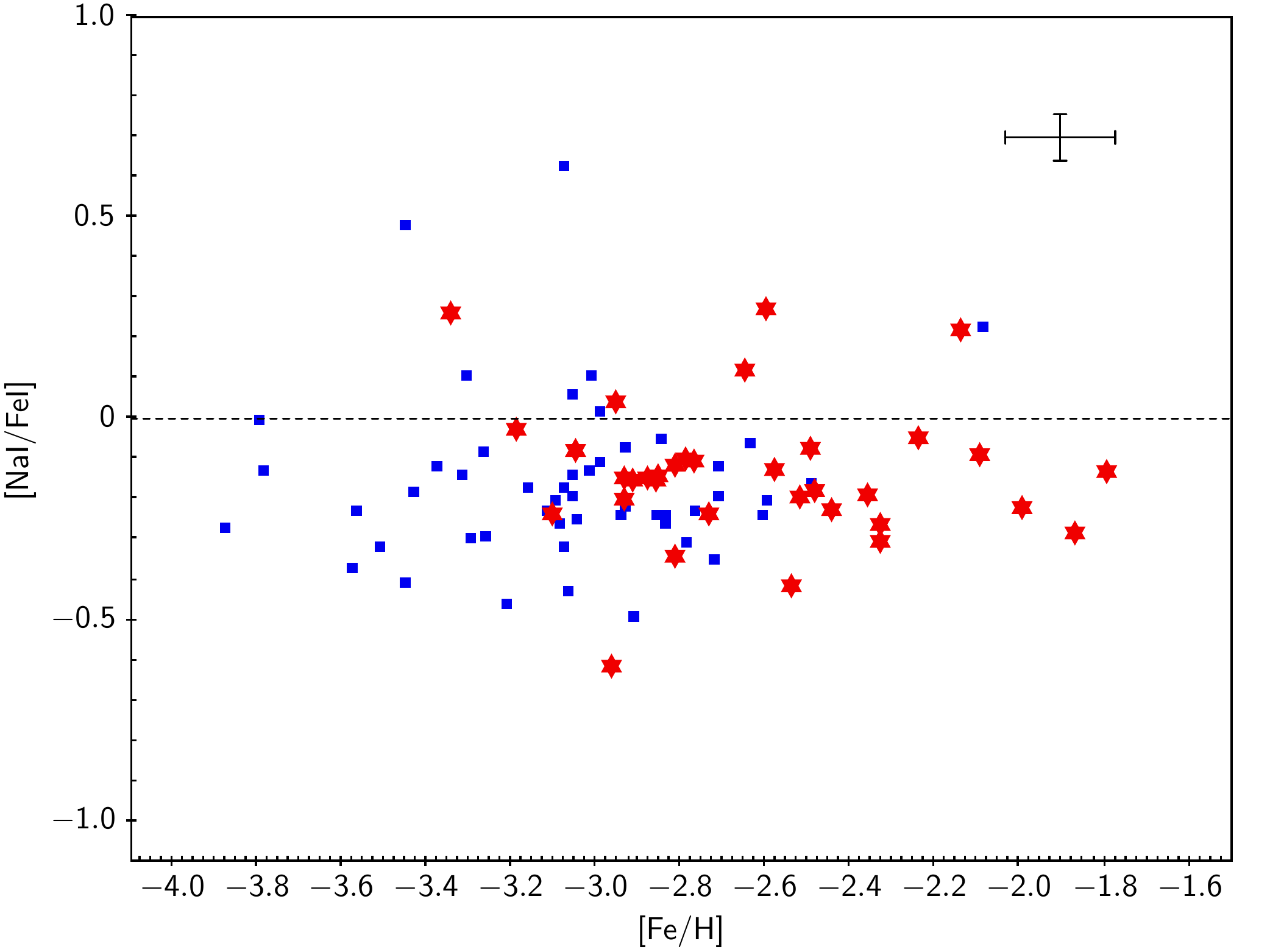}
   \includegraphics[width=9.1 cm]{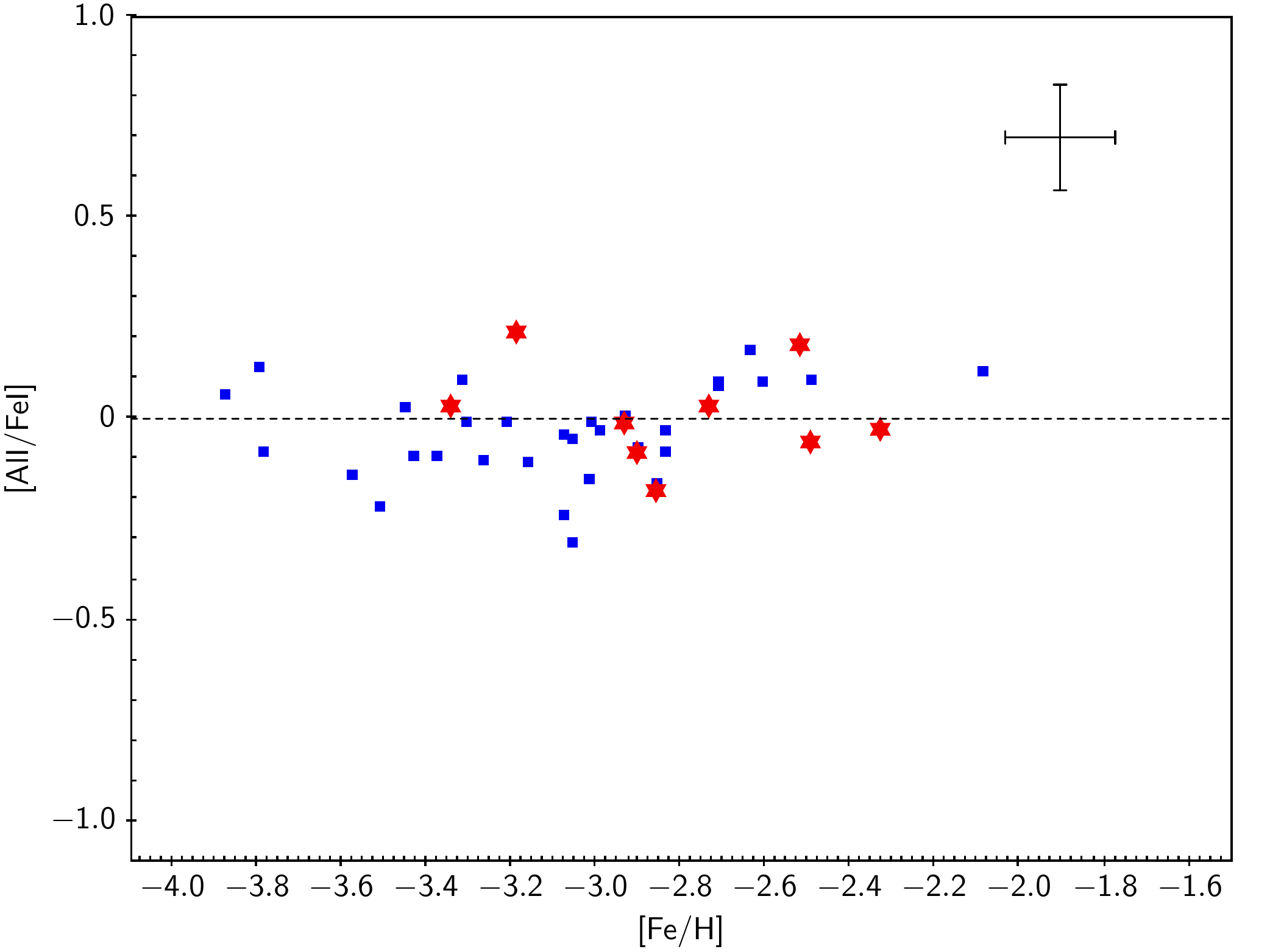}
      \caption{NLTE abundance ratios of Na and Al as a function of [Fe/H] for stars in our sample (red star symbols). Blue squares are stars from \citet{Andrievsky2007,Andrievsky2008} for Na and Al, respectively. A representative error is plotted in the upper-right corner.}
         \label{odd_fe}
   \end{figure*}

Sodium abundances were derived from the \ion{Na}{I} 
lines at 498.3\,nm, 568.2\,nm, 568.8\,nm, and 616.0\,nm and from the Na D resonance lines at 588.9\,nm (D1) and 589.5\,nm (D2). As \ion{Na}{i} lines, especially the \ion{Na}{i} D doublet, are known for being strongly sensitive to NLTE effects \citep[e.g.][]{Mashonkina1993,Baumueller1998}, we applied the NLTE corrections provided by \citet{Lind2011}\footnote{\url{http://www.inspect-stars.com}} to each line to obtain a more accurate measurement 
of Na abundances. The NLTE corrections for \ion{Na}{i} lines are all negative, with an average of --0.26 dex and down to --0.57 dex for the \ion{Na}{i} D doublet. 

Aluminium abundances were derived from the \ion{Al}{I} resonance lines at 394.4\,nm and 396.1\,nm. Similarly to the Na D doublet, the Al resonance doublet is sensitive to NLTE effects, and when these lines are used in LTE approximation, the derived Al abundances are 
severely underestimated \citep{BaumuellerGehren1997,Norris2001}. To avoid this, we applied the NLTE corrections by \citet{Andrievsky2008}. 
Nine stars in our sample have \teff\ in the temperature range of the \citeauthor{Andrievsky2008} grid, so we provide NLTE Al abundances only for these stars. The NLTE corrections are all positive, with an average of +0.63 dex.

The derived abundance ratios of Na and Al over Fe as a function of [Fe/H] are shown in Fig. \ref{odd_fe}. 
These are compared to the NLTE values obtained for the First Stars Large Program stars in \citet{Andrievsky2007} and \citet{Andrievsky2008}, respectively. 
Again, our results appear in line with previous studies. The abundance ratios show a large scatter  (0.17 dex for \ion{Na}{I}, and 0.12 dex for \ion{Al}{I}), and no clear trend with [Fe/H]. 

\subsection{Iron-peak elements}
\subsubsection{Sc and V}

\begin{figure*}
   \centering
   \includegraphics[width=9.1 cm]{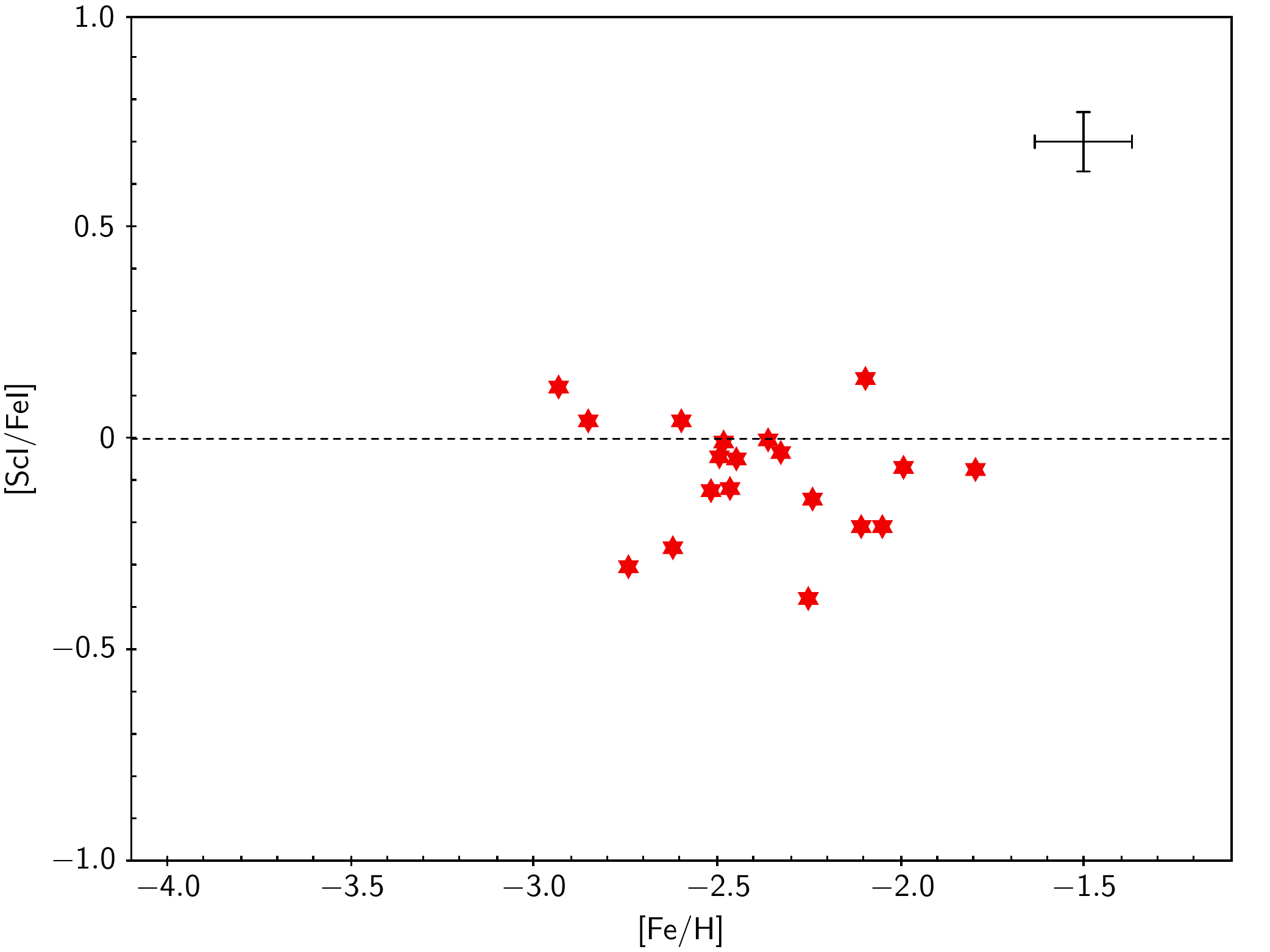}
   \includegraphics[width=9.1 cm]{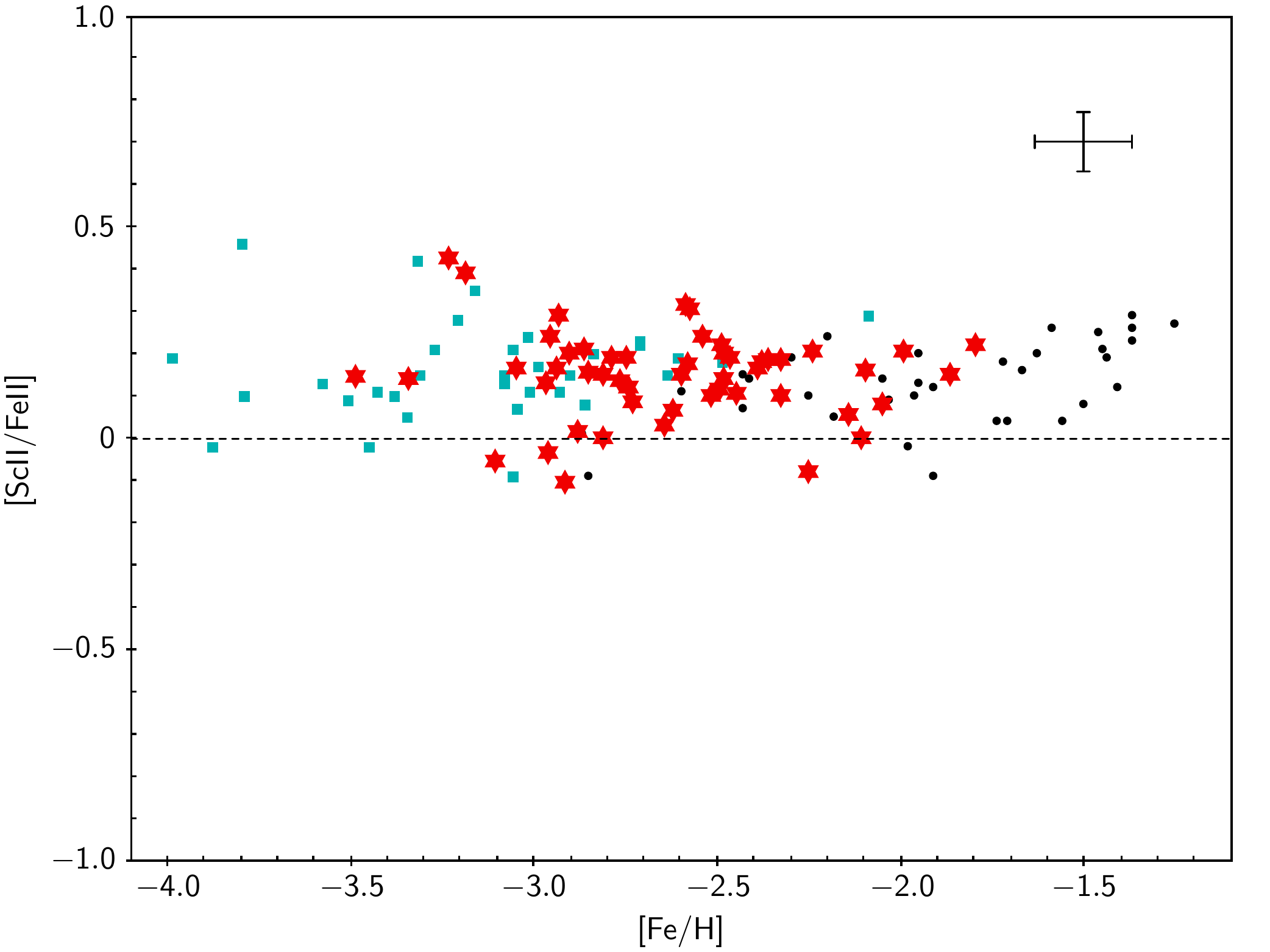}
   \includegraphics[width=9.1 cm]{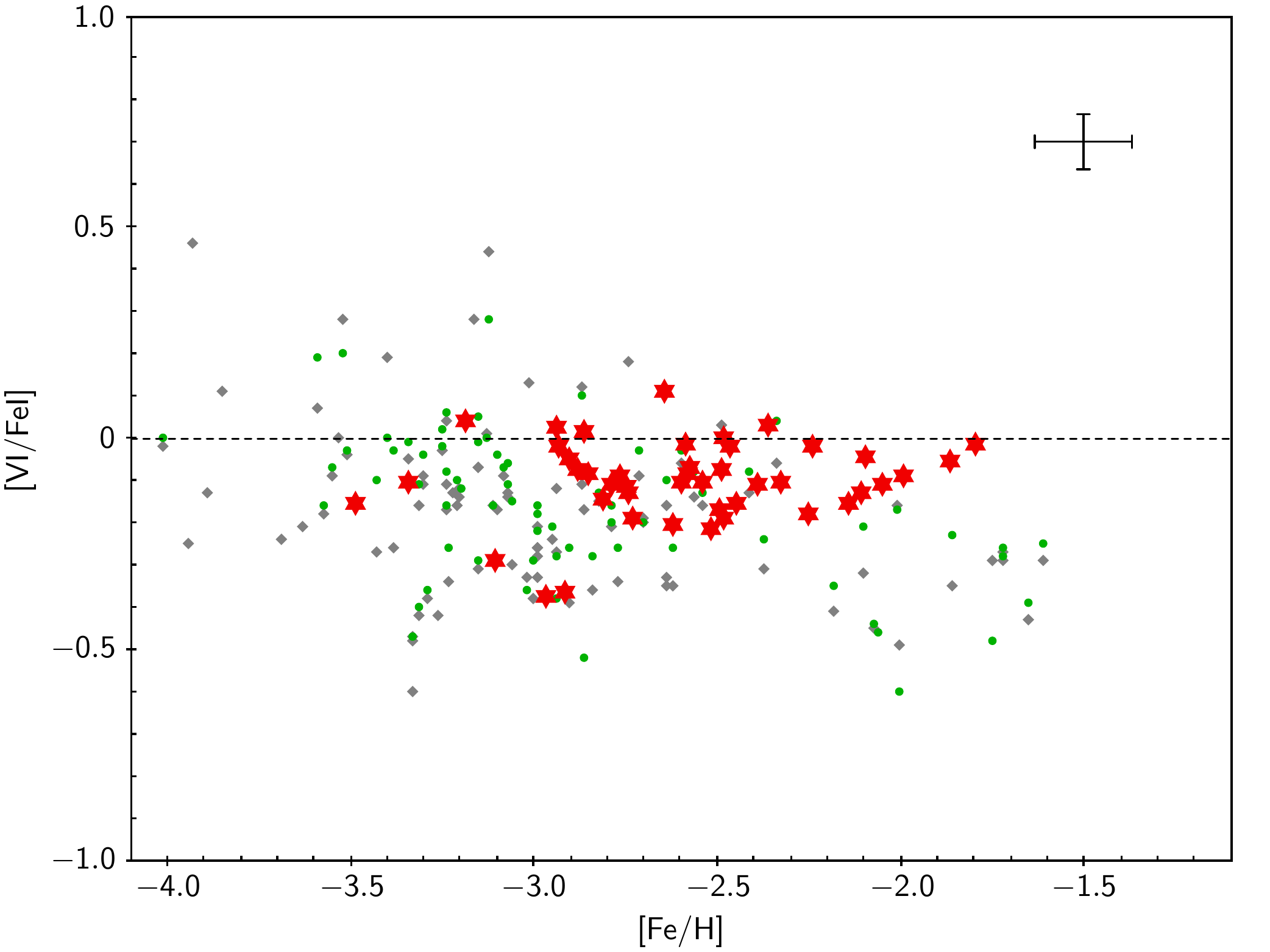}
   \includegraphics[width=9.1 cm]{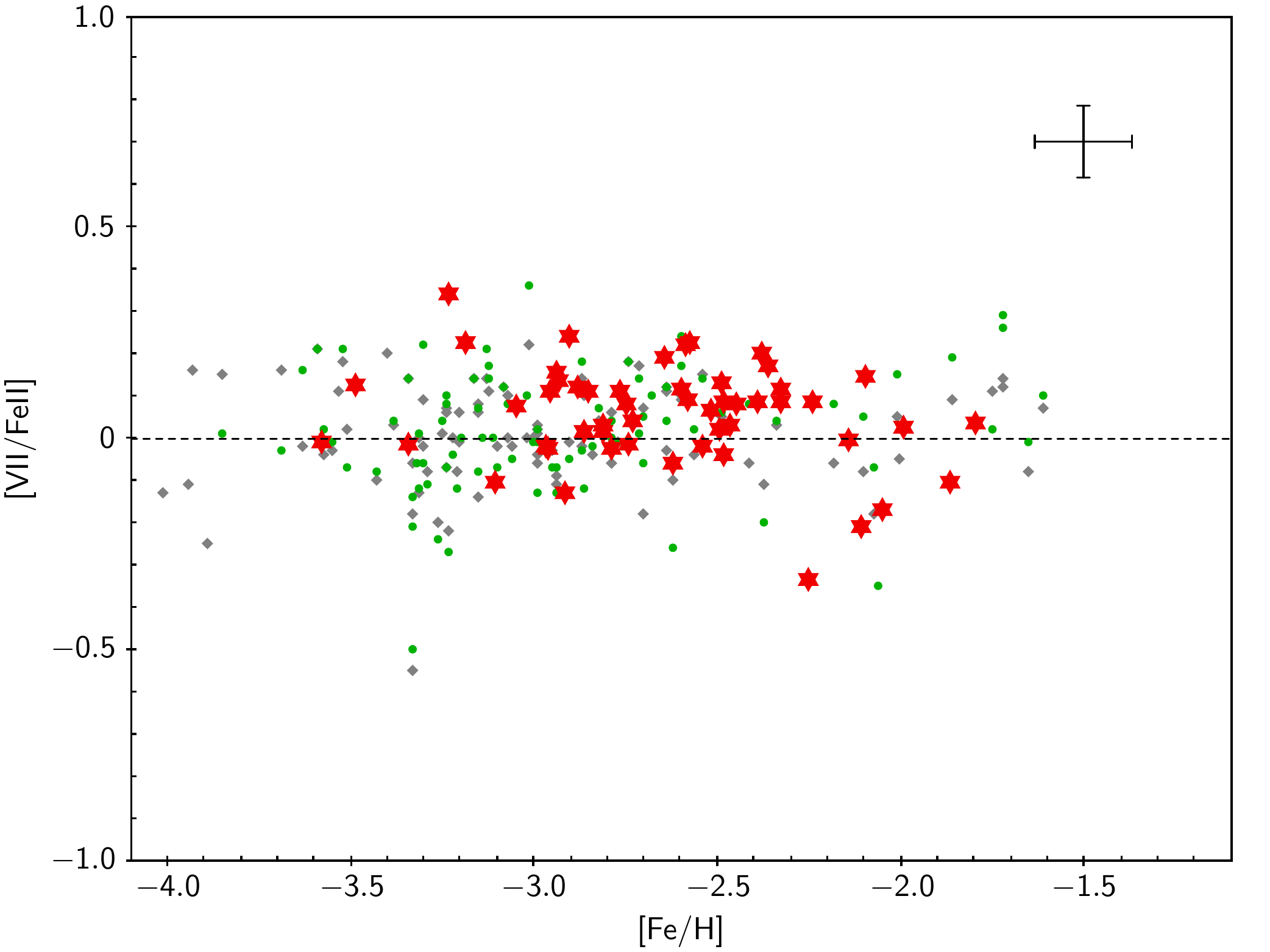}
      \caption{Elemental abundance ratios of Sc and V as a function of [Fe/H] for stars in our sample (red stars).
      Cyan squares and black dots represent Sc abundances for stars in the Large Program \citep{Cayrel2004} and in \citet{Ishigaki2013}, respectively. Green dots and grey diamonds represent V abundances for stars in \citet{Roederer2014} and \citet{Ou2020}, respectively. A representative error is plotted in the upper-right corner of each panel. }
         \label{sc_v}
   \end{figure*}

The derived abundance ratios of [\ion{Sc}{I}/\ion{Fe}{I}], [\ion{Sc}{II}/\ion{Fe}{II}], [\ion{V}{I}/\ion{Fe}{I}], and [\ion{V}{II}/\ion{Fe}{II}] as a function of [Fe/H] are shown in Fig. \ref{sc_v}, and compared to literature values for giant stars of similar metallicity.
The mean abundance ratios and standard deviations are $-0.09\pm0.13$ for [\ion{Sc}{I}/\ion{Fe}{I}], $-0.10\pm0.10$ for [\ion{V}{I}/\ion{Fe}{I}], $0.15\pm0.10$ for [\ion{Sc}{II}/\ion{Fe}{II}], and $0.05\pm0.12$\,dex for [\ion{V}{II}/\ion{Fe}{II}].
The trend appears flat down to the lowest measured [Fe/H] for both elements. 
This is in agreement with the results obtained by other authors. 

We find a mean difference between [\ion{Sc}{II}/H] and [\ion{Sc}{I}/H] of $0.37\pm0.16$ dex, and a mean difference between [\ion{V}{II}/H] and [\ion{V}{I}/H] of $0.27\pm0.11$ dex. These large discrepancies seem to suggest that NLTE effects on Sc and V are important. 
\citet{Zhang2008Sc} and \citet{Zhao2016Sc} provide NLTE corrections for Sc in cool dwarf stars, but we are not currently aware of any NLTE studies of scandium conducted on metal-poor giant stars. 
Similarly, we are not aware of any studies that have performed NLTE corrections for V. 
However, we note that \citet{Ou2020} found a difference between [\ion{V}{II}/H] and [\ion{V}{I}/H] of $0.25\pm0.01$ dex, which is in excellent agreement with that found in this study. Similar results have been obtained by \citet{Roederer2014} and \citet{Hansen2020}, who found higher \ion{V}{II} than \ion{V}{I}. 
We would like to stress that hyperfine splitting for Sc and V lines is not taken into account in the adopted GES line list. 
\citet{Roederer2014} pointed out that the lack of hyperfine splitting for \ion{V}{II} might lower \ion{V}{II} abundances by $<0.1$ dex. We expect a similar behaviour for \ion{Sc}{II} abundances. This may partially explain why we do not find an ionisation equilibrium for scandium and vanadium. 

\subsubsection{Cr and Mn}

   \begin{figure*}
   \centering
   \includegraphics[width=9.1 cm]{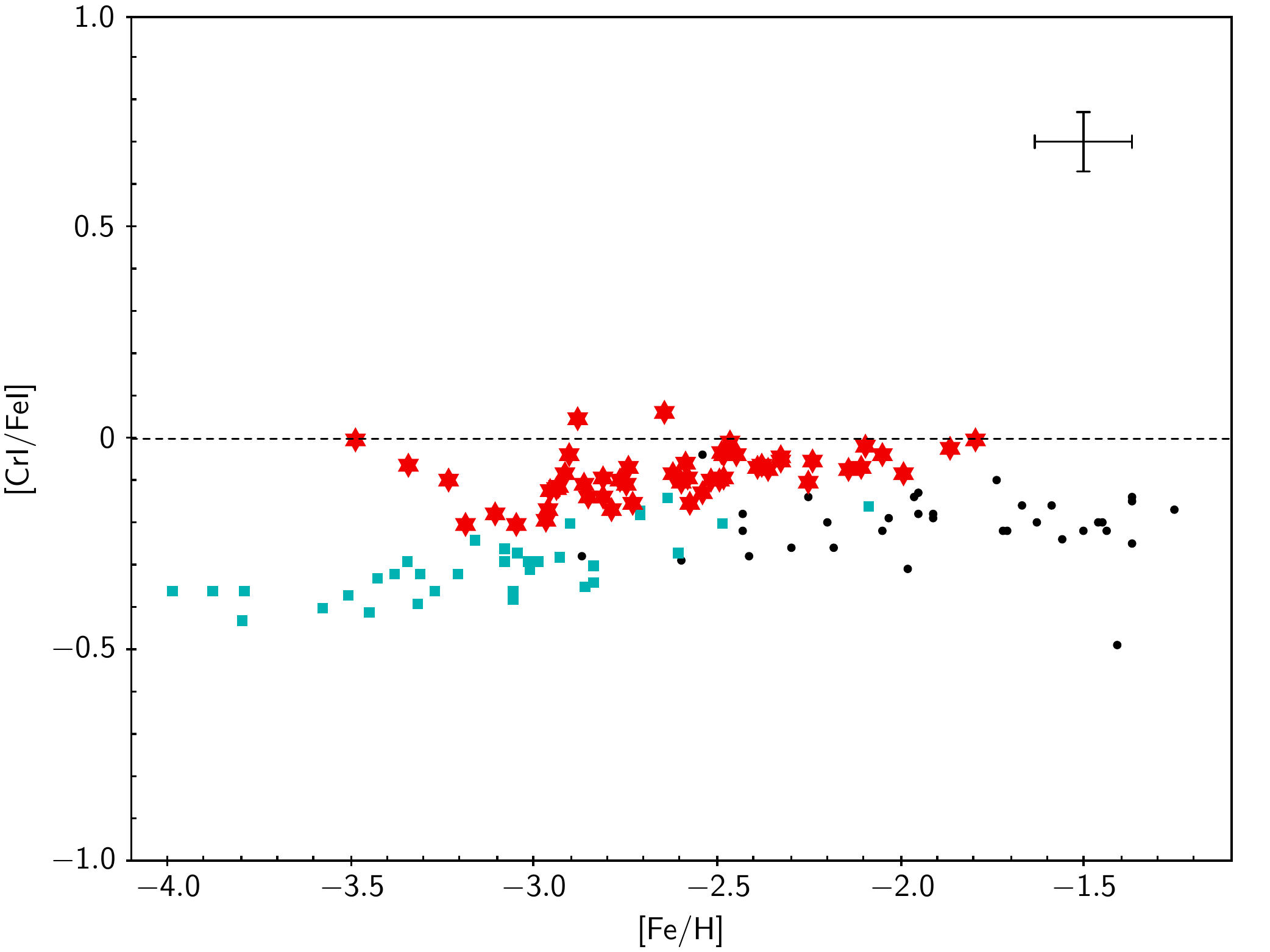}
   \includegraphics[width=9.1 cm]{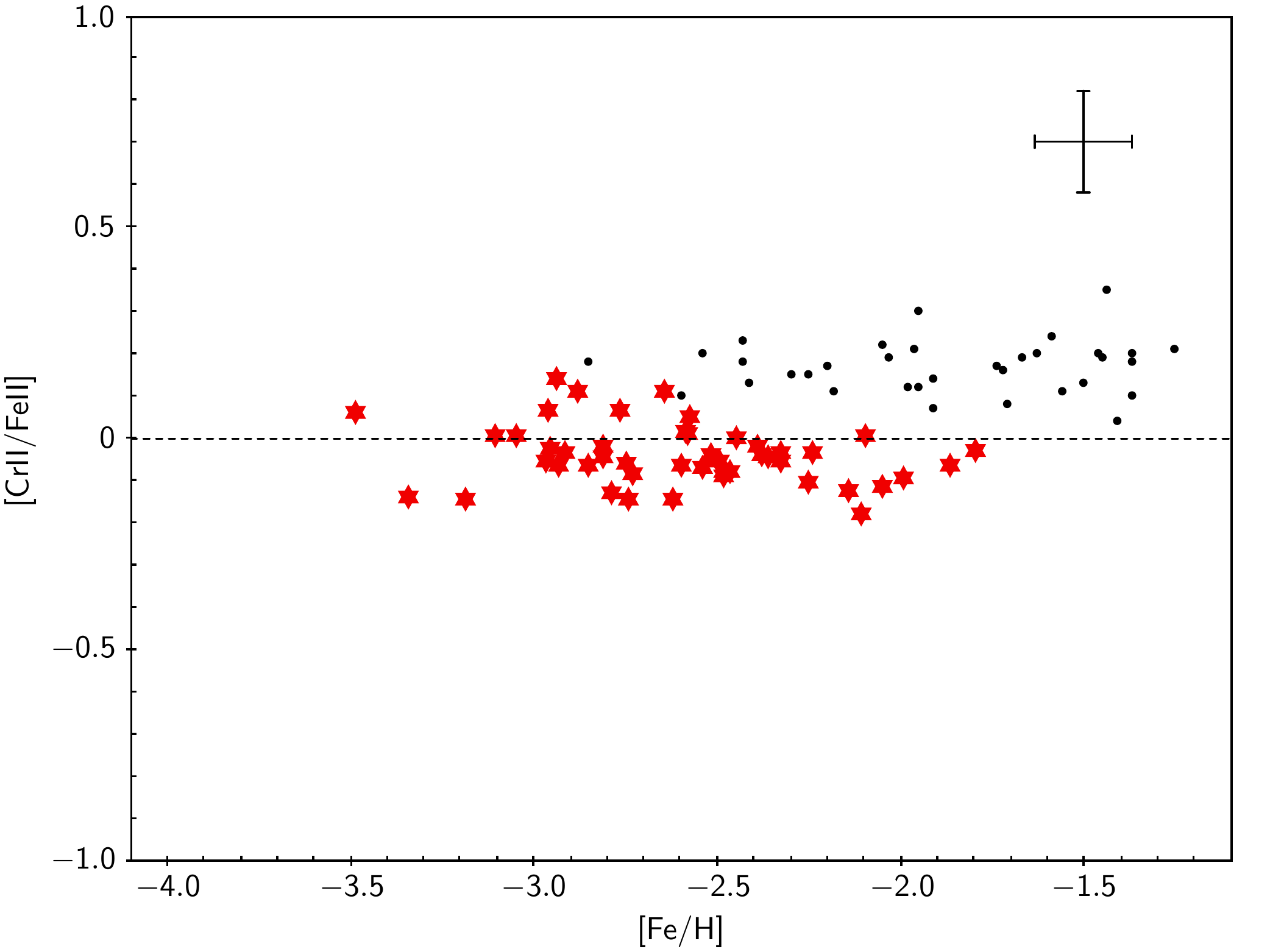}
   \includegraphics[width=9.1 cm]{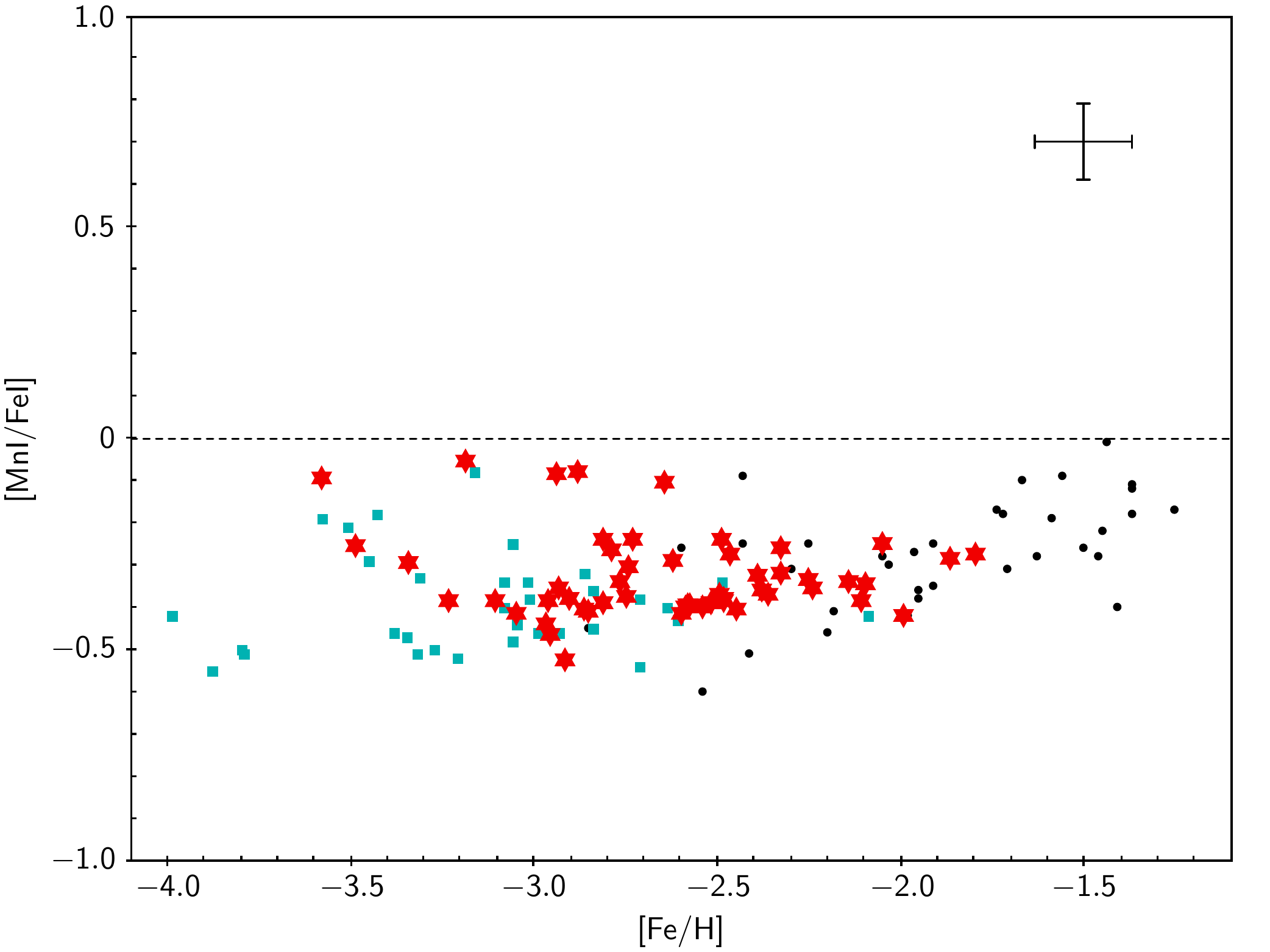}
   \includegraphics[width=9.1 cm]{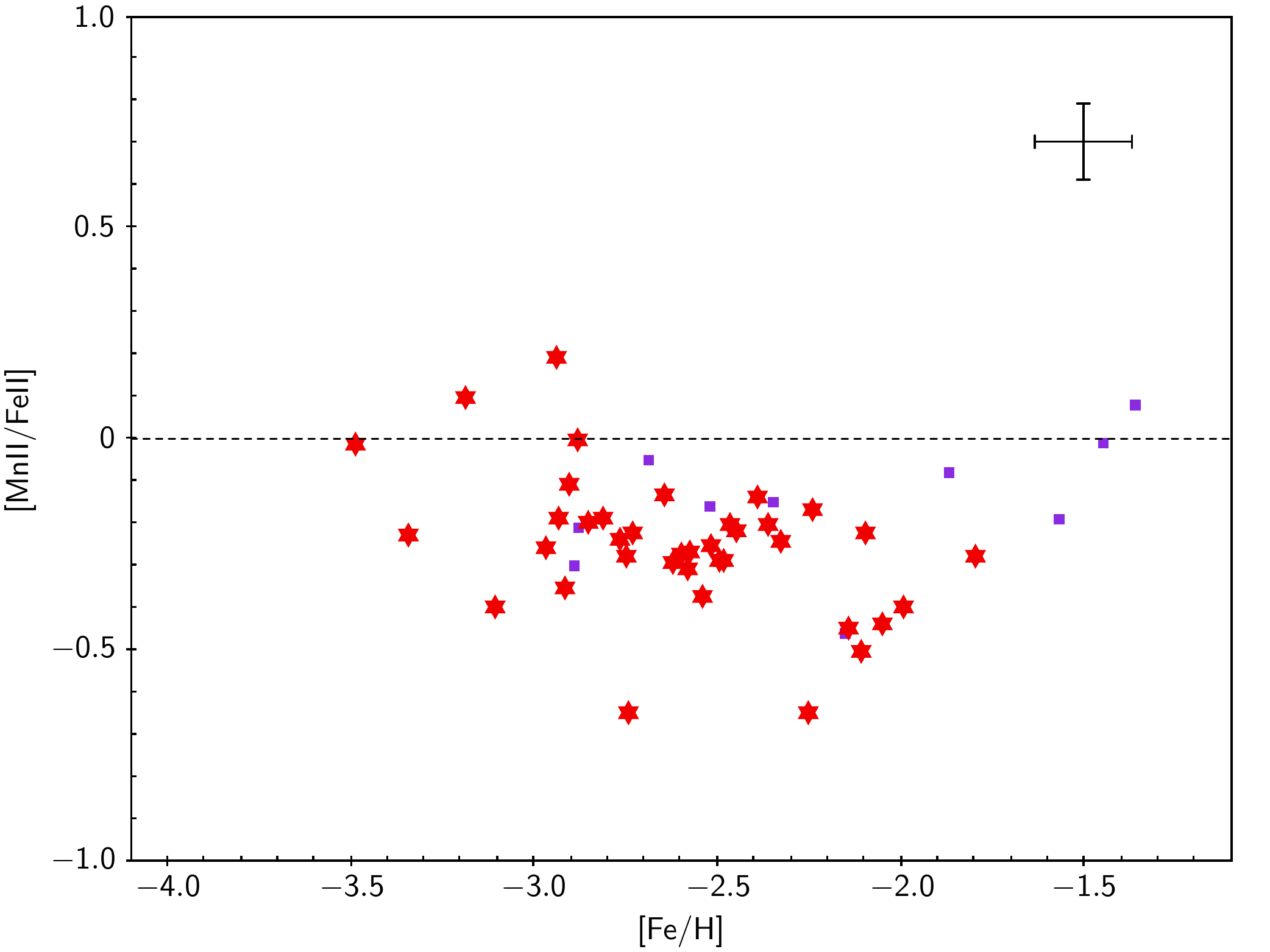}
      \caption{Elemental abundance ratios of Cr and Mn as a function of [Fe/H] for stars in our sample (red star symbols).  
      Cyan squares and black dots are from \citet{Cayrel2004} and \citet{Ishigaki2013}, respectively. Purple squares represent NLTE [Mn/Fe] abundance ratios for stars in \citet{eitner}. A representative error is plotted in the upper-right corner of each panel. 
      }
         \label{cr_mn}
   \end{figure*}

In Fig. \ref{cr_mn} we compare the derived [\ion{Cr}{I}/\ion{Fe}{I}],  [\ion{Cr}{II}/\ion{Fe}{II}], [\ion{Mn}{I}/\ion{Fe}{I}], and [\ion{Mn}{II}/\ion{Fe}{II}] abundance ratios to the values obtained in \citet{Cayrel2004} and in \citet{Ishigaki2013}. 
The mean abundance ratios and standard deviations are 
[\ion{Cr}{I}/\ion{Fe}{I}]=$-0.09\pm0.06$, [\ion{Cr}{II}/\ion{Fe}{II}]=$-0.04\pm0.07$, [\ion{Mn}{I}/\ion{Fe}{I}]=$-0.33\pm0.10$, and [\ion{Mn}{II}/\ion{Fe}{II}]=$-0.26\pm0.17$. 
Contrary to Mn, for which the values found are in agreement with the literature, both [\ion{Cr}{I}/\ion{Fe}{I}] and [\ion{Cr}{II}/\ion{Fe}{II}] abundance ratios appear to differ from previous results.

For the stars that we have in common with the other studies, the derived [\ion{Cr}{I}/\ion{Fe}{I}] ratios are 0.17\,dex and 0.12\,dex higher than the values in \citet{Cayrel2004} and \citet{Ishigaki2013}, respectively. 
On the contrary, the derived [\ion{Cr}{II}/\ion{Fe}{II}] ratios are 0.21\,dex lower than the values in \citet{Ishigaki2013}.
Another difference compared to literature \citep[see e.g.][]{Cayrel2004,Bonifacio2009} is that the [\ion{Cr}{I}/\ion{Fe}{I}] ratio does not decrease with metallicity in our sample, but instead seems to increase again at the lowest observed metallicities ([Fe/H] $<-3.2$). 

The difference in the [\ion{Cr}{I}/\ion{Fe}{I}] abundance ratios with \citet{Ishigaki2013} results seems to arise from the different \ion{Fe}{I} abundances, as the mean [\ion{Cr}{I}/H] is almost the same in both studies ($\Delta$[\ion{Cr}{I}/H]$\sim0.01$). 
However, we note that the mean [\ion{Cr}{I}/H] abundance in our sample is 0.2\,dex higher than in \citet{Cayrel2004}.
In this case, the difference in the abundance ratios is probably due to the different line selections. 
\citet{Cayrel2004} relied mainly on \ion{Cr}{I} resonance lines to derive Cr abundance. In \citet{Ishigaki2013} and in this study these lines were excluded from the line list, and the same set of \ion{Cr}{I} lines were used.
To test this hypothesis, we derived the \ion{Cr}{I} abundance for the star CES1942-6103 (CS22891-209) employing the same lines used in \citet{Cayrel2004}. We obtained A(\ion{Cr}{I}) $=2.02\pm0.14$ dex, which is in excellent agreement with the value found by \citet{Cayrel2004} (A(\ion{Cr}{I}) $=2.01\pm0.16$ dex). 
This confirms that the \ion{Cr}{I} abundance depends on the lines chosen for the analysis and that the observed trend with metallicity is probably due to NLTE effects, which affect each line differently.
We underline that for \ion{Cr}{I} we used $gf$ values taken from \citet{SLS}, which are more recent than those used in \citet{Cayrel2004}. 

Similarly to the \ion{Cr}{I}, we observe a discrepancy of --0.18\,dex between our mean [\ion{Cr}{II}/H] abundance and that of \citet{Ishigaki2013}, as they adopted a different set of \ion{Cr}{II} lines. 
For \ion{Cr}{ii} we used the $gf$ values of \citet{2006A&A...445.1165N}.
The accuracy of the lifetimes used by these authors to derive the oscillator strengths has been questioned by \citet{2015A&A...573A..26S} in their solar abundance analysis. However, \citet{Sneden_2016}, in their analysis of the metal-poor dwarf HD\,84937, showed that the use of the \citet{2006A&A...445.1165N} values considerably reduces the line-to-line scatter with respect to what was obtained using earlier $gf$ values. For this reason, we decided to keep the $gf$ values of \citet{2006A&A...445.1165N}.

We find a mean difference between [\ion{Cr}{II}/H] and [\ion{Cr}{I}/H] of $0.18\pm0.09$ dex, and a mean difference between [\ion{Mn}{II}/H] and [\ion{Mn}{I}/H] of $0.20\pm0.10$ dex. According to \citet{BergemannCescutti2010Cr} and \citet{BergemannGehren2008Mn}, these discrepancies are due to NLTE effects on neutral Cr and Mn. 
The NLTE corrections provided by \citet{BergemannCescutti2010Cr} and \citet{BergemannGehren2008Mn}\footnote{\url{https://nlte.mpia.de}} for stars with similar parameters to those of our targets are between +0.20 and +0.60 dex for \ion{Cr}{I}, and between +0.40 and +0.60 for \ion{Mn}{I}. No corrections are available for the \ion{Cr}{II} and \ion{Mn}{II} lines we used, but we expect them to be positive and $<0.1$ dex, similar to the corrections calculated for metal-poor dwarf stars \citep{BergemannGehren2008Mn,BergemannCescutti2010Cr,Bergemann2019Mn}. 
If we assume that  the ionised species, which are the majority species in these stars, are formed close to LTE, the result of applying the above NLTE corrections to our \ion{Cr}{I} and \ion{Mn}{I} abundances would be to worsen the ionisation balance for both elements.
A possible explanation is that hydrodynamical effects have not been taken into account. 

For many lines of several elements the 1D NLTE corrections are positive while the 3D NLTE corrections are negative, and the full 3D NLTE correction is lower than the 1D NLTE.
\citet{Bergemann2019Mn} have computed 3D NLTE corrections for several \ion{Mn}{I} lines and for one \ion{Mn}{II} line. 
The 3D NLTE corrections for \ion{Mn}{I} lines are positive and, surprisingly, larger 
than the 1D NLTE \citep[see Fig. 17 of][]{Bergemann2019Mn}, while the \ion{Mn}{II} 348.8\,nm line has the expected behaviour, with the 3D NLTE correction being smaller than the 1D NLTE correction.
It thus seems that if the 3D NLTE corrections were applied, the ionisation balance for Mn would be worse than in LTE.

In the lower-right panel of Fig. \ref{cr_mn}, we compare our LTE \ion{Mn}{ii} abundances with the NLTE abundances of \citet{eitner}. In their study, they found a difference between \ion{Mn}{II} and \ion{Mn}{I} between 0.1 and 0.45 dex in the LTE approximation, which is in agreement with our results. In the NLTE approximation, this discrepancy becomes lower than 0.1 dex. This is further evidence that the difference between the ionised and neutral species is due to NLTE effects that affect the \ion{Mn}{I} lines.

\subsubsection{Co, Ni, Cu, and Zn}

   \begin{figure*}
   \centering
   \includegraphics[width=9.1 cm]{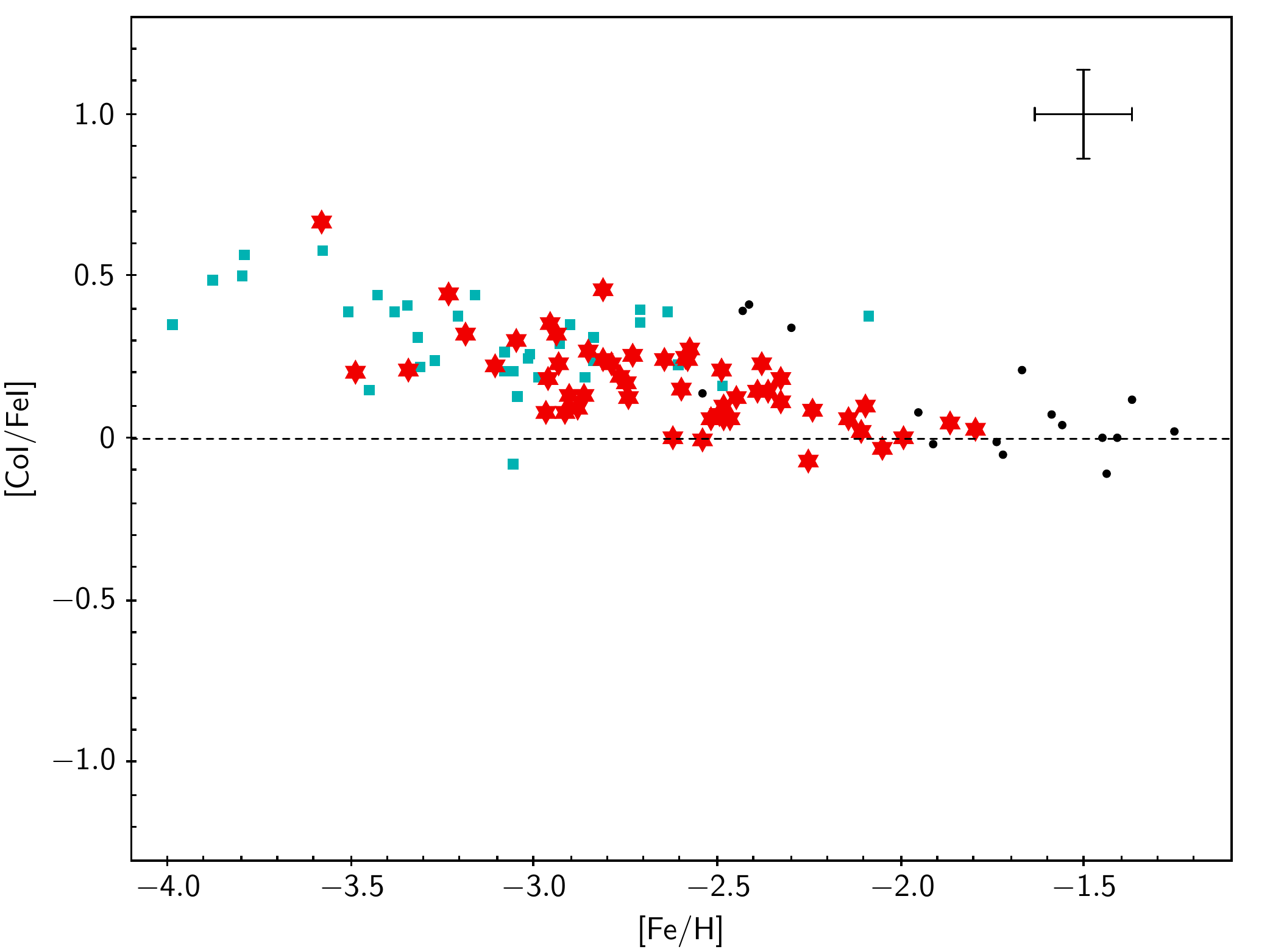}
   \includegraphics[width=9.1 cm]{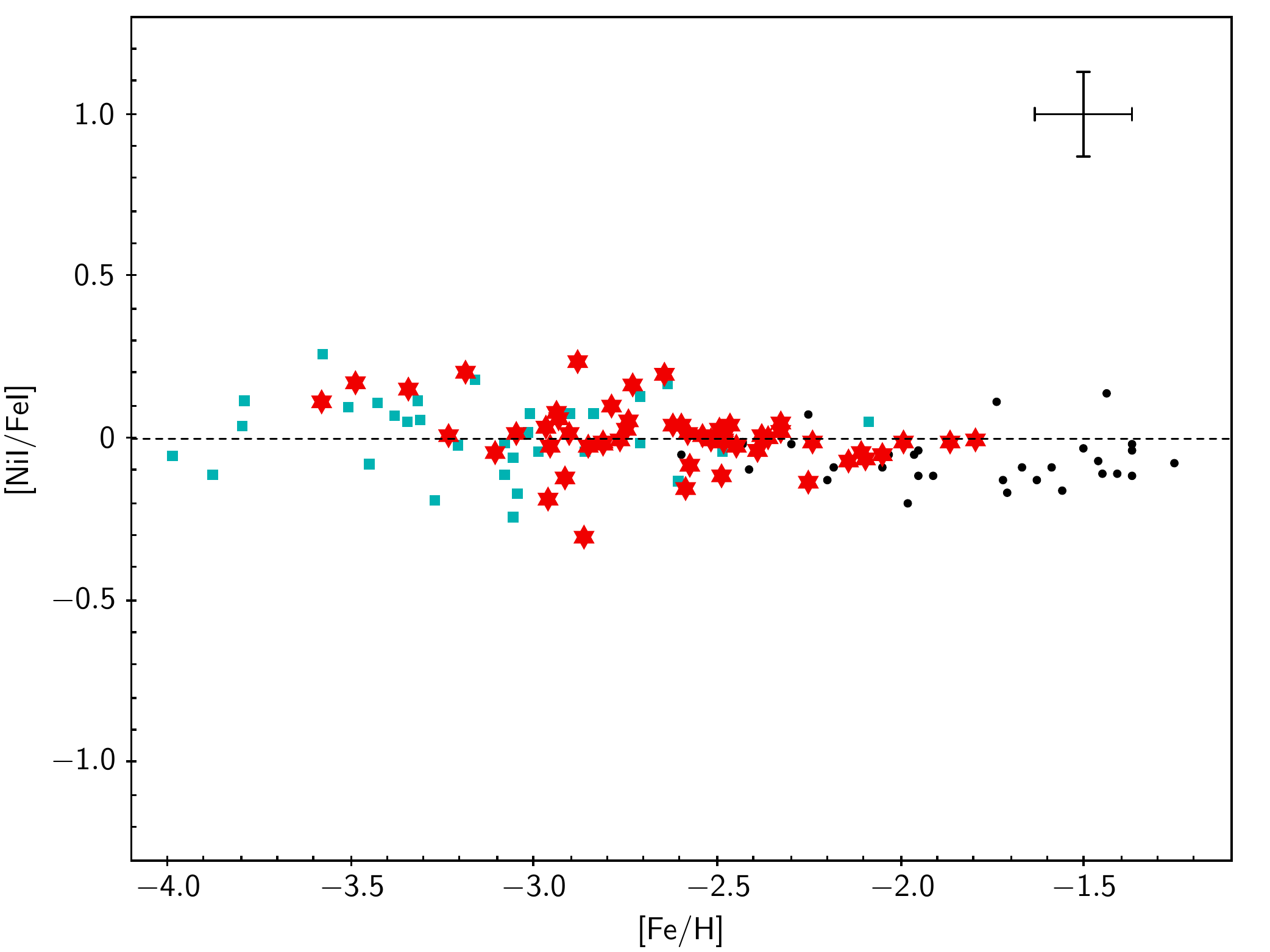}
   \includegraphics[width=9.1 cm]{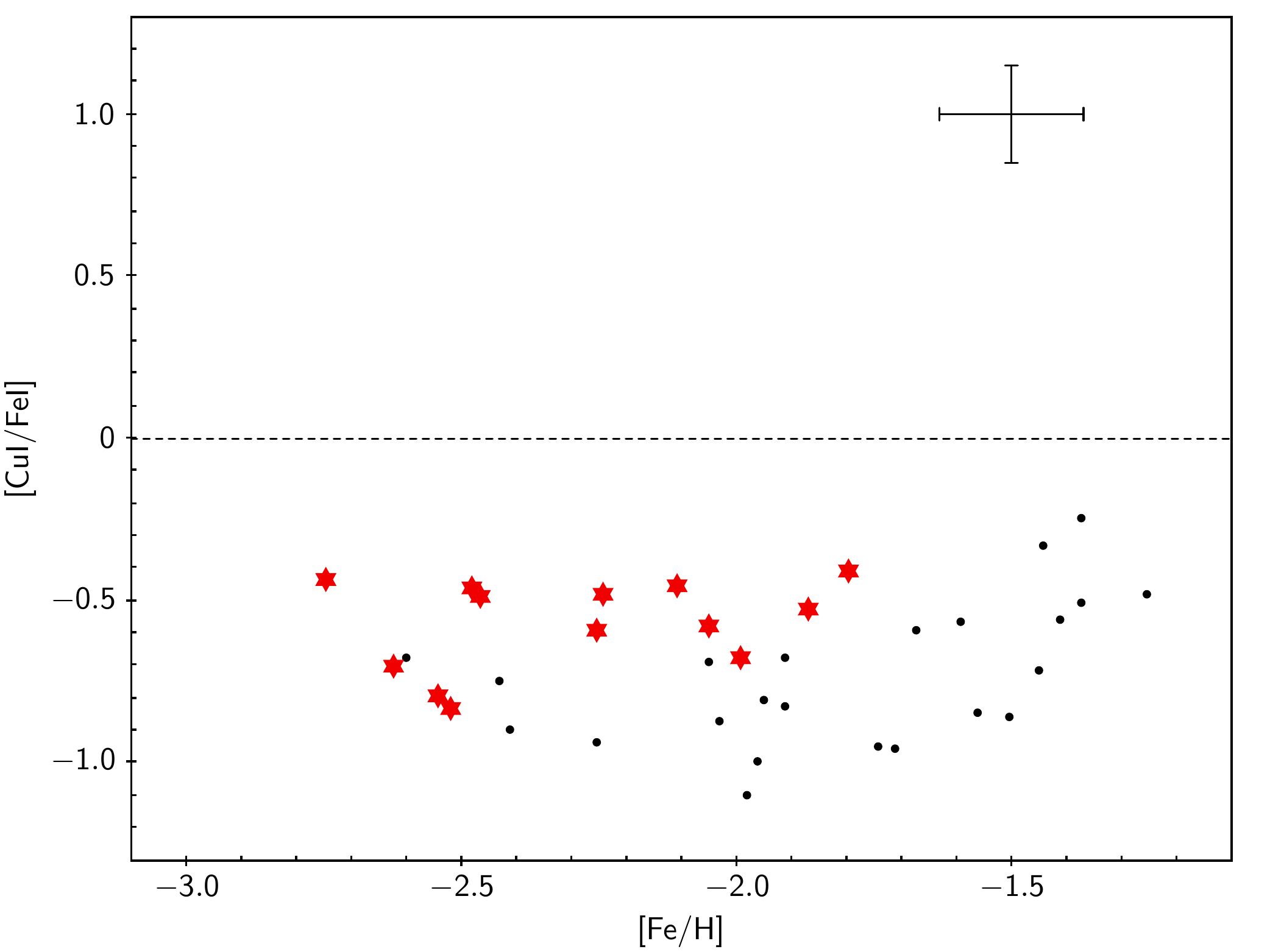}
   \includegraphics[width=9.1 cm]{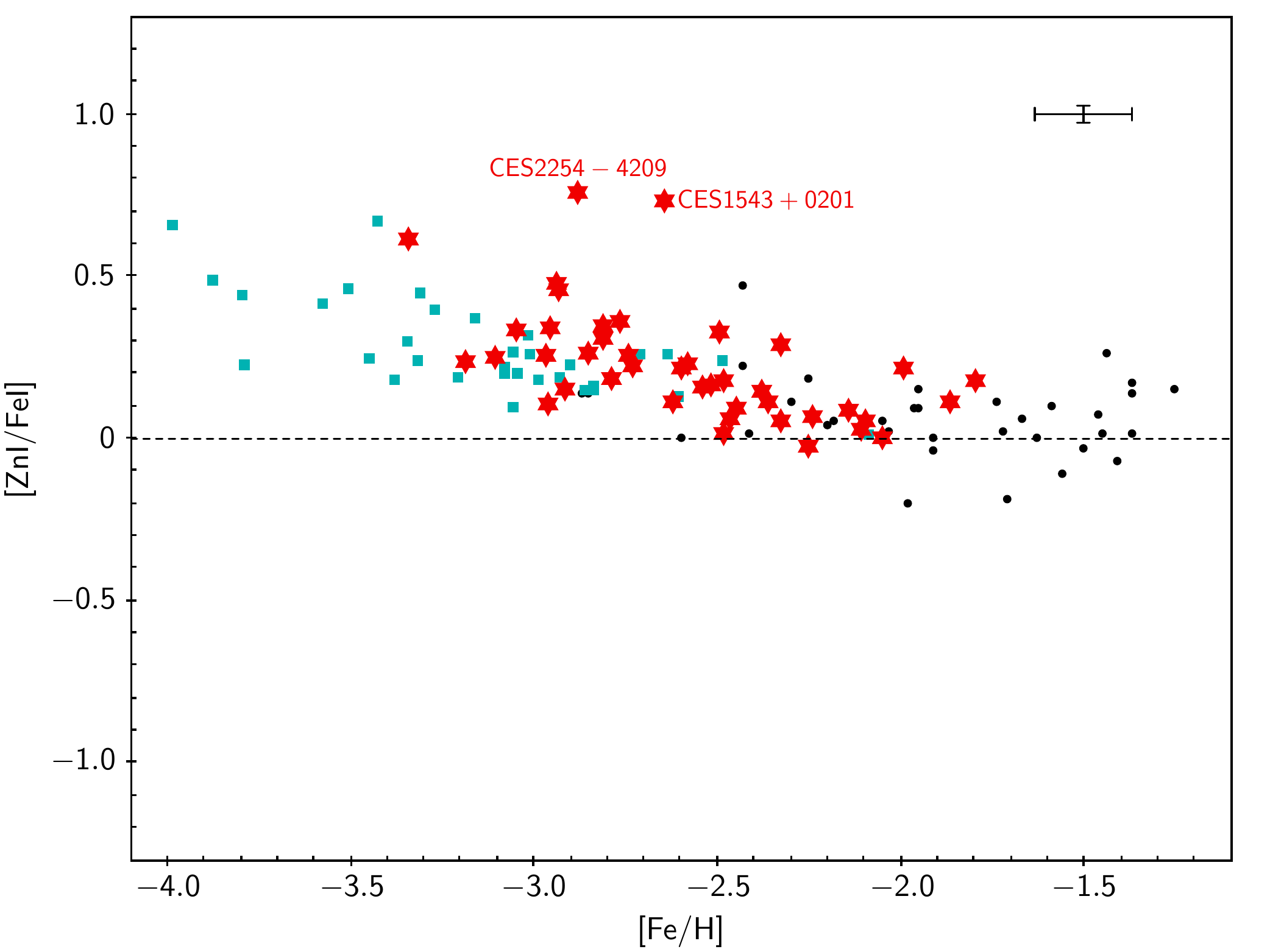}
      \caption{Elemental abundance ratios of Co, Ni, Cu, and Zn as a function of [Fe/H] for stars in our sample (red star symbols).  
      Cyan squares and black dots represent the same quantities for stars in the Large Program \citep{Cayrel2004} and in \citet{Ishigaki2013}, respectively. A representative error is plotted in the upper-right corner of each panel. }
         \label{fepeak_fe}
   \end{figure*}

Figure \ref{fepeak_fe} shows the comparison between [\ion{Co}{I}/\ion{Fe}{I}], [\ion{Ni}{I}/\ion{Fe}{I}], [\ion{Cu}{I}/\ion{Fe}{I}], and [\ion{Zn}{I}/\ion{Fe}{I}] abundance ratios as a function of [Fe/H] for our stars and the same quantities in literature. 
We observe a decreasing trend with metallicity for Co and Zn, and a flat trend with a mean value around zero for Ni.
Our results appear in agreement with previous studies and confirm the trend with metallicity found by other authors \citep[and references therein]{Cayrel2004,Takeda2005Zn,Lai2008,Ishigaki2013}. 

Previous studies have found that the [Cu/Fe] abundance ratio decreases with decreasing metallicity
in LTE \citep[see e.g.][]{Andrievsky2018Cu,Roederer2018CuZn,Shi2018Cu}.
However, the NLTE analysis of \citet{Andrievsky2018Cu} found that the decrease is much
smaller, with [Cu/Fe]$\ga -0.3$, and for the extremely metal-poor giant
CD\,--38 245 ([Fe/H]=--4.19) they found [Cu/Fe]=-0.07. 
Both our results and those of \citet{Ishigaki2013} 
are based on LTE and they seem to suggest that the trend with metallicity flattens out at [Fe/H]$<-1.8$ for giant stars. We note that our derived [\ion{Cu}{I}/\ion{Fe}{I}] ratios are $\sim 0.25$ dex higher than the values found by \citet{Ishigaki2013} for stars in the same range of metallicity. The origin of this discrepancy is not clear, since we do not have any Cu measurement in common with \citet{Ishigaki2013}. 

\subsubsection{Zn-rich stars: CES1543+0201 and CES2254-4209}

   \begin{figure*}
   \centering
   \includegraphics[width=18 cm]{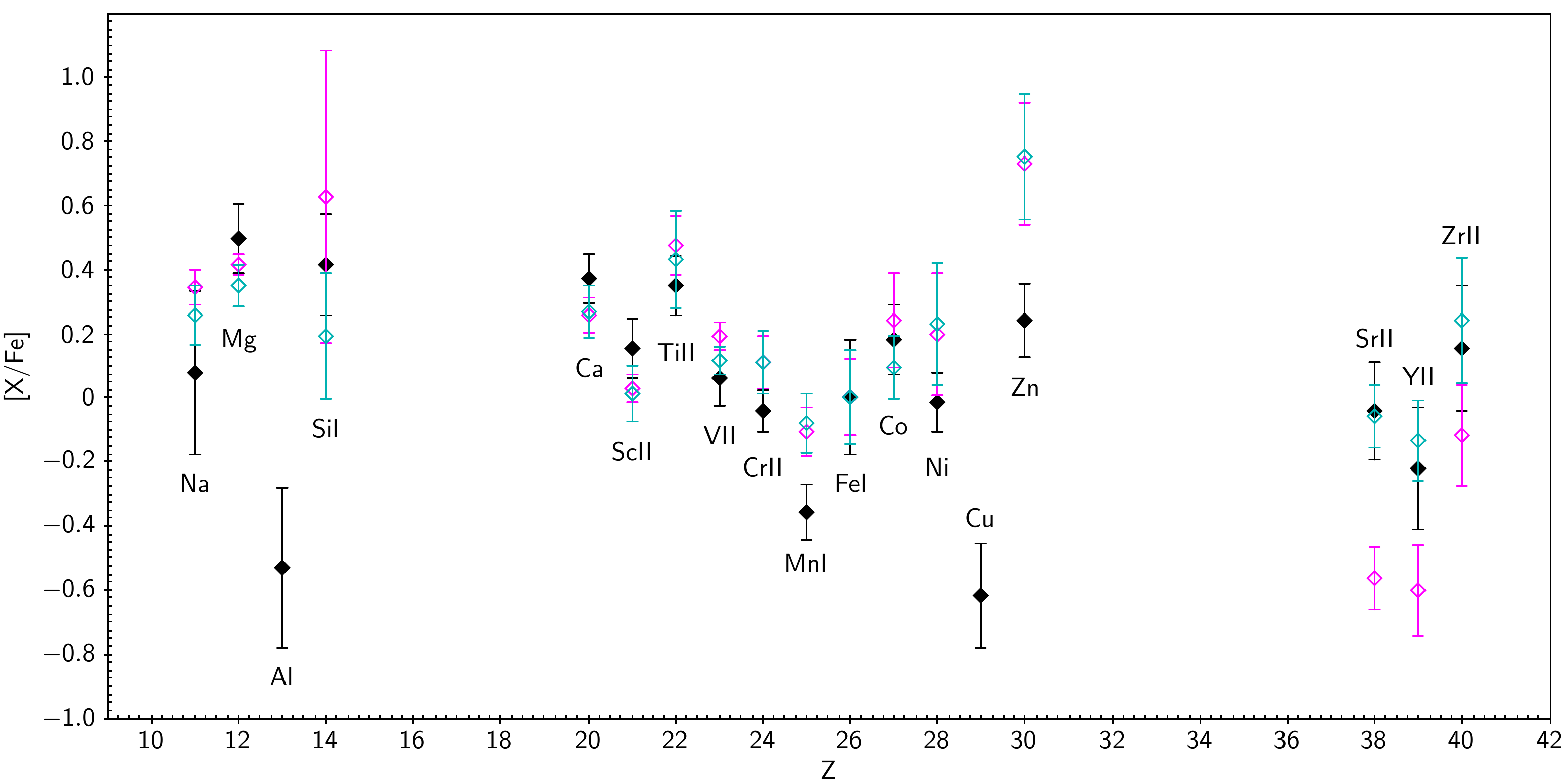}
      \caption{Elemental abundance ratios versus atomic number, Z, for stars in the sample with  --3.1\,$<$\,[Fe/H]\,$<$\,--2.4. Coloured symbols indicate abundance patterns for stars CES1543+0201 (magenta) and CES2254-4209 (cyan) with errors. Black symbols indicate the average abundance ratios of Zn-normal stars. Black error bars represent the standard deviation around the mean abundance. Abundance ratios of neutral and ionised species are scaled to their Fe counterpart. }
         \label{abu_pattern}
   \end{figure*}

   \begin{figure}
   \centering
   \includegraphics[width=9.1 cm]{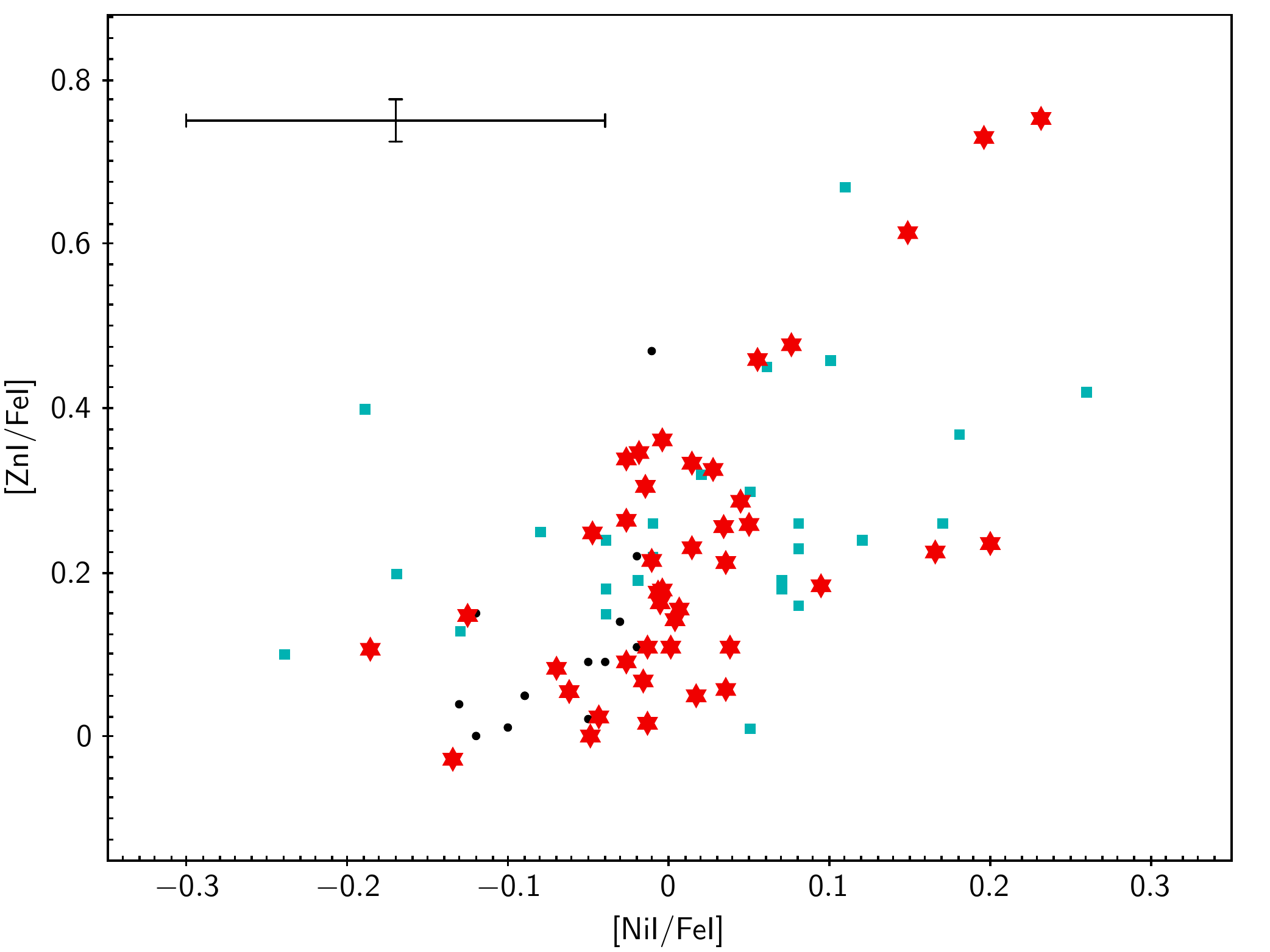}
      \caption{[Zn/Fe] versus [Ni/Fe] abundance ratios for stars in our sample (red star symbols).  
      Cyan squares and black dots represent the same quantities for stars of similar metallicity in the Large Program \citep{Cayrel2004} and in \citet{Ishigaki2013}, respectively. A representative error is plotted in the upper-left corner. }
         \label{znfe_nife}
   \end{figure}

In Fig. \ref{fepeak_fe}, in the panel showing [Zn/Fe],
two stars stand out from the trend defined by the others and by the measurements in the literature, showing [Zn/Fe] $\sim+0.7$.
CES2254--4209 (also known as HE2252--4225) was discovered
in the context of the HERES survey 
by \citet{Mashonkina2014} as an $r-$enhanced star ([$r$/Fe]=+0.80). 
\citet{Mashonkina2014} found a lower enhancement for CES2254-4209 ([Zn/Fe]=+0.43), mostly due to the fact that both Fe and Zn abundances have been derived by applying NLTE corrections. 
The measured abundance of Th and the anomalously high Th/Eu ratio, corresponding to a radioactive age of $1.5\pm1.5$ Gyr, classify it as an actinide boost star, a class that still only contains a handful of stars and whose prototype is CS\,31082--001 \citep{2001Natur.409..691C,2002A&A...387..560H}.
Such an occurrence may suggest that the overabundance of Zn is in fact due to its production through the $r$ process.
However, the other Zn-rich star, CES1543+0201 (also known as CS\,30312-100), initially discovered by \citet{2002ApJ...567.1166A}, has been classified as a CEMP star ([C/Fe]=+0.98) with no enhancement of n-capture elements (CEMP-no) by \citet{2007ApJ...655..492A}. This star has been analysed in detail in \citet{Roederer2014}, who found a similar enhancement as in our study ([Zn/Fe]=+0.71).

In Fig. \ref{abu_pattern} we show abundance patterns for stars CES1543+0201 (magenta) and CES2254--4209 (cyan). 
We tried to fit these patterns with different SN yields using STARFIT\footnote{\url{http://starfit.org/}} 
\citep{HegerWoosley2010}.
For the lighter elements, the low mass SNe ($\sim$12 \Msol) derived by STARFIT provide reasonable fits to the observations, as quantified by $\chi ^2$ statistics, while they fail to reproduce the observed Zn abundance, and some models also fail at explaining the Sc abundances. This is expected since such models cannot account for the neutrino-driven ejecta, where the weak $r$ process and/or the $\nu$p process may contribute to these and heavier elements. 

From the abundance patterns in Fig. \ref{abu_pattern}, we notice that the two Zn-rich stars also show a slight enhancement in Ni compared to the mean value observed for the sample stars of similar metallicity. 
In their study of Zn abundances in RGB stars of the Sculptor dwarf galaxy, \citet{Sk2017A&A...606A..71S} found a correlation between Zn and Ni abundances. We checked whether this correlation was also present in our sample by performing a non-parametric Kendall's $\tau$ test, and we found a correlation probability of 99.9\%.  Looking at Fig. \ref{znfe_nife}, we note that the stars seem to follow two different branches in the plane for [Ni/Fe]$>$0.05. For some stars [Zn/Fe] increases with [Ni/Fe], while for the others [Zn/Fe] remains approximately constant as [Ni/Fe] increases.  
In our opinion, this pattern could be due to a different explosion energy of the SN for a given mass of the progenitor star \citep[see e.g.][and references therein]{Nomoto2013ARA&A..51..457N}.
Hypernovae (HNe), a type of core collapse SNe with extremely large explosion energies ($\gtrsim 10^{52}$ erg), are able to produce a much larger amount of iron-peak elements, especially Zn, than classical SNe via $\alpha-$rich freeze-out \citep[e.g.][]{Galama1998Natur.395..670G,Iwamoto1998Natur.395..672I,Nomoto2001ASSL..264..507N,Umeda2002ApJ...565..385U}. 
It is therefore possible that the Zn-rich stars in our sample formed in a gas cloud pre-enriched by HNe, while stars with approximately constant [Zn/Fe] formed from gas enriched by SNe with lower explosion energies.

\subsection{Light n-capture elements: Sr, Y, and Zr}
   \begin{figure*}
   \centering
   \includegraphics[width=9.1 cm]{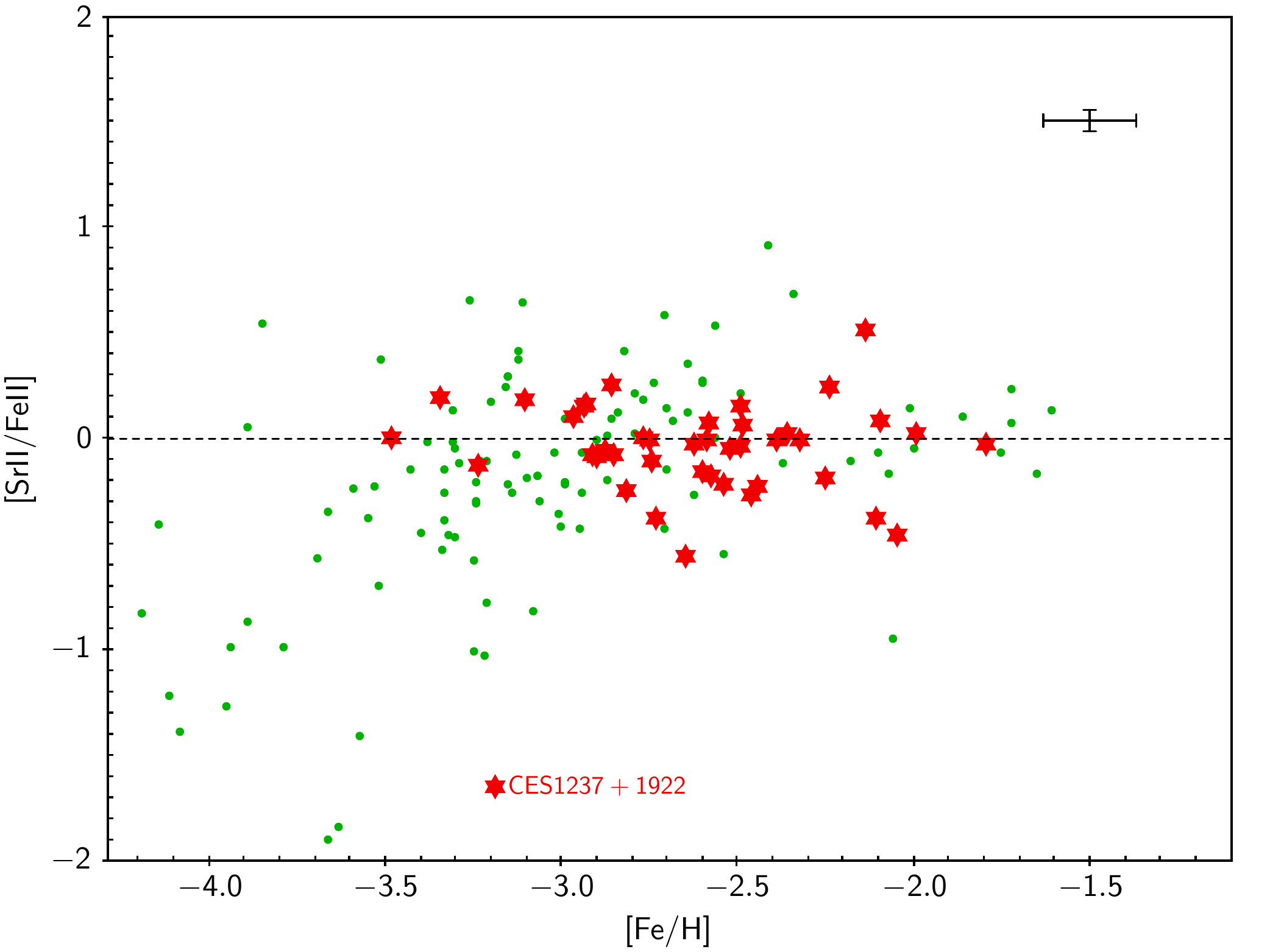}
   \includegraphics[width=9.1 cm]{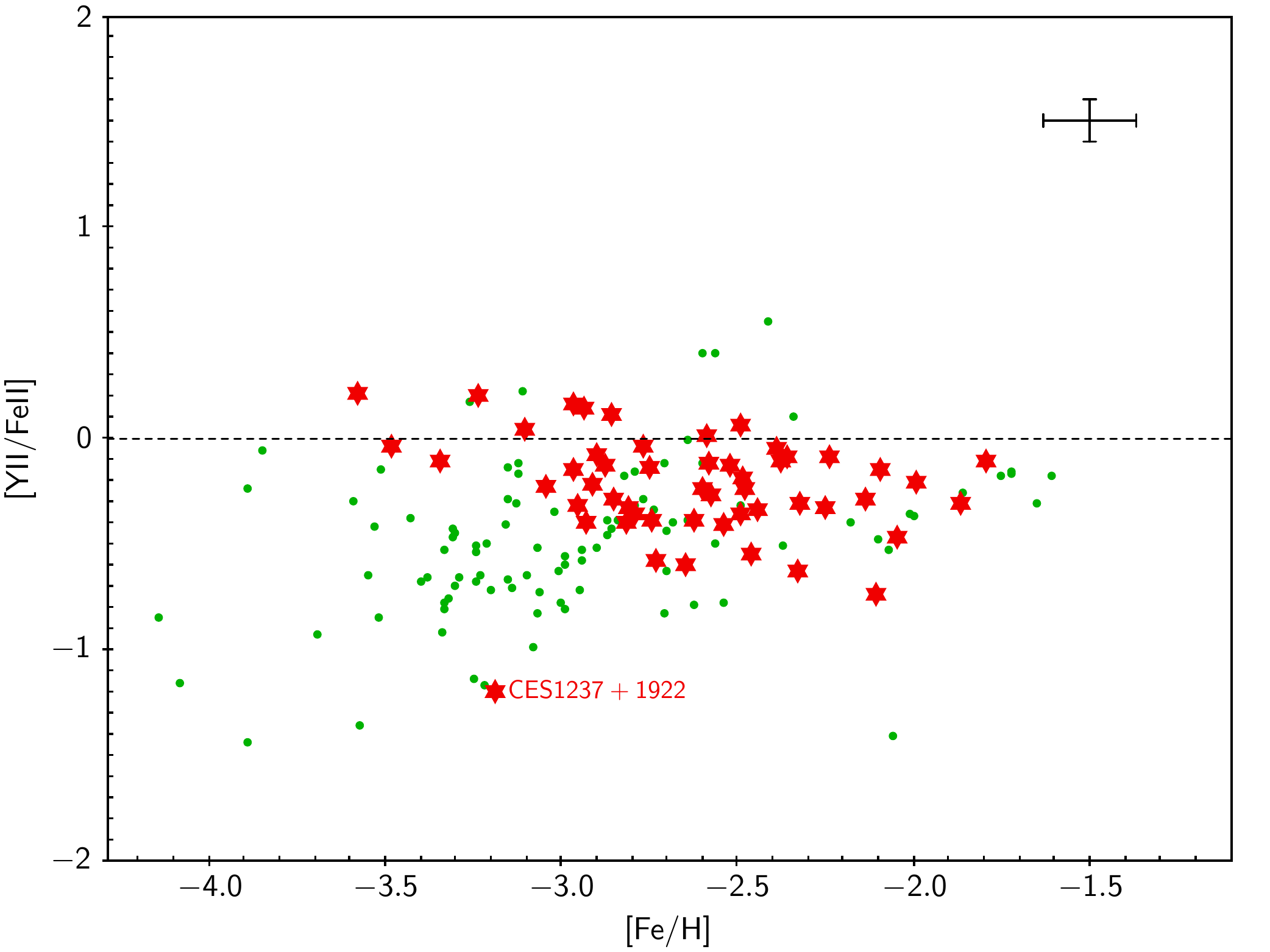}
   \includegraphics[width=9.1 cm]{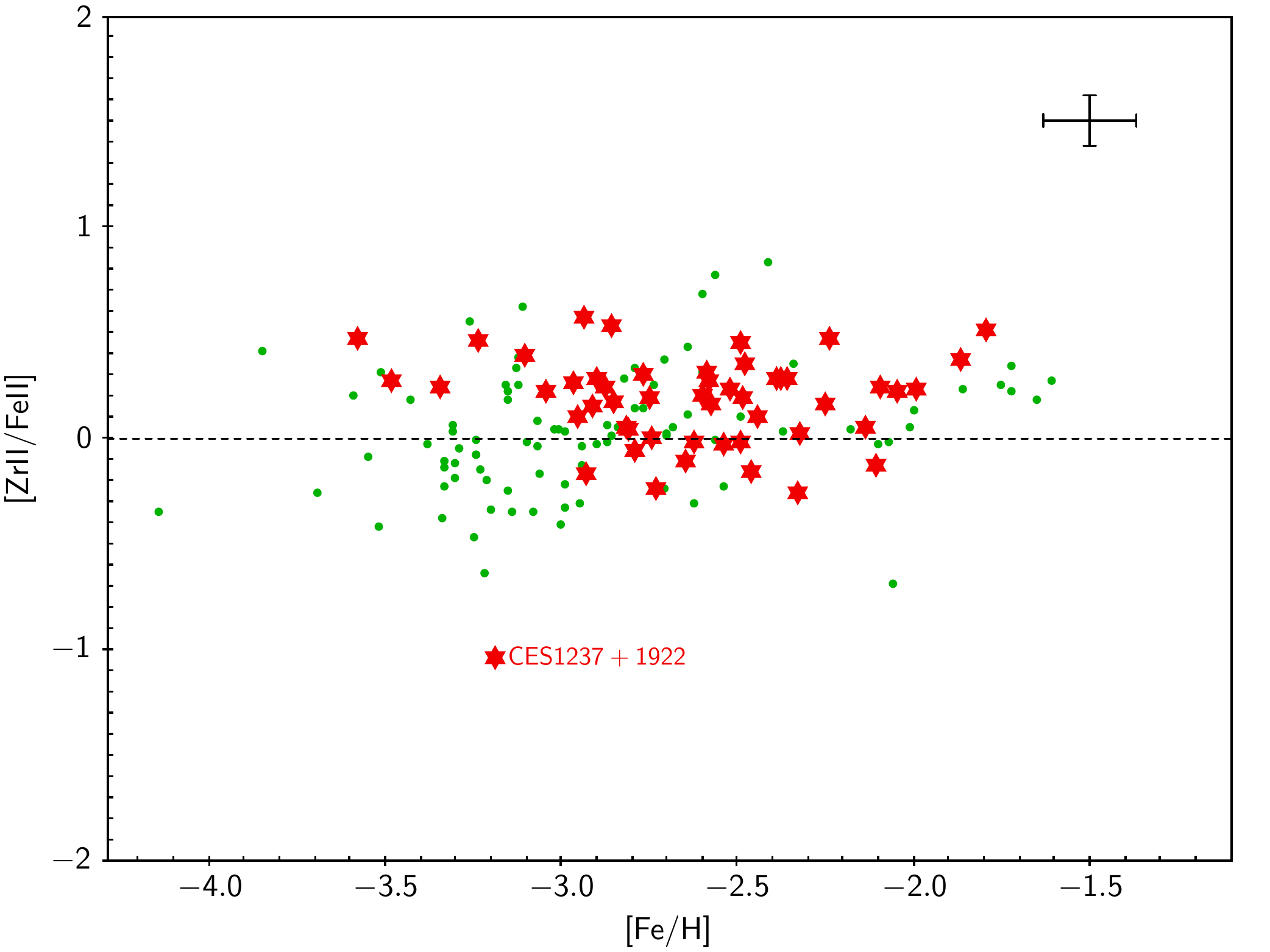}
      \caption{Elemental abundance ratios of Sr, Y, and Zr as a function of [Fe/H] for stars in our sample (red star symbols).  
      Green dots represent Sr, Y, and Zr abundances derived in \citet{Roederer2014}. A representative error is plotted in the upper-right corner of each panel. }
         \label{ncapt_fe}
   \end{figure*}

The derived [\ion{Sr}{II}/\ion{Fe}{II}], [\ion{Y}{II}/\ion{Fe}{II}], and [\ion{Zr}{II}/\ion{Fe}{II}] abundance ratios as a function of [Fe/H] are shown in Fig. \ref{ncapt_fe}, and compared to literature values for giant stars of similar metallicity.
Our results appear in general agreement with previous studies. For stars in \citet{Roederer2014}, we note that for [Fe/H]<$-3.0$ the dispersions around the mean value become larger and the abundance ratios of Sr and Y seem to decrease with metallicity. These trends have also been observed in other studies \citep[see e.g.][]{Francois2007}. This is not the case for our targets, for which the trend remains approximately flat at these metallicities, with sample averages and standard deviations of [\ion{Sr}{II}/\ion{Fe}{II}]=$-0.08\pm0.32$, [\ion{Y}{II}/\ion{Fe}{II}]=$-0.24\pm0.25$, and [\ion{Zr}{II}/\ion{Fe}{II}]=$+0.16\pm0.26$. 
We stress the fact that, at metallicities below --3, we only have measurements for about five 
stars in our sample, so it is possible that this difference in trend with the literature could be due to poor statistics.

For the seven stars for which we were able to measure \ion{Zr}{I}, we find a mean difference between [\ion{Zr}{II}/H] and [\ion{Zr}{I}/H] of $0.57 \pm 0.09$ dex. 
We suspect that this difference is due to strong NLTE effects. 
According to \citet{Velichko2010AstL...36..664V}, NLTE corrections for \ion{Zr}{II} are $\sim$\,+0.2\,dex for giant stars with [Fe/H]\,$\sim$\,--3. At solar metallicity, the NLTE corrections for \ion{Zr}{I} are about +0.3 dex, while no correction is available for metal-poor stars. 

%
\section{Discussion }
   
   \begin{figure*}
   \centering
   \includegraphics[width=9.1 cm]{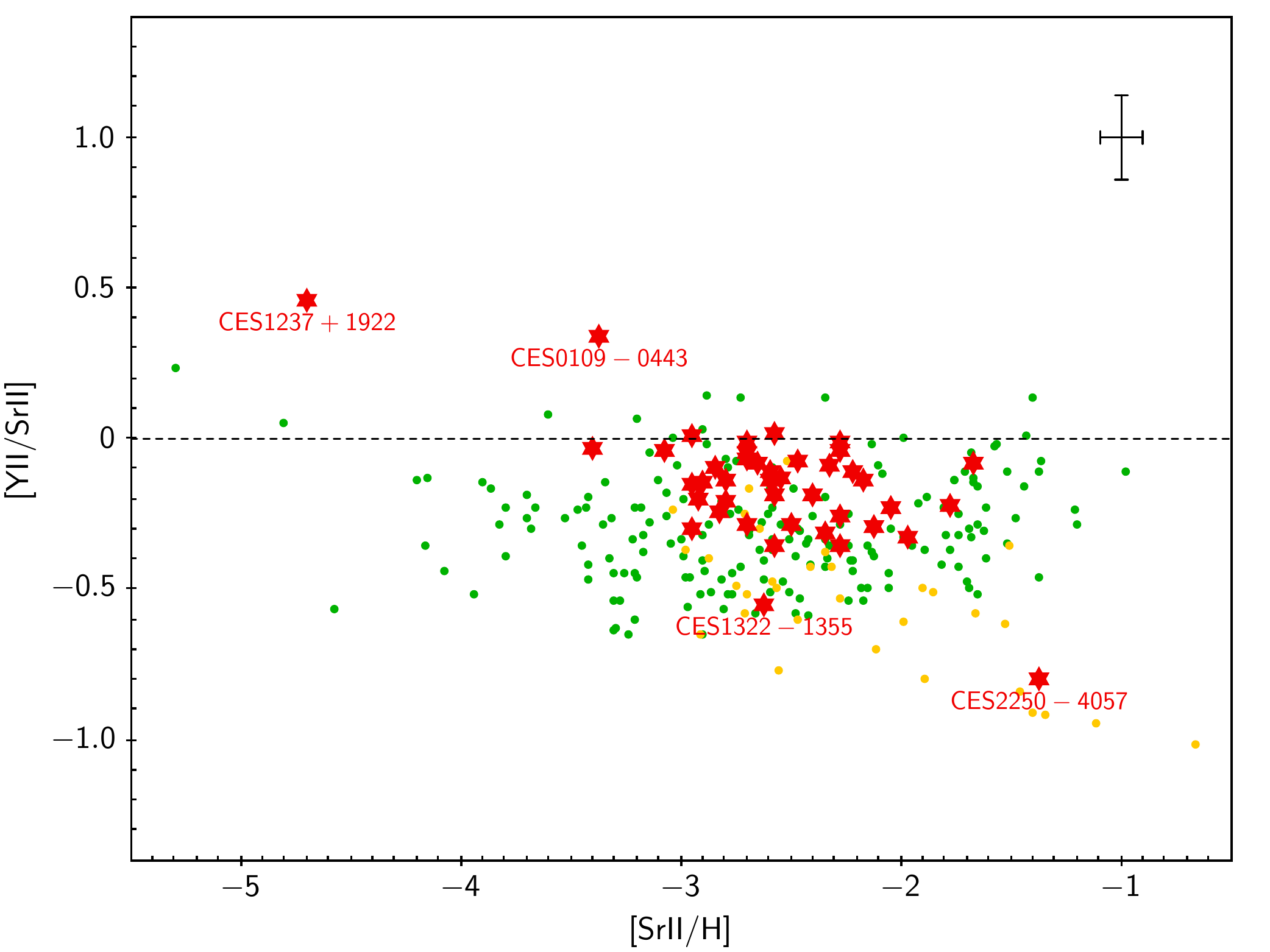}
   \includegraphics[width=9.1 cm]{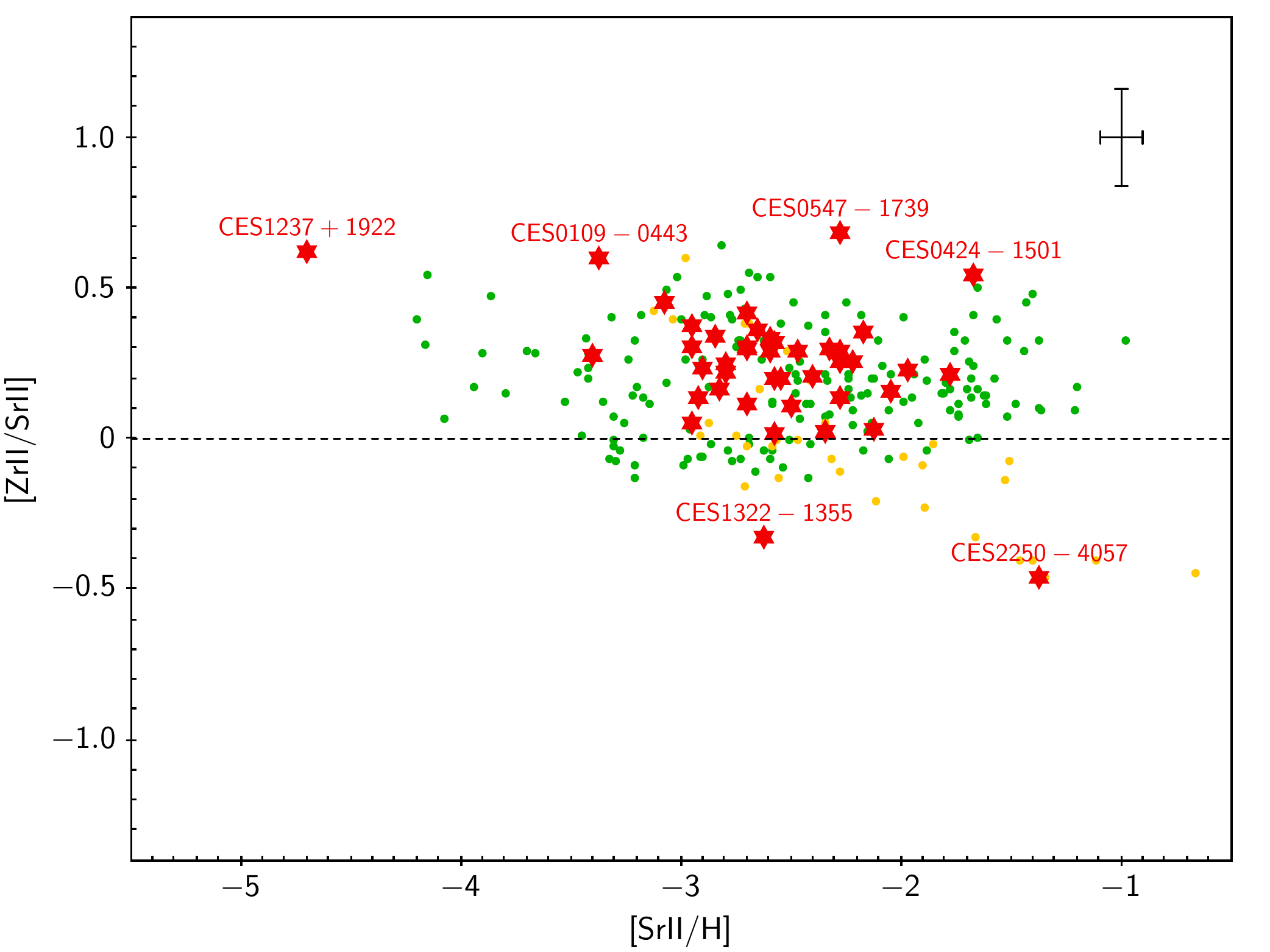}
      \caption{[\ion{Y}{II}/\ion{Sr}{II}] and [\ion{Zr}{II}/\ion{Sr}{II}] as a function of [\ion{Sr}{II}/H] for stars in our sample (red star symbols). Green dots represent the same quantities derived for main sequence, RGB, and sub-giant stars in the \citet{Roederer2014} sample. Yellow dots are the same quantities derived for horizontal branch stars in the \citet{Roederer2014} sample. A representative error is plotted in the upper-right corner of each panel. }
         \label{ysr_zrsr_srh}
   \end{figure*}

   \begin{figure*}
   \centering
   \includegraphics[width=16 cm]{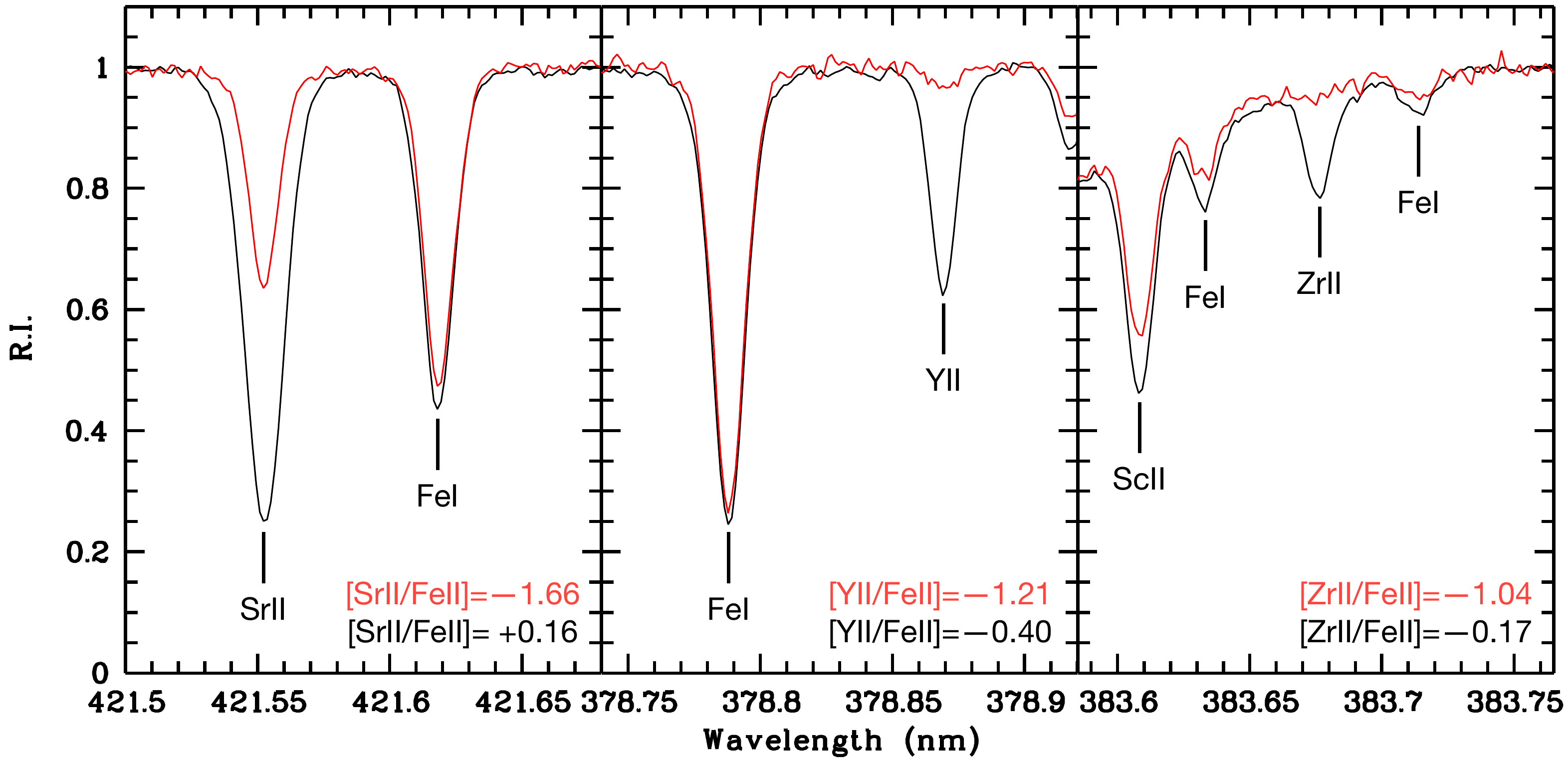}
      \caption{Normalised spectra of stars CES1237+1922 (red) and CES1322–1355 (black) around \ion{the Sr}{II} line at 421.5 nm (left panel), \ion{the Y}{II} line at  378.8 nm (central panel), and \ion{the Zr}{II} line at 383.6 nm (right panel). The stars have similar stellar parameters (\teff=4960, \logg=1.8, \vt=1.9, [Fe/H]=--3).}
         \label{ces1237_ces1322}
   \end{figure*}

The abundances of Sr, Y, and Zr are critical to constrain the astrophysical conditions of their production site once nuclear physics uncertainties are reduced \citep{Psaltis2022arXiv220407136P}. The variations in the observationally derived abundances could be due to differences in the analysis as described above; therefore, it is important to have homogeneously derived abundances for a large number of stars, as we presented here. 
In homogeneous analyses, the variations found in the observations can be linked to different astrophysical conditions and thus be used to constrain the site. 
A potential contribution to Sr, Y, and Zr are neutrino-driven ejecta in core-collapse SNe, where changes in the conditions (entropy, neutron richness, and expansion timescale) result in different patterns \citep[see e.g.][]{Hansen2014b}. 
Since Sr, Y, and Zr are produced in SNe by a process that runs close to stability, the nuclear physics uncertainties are small or can be constrained by experiments in the near future. 
Therefore, homogeneous abundances of the lighter heavy elements are a strong diagnostic for the conditions in SN explosions.

In Fig. \ref{ysr_zrsr_srh} we note that there are stars that show a [\ion{Y}{II}/\ion{Sr}{II}] and [\ion{Zr}{II}/\ion{Sr}{II}] ratio that is different, higher or lower, from the average of the other stars in the sample.
In the following subsections we present these peculiar stars, and defer a deeper discussion on their abundance pattern in future articles, when a more complete inventory of the n-capture element abundances will be available.

\subsection{CES1237+1922}

The star CES1237+1922 (also known as BS\,16085-0050) is deficient in Sr, Y and Zr compared to stars of similar metallicity (see Fig. \ref{ncapt_fe} and \ref{ces1237_ces1322}), with A(\ion{Sr}{II})=$-1.78\pm0.04$ ([\ion{Sr}{II}/\ion{Fe}{II}]=$-1.66$), A(\ion{Y}{II})=$-2.03\pm0.05$ ([\ion{Y}{II}/\ion{Fe}{II}]=$-1.21$), and A(\ion{Zr}{II})=$-1.46\pm0.12$ ([\ion{Zr}{II}/\ion{Fe}{II}]=$-1.04$).
The chemical composition of this star was first studied by \citet{2001PASP..113..519G}, who noted it for being rich in $\alpha$-elements. This finding is confirmed by our analysis, as for this star we derived A(\ion{Mg}{I})=$+5.12\pm0.04$ ([\ion{Mg}{I}/\ion{Fe}{I}]=$+0.77$) and A(\ion{Ca}{I})=$+3.68\pm0.10$ ([\ion{Ca}{I}/\ion{Fe}{I}]=$+0.54$).
Sr abundance derived by \citet{2001PASP..113..519G} was low, and so was that in \citet{2004ApJ...607..474H}. This is essentially in line with our analysis, once the lower gravity adopted by  \citet{2001PASP..113..519G} is accounted for, with respect to ours and that of \citet{2004ApJ...607..474H}. 
There are very few stars that have such low abundances of all three elements that populate the first peak of n-capture elements.
We looked for stars with similar light n-capture elements abundances as CES1237+1922 in the SAGA database \citep{Suda2008}. We found that there are only seven stars with similar Sr and Y abundances, and only two of these have a measurement of Zr.

Figure~\ref{ysr_zrsr_srh} shows how CES1237+1922 stands out with respect to the rest of our sample \citep[see also Figs. 1 and 3 of][]{Francois2007}. 
\citet{2008ApJ...687..272Q} noted the difficulty in explaining the stars with low abundances of first peak elements.
In a scenario in which all n-capture elements are formed via the $r$ process, they invoked three distinct $r-$process sites to explain the observations.
Faint SNe, with mixing and fall-back \citep[][]{2005Sci...309..451I}, are appealing sites since they show an excess of hydrostatic burning products with respect to the explosive products, thus explaining the exceptional stars with very low [Sr,Y,Zr/Fe].

Several studies have shown that the observed abundances of the first peak elements and that of second peak elements in low metallicity stars require the existence of at least two sites for n-capture nucleosynthesis \citep[e.g.][]{Hansen2012,Hansen2014a,Hansen2014b,Spite2018}.
The Galactic chemical evolution models of \citet{2018MNRAS.476.3432P} seem incapable of producing low [Sr/Fe], [Y/Fe] and [Zr/Fe] ratios as observed in CES1237+1922, even in the unrealistic hypothesis of switching off completely the $r-$process contribution. 

Another possibility would be that the  $\alpha-$elements and Sr, Y, and Zr are produced in a neutrino-driven SN, while heavy n-capture elements (second peak and heavier) come from an $r-$process event where first peak elements are underproduced compared to heavier ones, like, for example,  neutron star merger dynamical ejecta \citep{Korobkin2012} and disk ejecta \citep{Wu2016}. The low abundance of Sr, Y, and Zr relative to the iron-group elements could point to proton-rich ejecta from SN explosions. Under such conditions, the $\nu$p process can also produce Sr, Y, and Zr; however, their abundances are low relative to the iron-group nuclei \citep{Hansen2014b, Froehlich2006, Wanajo2006, Pruet2006}.

\subsection{CES0109-0443}
CES0109-0443 (also known as CS\,22183-031) is underabundant in Sr with respect to Y and Zr, with [\ion{Y}{II}/\ion{Sr}{II}]=+0.34 and [\ion{Zr}{II}/\ion{Sr}{II}]=+0.59  (Fig. \ref{ysr_zrsr_srh}). 
This star was identified for the first time as an $r$-process-enhanced metal-poor star by \citet{2004ApJ...607..474H}, with [Eu/Fe]=+1.2. It has also been analysed in detail by \citet{Roederer2014}, who confirmed the high enhancement in $r-$process material \citep{Roederer2014, Roederer2014MNRAS}.  
For this star \citet{Roederer2014} derived A(Sr)=$-0.84$ dex, A(Y)=$-1.61$ dex, and A(Zr)=$-0.7$ dex, which are significantly lower than those found in this study (A(Sr)=$-0.45 \pm 0.07$ dex, A(Y)=$-0.82 \pm 0.14$ dex, and A(Zr)=$-0.16 \pm 0.06$ dex). 
Since the lines used in our study and in \citet{Roederer2014} are approximately the same but the parameters are different, we derived the abundances of these elements using the parameters obtained in \citet{Roederer2014}, and we found that our results are consistent with theirs.  
We conclude that the origin of the discrepancy is only due to the different choice of stellar parameters. 

\subsection{CES2250--4057 and CES1322--1355}
The star CES2250--4057 (also known as CD-41 15048 also known as HE\,2247-4113) is overabundant in Sr with respect to Y and Zr, with [\ion{Y}{II}/\ion{Sr}{II}]=$-0.80$ and [\ion{Zr}{II}/\ion{Sr}{II}]=$-0.46$ (Fig. \ref{ysr_zrsr_srh}).
For this star we derived A(\ion{Sr}{II})=$+1.55 \pm 0.07$ ([\ion{Sr}{II}/\ion{Fe}{II}]=$+0.51$), A(\ion{Y}{II})=$+0.04\pm0.09$ ([\ion{Y}{II}/\ion{Fe}{II}]=$-0.29$), and A(\ion{Zr}{II})=$+0.79\pm0.06$ ([\ion{Zr}{II}/\ion{Fe}{II}]=$+0.05$).
This pattern is striking in our sample, and more generally in the literature, with about a dozen stars showing
a similar behaviour. 
This star is one of the  Bidelman-MacConnell `weak-metal' stars \citep{1973AJ.....78..687B,1985ApJS...58..463N}.
It is a red horizontal branch star and has been studied by \citet{Pereira2013} as a potential star that is escaping the Galaxy. They do not measure Sr, but their Y and Zr abundances are compatible with ours, within errors, especially after the 0.6\,dex lower gravity is accounted for.
So this overabundance of Sr has not been noted before.
In Fig. \ref{ysr_zrsr_srh} we highlighted with yellow dots the horizontal branch stars in \citet{Roederer2014}. The star CES2250--4057 lies in the region occupied by other red horizontal branch stars. 

There is another, slightly milder, case of this behaviour in our sample (Fig. \ref{ysr_zrsr_srh}), CES1322--1355 (also known as HE\,1320--1339), for which we derived A(\ion{Sr}{II})=$+0.30\pm0.05$ ([\ion{Sr}{II}/\ion{Fe}{II}]=$+0.16$), A(\ion{Y}{II})=$-0.96\pm0.13$ ([\ion{Y}{II}/\ion{Fe}{II}]=$-0.40$), and A(\ion{Zr}{II})=$-0.33\pm0.07$ ([\ion{Zr}{II}/\ion{Fe}{II}]=$-0.17$).
This star was initially studied within the HERES survey \citep{Barklem2005}, who found [Sr/Fe]$=+0.23\pm0.16$ and [Y/Fe]$=-0.13\pm0.15$. 
This star has been analysed also 
by \citet{Sakari2018} who provide
[Sr/Fe]$=+0.50\pm0.14$.
Our results are in line with previous studies once we take the different choice of stellar parameters into account.
\citet{Sakari2018} classified
this star as limited-r, a classification introduced by
\citet{2018ARNPS..68..237F} ([Eu/Fe] $< 0.3$, [Sr/Ba] $> 0.5$, and 
[Sr/Eu] $> 0.0$) with the aim of capturing the stars enriched 
under the condition of low neutron-to-seed ratios, a process more
often referred to as  `weak-r' process.
Yet since they do not have measurements of Y and Zr,
\citet{Sakari2018} did not note
the overabundance of Sr ([\ion{Y}{II}/\ion{Sr}{II}]=$-0.55$ and [\ion{Zr}{II}/\ion{Sr}{II}]=$-0.33$).

\subsection{CES0547-1739 and CES0424-1501}
Looking at the right panel in Fig. \ref{ysr_zrsr_srh}, we note that there are two stars (CES0547-1739 and CES0424-1501) with an overabundance of Zr with respect to Sr and Y, with 
[\ion{Zr}{II}/\ion{Sr}{II}]=$+0.68$ and 
[\ion{Zr}{II}/\ion{Y}{II}]=$+0.70$ for CES0547-1739, and 
[\ion{Zr}{II}/\ion{Sr}{II}]=$+0.54$ and 
[\ion{Zr}{II}/\ion{Y}{II}]=$+0.62$ for CES0424-1501.
Star CES0547-1739 (also known as TYC 5922-517-1) has previously been observed by the GALAH survey, which provided stellar parameters and chemical abundances for some elements \citep{GALAH2017,GALAH2018}. As no Sr and Zr abundance is present in the literature for this star, we provide them in this study for the first time.
Star CES0424-1501 (also known as BD-15 779) has been studied in detail by \citet{Hansen2020}, who found [Zr/Sr]$=+0.38$ and [Zr/Y]$=+0.68$. These results are compatible with ours taking into account the uncertainties and the difference in stellar parameters.

\section{Conclusions}

In this study we present a homogeneous set of stellar parameters and a chemical abundance analysis of elements from Na to Zr for a sample of 52 Galactic halo giant stars with $-3.58\leq$ [Fe/H] $\leq-1.79$.  
We compared our results to the ones in the literature and find a good agreement with previous studies. For stars that have a few chemical abundances in the literature, we completed the chemical inventory of light elements.  \\

The main conclusions of this study are the following:
\begin{itemize}

\item For 22 stars we were able to measure the \ion{Si}{II}
abundances thanks
to the high S/N and resolution of our dataset.
Similar measures are not available in the literature for these stars. Quite interestingly, Si is very close to ionisation balance, with our parallax-based gravities, which suggests that either the NLTE effects are not large or they are similar for the neutral and singly ionised lines.
This is in agreement with the results obtained by \citet{Amarsi2020A&A...642A..62A}, who find that NLTE effects on Si are small for stars in the GALAH survey, although the metallicity of the stars in their sample is higher than ours. 

\item {We were able to measure} \ion{Sc}{I} in 19 metal-poor giants,
again a novelty with respect to previous studies.
The ionisation imbalance
is on average 0.37\,dex. It would be interesting to study NLTE effects on \ion{Sc}{I}
to see if this is the cause of the ionisation imbalance.

\item For Cr we have a very important result: for our choice
of lines we find a flat trend of [Cr/Fe] $\approx 0$, both for
the neutral and the singly ionised species, in our LTE treatment.
This trend is theoretically expected  since Cr and Fe are both formed
under the same physical conditions. Previous investigations 
\citep[e.g.][]{mcwilliam95,Cayrel2004} relied heavily on 
\ion{Cr}{I} resonance lines, which are strongly affected 
by NLTE effects.
{This has been observationally confirmed by \citet{Bonifacio2009}, who find that, for their sample of extremely metal-poor giants, the [\ion{Cr}{ii}/Fe] is around zero, while [\ion{Cr}{i}/Fe] decreases with decreasing metallicity, suggesting NLTE effects on the \ion{Cr}{i} lines. }
This is supported by the large scatter in Cr abundances, both from ionised and neutral species, from chemical studies of individual stars \citep{Sneden_2016}.

\item We were able to measure \ion{Mn}{II} abundances for
38 stars; again, this measure is not generally available.
Existing NLTE and 3D NLTE computations do not seem able to explain
the ionisation imbalance and in fact seem to worsen it.
Further theoretical investigation of the line formation of 
\ion{Mn}{I} and \ion{Mn}{II} stars is desirable. 

\item We have roughly doubled the number of measurements
of Cu in the metallicity regime [Fe/H] $\leq -2.5$.

\item 
We highlighted the existence of two Zn-rich stars in the sample. 
Both stars were previously known, and our measurements are
consistent with the literature.
The fact that one of the two stars (CES2254--4209)
is rich in $r$-process elements (and is in fact an actinide boost star) 
while the other (CES1543+0201) is not essentially rules
out the hypothesis that this enhancement is due
to the production of Zn through neutron captures.

\item 
We noted the existence of two branches in the [Zn/Fe] versus [Ni/Fe]
plane and suggest that the high [Zn/Fe] branch is due to a contribution
of HNe to the chemical enrichment of the gas out of which the stars
were formed. These two branches are also present in the \citep{Cayrel2004} sample, 
but they were not reported before. Our data help make the presence
of the two branches more obvious.

\item The measurement of Sr, Y, and Zr highlighted several stars
with a peculiar pattern. 
CES1237+1922 is deficient in all three elements compared to the other stars in the sample.
CES2250-4057 shows abundances of Y and Zr that are
compatible with those of other stars with similar
metallicity; however, [Sr/H] is about 1 dex more abundant than [Y/H] and [Zr/H], making the production site of Sr - Zr far from clear. 

\end{itemize}

The high quality of our spectra, both in terms of S/N and resolving power, allowed us to measure many weak lines that are in general not measured in stars in this metallicity range. In this way, we provided a unique sample of measures of \ion{Si}{II}, \ion{Sc}{I}, and \ion{Mn}{II}. 
These observations can provide important guidelines in the development of line formation computations, more sophisticated than those presented here, that take deviations from LTE and hydrodynamical effects into account. 

The homogeneity of our analysis was crucial in highlighting any chemical peculiarities in the stars of our sample. 
Stellar parameters derived in this study will be used to determine the heavy n-capture element abundances in future papers of this project, and this will allow us to draw more conclusions about the chemistry of these peculiar stars.

\begin{acknowledgements}
      We gratefully acknowledge A. Heger and C. Chan for allowing us to show the results obtained with STARFIT code. 
      We gratefully acknowledge support from the French National Research Agency (ANR) funded project “Pristine” (ANR-18- CE31-0017). 
      We acknowledge support from the Deutsche Forschungsgemeinschaft (DFG, German Research Foundation)—Project No. 279384907—SFB 1245, the European Research Council Grant No. 677912 EUROPIUM, and  the State of Hesse within the Research Cluster ELEMENTS (Project ID 500/10.006). 
      This project has received funding from the European Research Council (ERC) under the European Union’s Horizon 2020 research and innovation programme (grant agreement No. 804240).
      M.R. acknowledges the support by the Spanish Ministry of Science, Education and Universities (PGC2018-095984-B-I00) and the Valencian Community (PROMETEU/2019/071).
      AJKH gratefully acknowledges funding by the Deutsche Forschungsgemeinschaft (DFG, German Research Foundation) -- Project-ID 138713538 -- SFB 881 (``The Milky Way System''), subprojects A03, A05, A11. 
      This work has made use of data from the European Space Agency (ESA) mission
      {\it Gaia} (\url{https://www.cosmos.esa.int/gaia}), processed by the {\it Gaia}
      Data Processing and Analysis Consortium (DPAC,
      \url{https://www.cosmos.esa.int/web/gaia/dpac/consortium}). Funding for the DPAC
      has been provided by national institutions, in particular the institutions
      participating in the {\it Gaia} Multilateral Agreement. 
      This article is based upon work from the “ChETEC” COST Action (CA16117), supported by COST (European Cooperation in Science and Technology).
     \end{acknowledgements}

\bibliographystyle{aa} 
\bibliography{bibceres} 


\begin{appendix}
\onecolumn


\begin{landscape}
\section{Observation log}
\begin{longtable}{llllcrrccrrr}
\caption{\label{tab:obs_log}Observation log.}\\
\hline
\hline
Name ID & CERES name & RA2000 & DEC2000 & UVES arm & R & \vbroad & Date & MJD & exp. time & S/N & slit \\
 &  &   &   &   &   & \kms &  &  & s & px$^{-1}$ & " \\
\hline
\endfirsthead
\caption{Continued.} \\
\hline
Name ID & CERES name & RA2000 & DEC2000 & UVES arm & R & \vbroad & Date & MJD & exp. time & S/N & slit \\
 &  &   &   &   &   & \kms &  &  & s & px$^{-1}$ & " \\
\hline
\endhead
\hline
\endfoot
\hline
\endlastfoot

  HD2796 & CES0031-1647 & 00:31:16.91 & -16:47:40.8 & BLU390 & 71050 & 7.2 & 2005-11-18 & 53692.01 & 300 & 123 & 0.4\\
   &  &  &  & RED580 & 107200 & 7.3 & 2005-11-18 & 53692.01 & 300 & 259 & 0.3\\
  HD4306 & CES0045-0932 & 00:45:27.16 & -09:32:39.8 & BLU346 & 40970 & 7.7 & 2001-10-09 & 52191.14 & 697 & 131 & 1.0\\
   &  &  &  &  &  &  & 2001-10-09 & 52191.12 & 660 & 126 & 1.0\\
   &  &  &  &  &  &  & 2001-10-09 & 52191.12 & 660 & 134 & 1.0\\
   &  &  &  &  &  &  & 2001-10-09 & 52191.13 & 697 & 129 & 1.0\\
   &  &  &  &  &  &  & 2001-10-09 & 52191.11 & 660 & 129 & 1.0\\
   &  &  &  & RED580 & 56990 & 6.9 & 2001-10-09 & 52191.11 & 305 & 241 & 0.7\\
   &  &  &  &  &  &  & 2001-10-09 & 52191.12 & 305 & 233 & 0.7\\
   &  &  &  &  &  &  & 2001-10-09 & 52191.13 & 305 & 261 & 0.7\\
   &  &  &  &  &  &  & 2001-10-09 & 52191.11 & 305 & 244 & 0.7\\
   &  &  &  &  &  &  & 2001-10-09 & 52191.12 & 305 & 240 & 0.7\\
   &  &  &  &  &  &  & 2001-10-09 & 52191.12 & 305 & 239 & 0.7\\
  BD-11\_145 & CES0048-1041 & 00:48:24.31 & -10:41:30.9 & BLU390 & 40970 & 8.8 & 2019-11-19 & 58806.05 & 2200 & 128 & 1.0\\
   &  &  &  & RED564 & 42310 & 8.0 & 2019-11-19 & 58806.05 & 2200 & 0 & 1.0\\
  HD5426 & CES0055-3345 & 00:55:41.05 & -33:45:11.5 & BLU390 & 40970 & 7.0 & 2019-11-20 & 58807.05 & 900 & 160 & 1.0\\
   &  &  &  & RED564 & 42310 & 6.4 & 2019-11-20 & 58807.05 & 900 & 360 & 1.0\\
  HE0057-4541 & CES0059-4524 & 00:59:59.28 & -45:24:53.4 & BLU390 & 40970 & 6.7 & 2007-11-07 & 54411.04 & 3600 & 24 & 1.0\\
   &  &  &  &  &  &  & 2007-11-07 & 54411.14 & 3600 & 24 & 1.0\\
   &  &  &  &  &  &  & 2007-11-07 & 54411.10 & 3600 & 26 & 1.0\\
   &  &  &  &  &  &  & 2007-11-07 & 54411.18 & 3600 & 23 & 1.0\\
   &  &  &  &  &  &  & 2007-11-08 & 54412.23 & 3600 & 21 & 1.0\\
   &  &  &  &  &  &  & 2007-11-08 & 54412.18 & 3600 & 26 & 1.0\\
  BPSCS22953-003 & CES0102-6143 & 01:02:15.87 & -61:43:45.8 & BLU390 & 49620 & 8.2 & 2002-08-03 & 52489.18 & 1800 & 18 & 0.8\\
   &  &  &  &  &  &  & 2002-08-03 & 52489.20 & 1800 & 20 & 0.8\\
   &  &  &  & RED564 & 87410 & 6.8 & 2002-08-03 & 52489.18 & 1800 & 19 & 0.4\\
   &  &  &  &  &  &  & 2002-08-03 & 52489.20 & 1800 & 21 & 0.4\\
  HE0105-6141 & CES0107-6125 & 01:07:37.85 & -61:25:17.7 & BLU390 & 40970 & 6.6 & 2007-11-09 & 54413.22 & 4500 & 30 & 1.0\\
  BPSCS22183-031 & CES0109-0443 & 01:09:05.09 & -04:43:21.3 & BLU390 & 40970 & 7.1 & 2017-09-26 & 58022.22 & 3005 & 62 & 1.0\\
  HD13979 & CES0215-2554 & 02:15:20.85 & -25:54:54.9 & BLU390 & 40970 & 8.4 & 2019-11-19 & 58806.18 & 600 & 147 & 1.0\\
   &  &  &  & RED564 & 42310 & 8.6 & 2019-11-19 & 58806.18 & 600 & 330 & 1.0\\
  BD-22\_395 & CES0221-2130 & 02:21:57.94 & -21:30:43.0 & BLU390 & 40970 & 8.7 & 2019-11-20 & 58807.07 & 2200 & 141 & 1.0\\
   &  &  &  & RED564 & 42310 & 8.8 & 2019-11-20 & 58807.07 & 2200 & 357 & 1.0\\
  HE0240-0807 & CES0242-0754 & 02:42:57.73 & -07:54:35.4 & BLU390 & 40970 & 8.0 & 2007-11-07 & 54411.23 & 3600 & 19 & 1.0\\
   &  &  &  &  &  &  & 2007-11-08 & 54412.27 & 4500 & 16 & 1.0\\
   &  &  &  &  &  &  & 2007-11-10 & 54414.07 & 3600 & 17 & 1.0\\
   &  &  &  &  &  &  & 2007-11-10 & 54414.03 & 3600 & 12 & 1.0\\
   &  &  &  &  &  &  & 2007-11-10 & 54414.11 & 3600 & 18 & 1.0\\
   &  &  &  &  &  &  & 2007-11-10 & 54414.21 & 3600 & 19 & 1.0\\
   &  &  &  &  &  &  & 2007-11-10 & 54414.17 & 3600 & 19 & 1.0\\
   &  &  &  &  &  &  & 2007-11-10 & 54414.25 & 3600 & 15 & 1.0\\
  BPSCS31078-018 & CES0301+0616 & 03:01:00.69 & +06:16:31.8 & BLU390 & 40970 & 6.7 & 2019-11-19 & 58806.08 & 4200 & 57 & 1.0\\
   &  &  &  &  &  &  & 2019-11-19 & 58806.13 & 4200 & 59 & 1.0\\
   &  &  &  & BLU390 & 40970 & 6.7 & 2017-10-01 & 58027.19 & 3005 & 65 & 1.0\\
   &  &  &  & RED564 & 42310 & 6.2 & 2019-11-19 & 58806.08 & 4200 & 141 & 1.0\\
   &  &  &  &  &  &  & 2019-11-19 & 58806.13 & 4200 & 142 & 1.0\\
  HE0336-2412 & CES0338-2402 & 03:38:41.49 & -24:02:50.3 & BLU346 & 40970 & 6.7 & 2001-12-13 & 52256.14 & 1175 & 122 & 1.0\\
   &  &  &  &  &  &  & 2001-12-13 & 52256.15 & 1175 & 122 & 1.0\\
   &  &  &  &  &  &  & 2001-12-13 & 52256.05 & 1475 & 121 & 1.0\\
   &  &  &  &  &  &  & 2001-12-13 & 52256.18 & 1175 & 111 & 1.0\\
   &  &  &  &  &  &  & 2001-12-13 & 52256.17 & 1175 & 119 & 1.0\\
   &  &  &  &  &  &  & 2001-12-13 & 52256.03 & 1475 & 124 & 1.0\\
   &  &  &  & RED580 & 56990 & 6.3 & 2001-12-13 & 52256.15 & 560 & 180 & 0.7\\
   &  &  &  &  &  &  & 2001-12-13 & 52256.16 & 560 & 165 & 0.7\\
   &  &  &  &  &  &  & 2001-12-13 & 52256.14 & 560 & 165 & 0.7\\
   &  &  &  &  &  &  & 2001-12-13 & 52256.15 & 560 & 180 & 0.7\\
   &  &  &  &  &  &  & 2001-12-13 & 52256.19 & 560 & 163 & 0.7\\
   &  &  &  &  &  &  & 2001-12-13 & 52256.17 & 560 & 166 & 0.7\\
   &  &  &  &  &  &  & 2001-12-13 & 52256.17 & 560 & 175 & 0.7\\
   &  &  &  &  &  &  & 2001-12-13 & 52256.18 & 560 & 159 & 0.7\\
  BD+06\_648 & CES0413+0636 & 04:13:13.11 & +06:36:01.8 & BLU390 & 40970 & 9.7 & 2019-11-20 & 58807.21 & 600 & 70 & 1.0\\
   &  &  &  & RED564 & 42310 & 8.4 & 2019-11-20 & 58807.21 & 600 & 322 & 1.0\\
  BPSCS22186-023 & CES0419-3651 & 04:19:45.53 & -36:51:36.0 & BLU390 & 40970 & 7.5 & 2006-10-18 & 54026.30 & 2400 & 60 & 1.0\\
   &  &  &  &  &  &  & 2019-11-19 & 58806.21 & 3600 & 64 & 1.0\\
   &  &  &  &  &  &  & 2019-11-19 & 58806.25 & 3600 & 52 & 1.0\\
   &  &  &  & RED564 & 42310 & 7.6 & 2019-11-19 & 58806.21 & 3600 & 144 & 1.0\\
   &  &  &  &  &  &  & 2019-11-19 & 58806.25 & 3600 & 122 & 1.0\\
   &  &  &  &  &  &  & 2006-10-18 & 54026.30 & 2400 & 137 & 1.0\\
  HD27928 & CES0422-3715 & 04:22:55.14 & -37:15:49.2 & BLU390 & 40970 & 7.2 & 2019-11-19 & 58806.20 & 900 & 148 & 1.0\\
   &  &  &  & RED564 & 42310 & 7.3 & 2019-11-19 & 58806.20 & 900 & 333 & 1.0\\
  BD-15\_779 & CES0424-1501 & 04:24:45.64 & -15:01:50.7 & BLU390 & 40970 & 8.0 & 2020-03-02 & 58910.99 & 2700 & 157 & 1.0\\
   &  &  &  & RED564 & 42310 & 7.6 & 2020-03-02 & 58910.99 & 2700 & 506 & 1.0\\
  HE0428-1340 & CES0430-1334 & 04:30:51.42 & -13:34:08.1 & BLU390 & 40970 & 7.2 & 2019-11-20 & 58807.13 & 1800 & 159 & 1.0\\
   &  &  &  & RED564 & 42310 & 7.7 & 2019-11-20 & 58807.13 & 1800 & 344 & 1.0\\
  HE0442-1234 & CES0444-1228 & 04:44:51.71 & -12:28:45.5 & BLU390 & 58640 & 7.4 & 2003-02-05 & 52675.11 & 3600 & 37 & 0.6\\
   &  &  &  &  &  &  & 2003-02-05 & 52675.15 & 3197 & 28 & 0.6\\
   &  &  &  &  &  &  & 2003-02-05 & 52675.06 & 3600 & 35 & 0.6\\
   &  &  &  &  &  &  & 2003-02-06 & 52676.06 & 3600 & 44 & 0.6\\
   &  &  &  &  &  &  & 2003-02-06 & 52676.11 & 3600 & 39 & 0.6\\
   &  &  &  &  &  &  & 2003-02-08 & 52678.12 & 3600 & 31 & 0.6\\
   &  &  &  & RED580 & 66320 & 7.3 & 2003-02-05 & 52675.06 & 3600 & 129 & 0.6\\
   &  &  &  &  &  &  & 2003-02-05 & 52675.15 & 3205 & 114 & 0.6\\
   &  &  &  &  &  &  & 2003-02-05 & 52675.11 & 3600 & 136 & 0.6\\
   &  &  &  &  &  &  & 2003-02-06 & 52676.06 & 3600 & 149 & 0.6\\
   &  &  &  &  &  &  & 2003-02-06 & 52676.11 & 3600 & 139 & 0.6\\
   &  &  &  &  &  &  & 2003-02-08 & 52678.12 & 3600 & 124 & 0.6\\
  HE0516-3820 & CES0518-3817 & 05:18:12.92 & -38:17:32.7 & BLU390 & 40970 & 6.7 & 2007-11-09 & 54413.33 & 3600 & 29 & 1.0\\
  HE0524-2055 & CES0527-2052 & 05:27:04.44 & -20:52:42.1 & BLU390 & 40970 & 8.3 & 2007-11-07 & 54411.32 & 3600 & 38 & 1.0\\
   &  &  &  &  &  &  & 2007-11-07 & 54411.28 & 3600 & 39 & 1.0\\
   &  &  &  &  &  &  & 2007-11-08 & 54412.33 & 3300 & 25 & 1.0\\
   &  &  &  &  &  &  & 2007-11-10 & 54414.30 & 3600 & 30 & 1.0\\
  TYC5922-517-1 & CES0547-1739 & 05:47:20.81 & -17:39:41.0 & BLU390 & 40970 & 10.0 & 2019-11-20 & 58807.23 & 4080 & 49 & 1.0\\
   &  &  &  &  &  &  & 2019-11-20 & 58807.27 & 4080 & 43 & 1.0\\
   &  &  &  & RED564 & 42310 & 8.4 & 2019-11-20 & 58807.23 & 4080 & 224 & 1.0\\
   &  &  &  &  &  &  & 2019-11-20 & 58807.27 & 4080 & 204 & 1.0\\
  TYC4840-159-1 & CES0747-0405 & 07:47:15.82 & -04:05:46.1 & BLU390 & 40970 & 11.2 & 2020-03-03 & 58911.03 & 2500 & 38 & 1.0\\
   &  &  &  &  &  &  & 2020-03-03 & 58911.06 & 2500 & 36 & 1.0\\
   &  &  &  &  &  &  & 2020-03-04 & 58912.02 & 1800 & 28 & 1.0\\
   &  &  &  &  &  &  & 2020-03-04 & 58912.04 & 1800 & 31 & 1.0\\
   &  &  &  & RED564 & 42310 & 10.0 & 2020-03-03 & 58911.03 & 2500 & 267 & 1.0\\
   &  &  &  &  &  &  & 2020-03-03 & 58911.05 & 2500 & 259 & 1.0\\
   &  &  &  &  &  &  & 2020-03-04 & 58912.02 & 1800 & 215 & 1.0\\
   &  &  &  &  &  &  & 2020-03-04 & 58912.04 & 1800 & 229 & 1.0\\
  TYC8931-1111-1 & CES0900-6222 & 09:00:52.59 & -62:22:52.8 & BLU390 & 40970 & 8.9 & 2019-11-19 & 58806.30 & 4200 & 53 & 1.0\\
   &  &  &  &  &  &  & 2019-11-20 & 58807.33 & 2200 & 25 & 1.0\\
   &  &  &  & RED564 & 42310 & 8.2 & 2019-11-19 & 58806.30 & 4200 & 278 & 1.0\\
   &  &  &  &  &  &  & 2019-11-20 & 58807.33 & 2200 & 161 & 1.0\\
  TYC8939-2532-1 & CES0908-6607 & 09:08:07.51 & -66:07:33.9 & BLU390 & 40970 & 8.8 & 2020-03-04 & 58912.19 & 3600 & 60 & 1.0\\
   &  &  &  & RED564 & 42310 & 9.1 & 2020-03-04 & 58912.19 & 3600 & 250 & 1.0\\
  TYC9200-2292-1 & CES0919-6958 & 09:19:16.29 & -69:58:39.9 & BLU390 & 40970 & 8.9 & 2020-03-04 & 58912.15 & 3600 & 50 & 1.0\\
   &  &  &  & RED564 & 42310 & 7.6 & 2020-03-04 & 58912.15 & 3600 & 242 & 1.0\\
  UCAC2\_1106907 & CES1116-7250 & 11:16:54.01 & -72:50:16.1 & BLU390 & 40970 & 10.3 & 2020-03-04 & 58912.23 & 3600 & 24 & 1.0\\
   &  &  &  & RED564 & 42310 & 9.4 & 2020-03-04 & 58912.23 & 3600 & 213 & 1.0\\
  HE1219-0312 & CES1221-0328 & 12:21:34.14 & -03:28:39.6 & BLU346 & 58640 & 5.6 & 2004-02-20 & 53055.33 & 3600 & 6 & 0.6\\
   &  &  &  &  &  &  & 2004-02-20 & 53055.37 & 3600 & 5 & 0.6\\
   &  &  &  &  &  &  & 2004-02-20 & 53055.28 & 3600 & 6 & 0.6\\
   &  &  &  &  &  &  & 2004-04-16 & 53111.07 & 1198 & 1 & 0.6\\
   &  &  &  &  &  &  & 2004-05-11 & 53136.02 & 3600 & 6 & 0.6\\
   &  &  &  &  &  &  & 2004-05-11 & 53136.06 & 3600 & 6 & 0.6\\
   &  &  &  &  &  &  & 2004-05-24 & 53149.09 & 3600 & 4 & 0.6\\
   &  &  &  &  &  &  & 2004-05-23 & 53149.00 & 3600 & 4 & 0.6\\
   &  &  &  &  &  &  & 2004-05-24 & 53149.04 & 3600 & 4 & 0.6\\
   &  &  &  &  &  &  & 2005-01-19 & 53389.32 & 3600 & 6 & 0.6\\
   &  &  &  &  &  &  & 2005-04-08 & 53468.12 & 3600 & 6 & 0.6\\
   &  &  &  &  &  &  & 2005-04-08 & 53468.16 & 3600 & 6 & 0.6\\
   &  &  &  &  &  &  & 2005-04-08 & 53468.21 & 3600 & 6 & 0.6\\
   &  &  &  &  &  &  & 2005-04-10 & 53470.18 & 3600 & 7 & 0.6\\
   &  &  &  &  &  &  & 2005-04-10 & 53470.13 & 3600 & 6 & 0.6\\
   &  &  &  &  &  &  & 2005-04-10 & 53470.09 & 3600 & 6 & 0.6\\
   &  &  &  &  &  &  & 2005-04-11 & 53471.10 & 3600 & 6 & 0.6\\
   &  &  &  &  &  &  & 2005-04-11 & 53471.15 & 3600 & 5 & 0.6\\
   &  &  &  & RED580 & 66320 & 6.4 & 2004-02-20 & 53055.28 & 3600 & 36 & 0.6\\
   &  &  &  &  &  &  & 2004-02-20 & 53055.33 & 3600 & 34 & 0.6\\
   &  &  &  &  &  &  & 2004-02-20 & 53055.37 & 3600 & 33 & 0.6\\
   &  &  &  &  &  &  & 2004-04-16 & 53111.07 & 1201 & 15 & 0.6\\
   &  &  &  &  &  &  & 2004-05-11 & 53136.02 & 3600 & 34 & 0.6\\
   &  &  &  &  &  &  & 2004-05-11 & 53136.06 & 3600 & 35 & 0.6\\
   &  &  &  &  &  &  & 2004-05-24 & 53149.09 & 3600 & 29 & 0.6\\
   &  &  &  &  &  &  & 2004-05-23 & 53149.00 & 3600 & 29 & 0.6\\
   &  &  &  &  &  &  & 2004-05-24 & 53149.04 & 3600 & 29 & 0.6\\
   &  &  &  &  &  &  & 2005-01-19 & 53389.32 & 3600 & 35 & 0.6\\
   &  &  &  &  &  &  & 2005-04-08 & 53468.16 & 3600 & 35 & 0.6\\
   &  &  &  &  &  &  & 2005-04-08 & 53468.12 & 3600 & 35 & 0.6\\
   &  &  &  &  &  &  & 2005-04-08 & 53468.21 & 3600 & 34 & 0.6\\
   &  &  &  &  &  &  & 2005-04-10 & 53470.13 & 3600 & 36 & 0.6\\
   &  &  &  &  &  &  & 2005-04-10 & 53470.09 & 3600 & 33 & 0.6\\
   &  &  &  &  &  &  & 2005-04-10 & 53470.18 & 3600 & 37 & 0.6\\
   &  &  &  &  &  &  & 2005-04-11 & 53471.10 & 3600 & 34 & 0.6\\
   &  &  &  &  &  &  & 2005-04-11 & 53471.15 & 3600 & 33 & 0.6\\
  HD107752 & CES1222+1136 & 12:22:52.72 & +11:36:25.5 & BLU390 & 40970 & 8.0 & 2020-03-03 & 58911.22 & 2400 & 163 & 1.0\\
   &  &  &  &  &  &  & 2020-03-03 & 58911.25 & 2400 & 163 & 1.0\\
   &  &  &  & RED564 & 42310 & 8.1 & 2020-03-03 & 58911.22 & 2400 & 426 & 1.0\\
   &  &  &  &  &  &  & 2020-03-03 & 58911.25 & 2400 & 424 & 1.0\\
  HD108317 & CES1226+0518 & 12:26:36.83 & +05:18:09.0 & BLU346 & 40970 & 7.7 & 2002-02-04 & 52309.29 & 260 & 154 & 1.0\\
   &  &  &  &  &  &  & 2002-02-04 & 52309.29 & 260 & 127 & 1.0\\
   &  &  &  &  &  &  & 2002-02-04 & 52309.30 & 260 & 160 & 1.0\\
   &  &  &  &  &  &  & 2002-02-04 & 52309.31 & 260 & 155 & 1.0\\
   &  &  &  &  &  &  & 2002-02-04 & 52309.29 & 250 & 158 & 1.0\\
   &  &  &  &  &  &  & 2002-02-04 & 52309.30 & 260 & 132 & 1.0\\
   &  &  &  &  &  &  & 2002-02-04 & 52309.28 & 250 & 161 & 1.0\\
   &  &  &  & RED580 & 56990 & 6.3 & 2002-02-04 & 52309.30 & 107 & 232 & 0.7\\
   &  &  &  &  &  &  & 2002-02-04 & 52309.30 & 107 & 170 & 0.7\\
   &  &  &  &  &  &  & 2002-02-04 & 52309.29 & 107 & 200 & 0.7\\
   &  &  &  &  &  &  & 2002-02-04 & 52309.30 & 107 & 182 & 0.7\\
   &  &  &  &  &  &  & 2002-02-04 & 52309.29 & 107 & 186 & 0.7\\
   &  &  &  &  &  &  & 2002-02-04 & 52309.30 & 107 & 202 & 0.7\\
   &  &  &  &  &  &  & 2002-02-04 & 52309.31 & 107 & 218 & 0.7\\
   &  &  &  &  &  &  & 2002-02-04 & 52309.30 & 107 & 168 & 0.7\\
   &  &  &  &  &  &  & 2002-02-04 & 52309.29 & 107 & 222 & 0.7\\
   &  &  &  &  &  &  & 2002-02-04 & 52309.31 & 107 & 205 & 0.7\\
  HD108577 & CES1228+1220 & 12:28:16.86 & +12:20:41.1 & BLU390 & 40970 & 8.8 & 2020-03-03 & 58911.20 & 1100 & 147 & 1.0\\
   &  &  &  & RED564 & 42310 & 8.9 & 2020-03-03 & 58911.20 & 1100 & 359 & 1.0\\
  BPSBS16085-0050 & CES1237+1922 & 12:37:46.68 & +19:22:49.6 & BLU390 & 40970 & 7.9 & 2020-03-04 & 58912.28 & 3000 & 76 & 1.0\\
   &  &  &  &  &  &  & 2020-03-04 & 58912.32 & 3000 & 76 & 1.0\\
   &  &  &  & RED564 & 42310 & 8.0 & 2020-03-04 & 58912.28 & 3000 & 182 & 1.0\\
   &  &  &  &  &  &  & 2020-03-04 & 58912.32 & 3000 & 189 & 1.0\\
  HE1243-2408 & CES1245-2425 & 12:45:53.85 & -24:25:02.4 & BLU390 & 40970 & 7.5 & 2020-03-03 & 58911.12 & 3250 & 133 & 1.0\\
   &  &  &  &  &  &  & 2020-03-03 & 58911.16 & 3250 & 138 & 1.0\\
   &  &  &  & RED564 & 42310 & 7.6 & 2020-03-03 & 58911.12 & 3250 & 342 & 1.0\\
   &  &  &  &  &  &  & 2020-03-03 & 58911.16 & 3250 & 346 & 1.0\\
  HE1320-1339 & CES1322-1355 & 13:22:44.11 & -13:55:31.4 & BLU390 & 40970 & 8.3 & 2020-03-03 & 58911.28 & 2600 & 141 & 1.0\\
   &  &  &  &  &  &  & 2020-03-03 & 58911.31 & 2600 & 149 & 1.0\\
   &  &  &  & RED564 & 42310 & 8.3 & 2020-03-03 & 58911.28 & 2600 & 340 & 1.0\\
   &  &  &  &  &  &  & 2020-03-03 & 58911.31 & 2600 & 358 & 1.0\\
  HD122563 & CES1402+0941 & 14:02:31.85 & +09:41:09.9 & BLU346 & 65030 & 7.0 & 2002-02-19 & 52324.39 & 86 & 84 & 0.5\\
   &  &  &  &  &  &  & 2002-02-19 & 52324.39 & 86 & 96 & 0.5\\
   &  &  &  & RED564 & 51690 & 8.0 & 2000-04-12 & 51646.27 & 60 & 431 & 0.8\\
   &  &  &  &  &  &  & 2000-04-12 & 51646.27 & 60 & 413 & 0.8\\
   &  &  &  &  &  &  & 2000-04-12 & 51646.27 & 60 & 407 & 0.8\\
  HD122956 & CES1405-1451 & 14:05:13.02 & -14:51:25.5 & BLU346 & 40970 & 8.6 & 2003-06-07 & 52797.97 & 250 & 109 & 1.0\\
   &  &  &  & RED564 & 42310 & 8.3 & 2000-04-12 & 51646.33 & 120 & 362 & 1.0\\
   &  &  &  &  &  &  & 2000-04-12 & 51646.34 & 120 & 376 & 1.0\\
  TYC9427-1414-1 & CES1413-7609 & 14:13:11.18 & -76:09:50.4 & BLU390 & 40970 & 8.1 & 2020-03-04 & 58912.36 & 2800 & 111 & 1.0\\
   &  &  &  & RED564 & 42310 & 7.7 & 2020-03-04 & 58912.36 & 2800 & 341 & 1.0\\
  HD126587 & CES1427-2214 & 14:27:00.36 & -22:14:39.0 & BLU346 & 40970 & 7.6 & 2002-03-22 & 52355.33 & 836 & 178 & 1.0\\
   &  &  &  &  &  &  & 2002-03-22 & 52355.32 & 836 & 170 & 1.0\\
   &  &  &  &  &  &  & 2002-03-22 & 52355.34 & 836 & 170 & 1.0\\
   &  &  &  &  &  &  & 2002-03-22 & 52355.31 & 836 & 172 & 1.0\\
   &  &  &  &  &  &  & 2002-03-23 & 52356.32 & 827 & 161 & 1.0\\
   &  &  &  &  &  &  & 2002-03-23 & 52356.31 & 827 & 164 & 1.0\\
   &  &  &  & RED580 & 56990 & 7.0 & 2002-03-22 & 52355.32 & 390 & 293 & 0.7\\
   &  &  &  &  &  &  & 2002-03-22 & 52355.31 & 390 & 303 & 0.7\\
   &  &  &  &  &  &  & 2002-03-22 & 52355.33 & 390 & 306 & 0.7\\
   &  &  &  &  &  &  & 2002-03-22 & 52355.34 & 390 & 301 & 0.7\\
   &  &  &  &  &  &  & 2002-03-22 & 52355.32 & 390 & 290 & 0.7\\
   &  &  &  &  &  &  & 2002-03-22 & 52355.34 & 390 & 288 & 0.7\\
   &  &  &  &  &  &  & 2002-03-22 & 52355.31 & 390 & 288 & 0.7\\
   &  &  &  &  &  &  & 2002-03-22 & 52355.33 & 390 & 307 & 0.7\\
  HD128279 & CES1436-2906 & 14:36:48.51 & -29:06:46.6 & BLU346 & 40970 & 5.9 & 2003-08-07 & 52858.99 & 800 & 252 & 1.0\\
   &  &  &  & BLU346 & 40970 & 5.9 & 2002-02-04 & 52309.36 & 250 & 190 & 1.0\\
   &  &  &  &  &  &  & 2002-02-04 & 52309.36 & 250 & 184 & 1.0\\
   &  &  &  &  &  &  & 2002-02-04 & 52309.36 & 250 & 191 & 1.0\\
   &  &  &  & RED564 & 42310 & 6.8 & 2000-04-12 & 51646.32 & 600 & 551 & 1.0\\
  BPSCS30312-100 & CES1543+0201 & 15:43:31.66 & +02:01:17.3 & BLU390 & 40970 & 6.9 & 2006-03-20 & 53814.39 & 900 & 27 & 1.0\\
   &  &  &  &  &  &  & 2006-03-20 & 53814.38 & 900 & 29 & 1.0\\
   &  &  &  &  &  &  & 2006-03-20 & 53814.40 & 900 & 28 & 1.0\\
   &  &  &  & RED580 & 42310 & 6.5 & 2006-03-20 & 53814.4 & 900 & 76 & 1.0\\
   &  &  &  &  &  &  & 2006-03-20 & 53814.38 & 900 & 77 & 1.0\\
   &  &  &  &  &  &  & 2006-03-20 & 53814.39 & 900 & 74 & 1.0\\
  BD+05\_3098 & CES1552+0517 & 15:52:17.26 & +05:17:44.3 & BLU390 & 40970 & 6.9 & 2020-03-03 & 58911.34 & 2200 & 148 & 1.0\\
   &  &  &  & RED564 & 42310 & 7.1 & 2020-03-03 & 58911.34 & 2200 & 359 & 1.0\\
  BD+23\_3130 & CES1732+2344 & 17:32:41.62 & +23:44:11.6 & BLU390 & 49620 & 6.3 & 2000-04-16 & 51650.35 & 600 & 151 & 0.8\\
   &  &  &  &  &  &  & 2000-04-16 & 51650.34 & 600 & 151 & 0.8\\
   &  &  &  &  &  &  & 2000-04-16 & 51650.33 & 600 & 155 & 0.8\\
  HD165195 & CES1804+0346 & 18:04:40.07 & +03:46:44.7 & BLU346 & 40970 & 9.3 & 2003-09-07 & 52889.02 & 625 & 108 & 1.0\\
   &  &  &  & RED564 & 42310 & 8.5 & 2000-04-12 & 51646.36 & 120 & 352 & 1.0\\
   &  &  &  &  &  &  & 2000-04-12 & 51646.36 & 120 & 370 & 1.0\\
  BPSCS22891-209 & CES1942-6103 & 19:42:02.18 & -61:03:44.5 & BLU390 & 40970 & 7.7 & 2006-10-15 & 54023.97 & 1200 & 44 & 1.0\\
   &  &  &  & RED580 & 42310 & 8.6 & 2006-10-15 & 54023.97 & 1200 & 133 & 1.0\\
  BPSCS22873-166 & CES2019-6130 & 20:19:22.04 & -61:30:15.1 & BLU390 & 40970 & 8.6 & 2006-10-17 & 54025.98 & 600 & 25 & 1.0\\
   &  &  &  & RED580 & 42310 & 8.9 & 2006-10-17 & 54025.98 & 600 & 95 & 1.0\\
  BPSCS22897-008 & CES2103-6505 & 21:03:11.86 & -65:05:08.9 & BLU346 & 53750 & 7.3 & 2008-04-22 & 54578.36 & 5200 & 30 & 0.7\\
   &  &  &  &  &  &  & 2008-05-12 & 54598.36 & 5200 & 30 & 0.7\\
   &  &  &  &  &  &  & 2008-05-22 & 54608.26 & 5200 & 28 & 0.7\\
   &  &  &  &  &  &  & 2008-06-24 & 54641.36 & 5200 & 18 & 0.7\\
   &  &  &  &  &  &  & 2008-06-24 & 54641.30 & 5200 & 19 & 0.7\\
   &  &  &  &  &  &  & 2008-06-27 & 54644.17 & 5200 & 19 & 0.7\\
  BPSCS29491-069 & CES2231-3238 & 22:31:02.19 & -32:38:36.5 & BLU390 & 71050 & 4.5 & 2005-11-20 & 53694.08 & 3600 & 40 & 0.4\\
   &  &  &  &  &  &  & 2005-11-20 & 53694.12 & 3600 & 39 & 0.4\\
   &  &  &  & RED580 & 51690 & 6.9 & 2004-10-03 & 53281.00 & 3600 & 119 & 0.8\\
  HE2229-4153 & CES2232-4138 & 22:32:49.05 & -41:38:25.2 & BLU390 & 40970 & 7.1 & 2007-11-08 & 54412.05 & 2700 & 52 & 1.0\\
   &  &  &  &  &  &  & 2007-11-09 & 54413.01 & 2700 & 27 & 1.0\\
   &  &  &  &  &  &  & 2007-11-09 & 54413.99 & 2700 & 46 & 1.0\\
   &  &  &  & RED580 & 51690 & 6.8 & 2008-04-28 & 54584.36 & 1800 & 73 & 0.8\\
  HE2247-4113 & CES2250-4057 & 22:50:14.02 & -40:57:42.8 & BLU390 & 40970 & 10.9 & 2019-11-19 & 58806.01 & 1500 & 160 & 1.0\\
   &  &  &  &  &  &  & 2019-11-19 & 58806.03 & 1500 & 170 & 1.0\\
   &  &  &  & RED564 & 42310 & 11.6 & 2019-11-19 & 58806.01 & 1500 & 328 & 1.0\\
   &  &  &  &  &  &  & 2019-11-19 & 58806.03 & 1500 & 342 & 1.0\\
  HE2252-4225 & CES2254-4209 & 22:54:58.57 & -42:09:19.4 & BLU390 & 49620 & 7.1 & 2004-10-20 & 53298.15 & 3900 & 18 & 0.8\\
   &  &  &  &  &  &  & 2005-05-18 & 53508.39 & 3300 & 19 & 0.8\\
   &  &  &  &  &  &  & 2005-06-14 & 53535.29 & 3600 & 18 & 0.8\\
   &  &  &  &  &  &  & 2005-06-14 & 53535.34 & 3600 & 19 & 0.8\\
   &  &  &  &  &  &  & 2005-07-04 & 53555.37 & 3600 & 23 & 0.8\\
   &  &  &  &  &  &  & 2005-07-04 & 53555.33 & 3600 & 23 & 0.8\\
   &  &  &  &  &  &  & 2005-07-04 & 53555.29 & 3600 & 21 & 0.8\\
   &  &  &  &  &  &  & 2005-07-06 & 53557.32 & 3600 & 22 & 0.8\\
   &  &  &  &  &  &  & 2005-07-06 & 53557.37 & 3600 & 23 & 0.8\\
   &  &  &  &  &  &  & 2005-07-10 & 53561.26 & 3600 & 25 & 0.8\\
   &  &  &  & RED580 & 51690 & 7.3 & 2005-05-18 & 53508.39 & 3300 & 54 & 0.8\\
   &  &  &  &  &  &  & 2005-06-14 & 53535.29 & 3600 & 55 & 0.8\\
   &  &  &  &  &  &  & 2005-06-14 & 53535.34 & 3600 & 56 & 0.8\\
   &  &  &  &  &  &  & 2005-07-04 & 53555.33 & 3600 & 63 & 0.8\\
   &  &  &  &  &  &  & 2005-07-04 & 53555.37 & 3600 & 62 & 0.8\\
   &  &  &  &  &  &  & 2005-07-04 & 53555.29 & 3600 & 60 & 0.8\\
   &  &  &  &  &  &  & 2005-07-06 & 53557.37 & 3600 & 63 & 0.8\\
   &  &  &  &  &  &  & 2005-07-06 & 53557.32 & 3600 & 61 & 0.8\\
   &  &  &  &  &  &  & 2005-07-10 & 53561.26 & 3600 & 67 & 0.8\\
  HE2327-5642 & CES2330-5626 & 23:30:37.09 & -56:26:14.4 & BLU390 & 49620 & 7.0 & 2004-11-15 & 53324.07 & 3600 & 46 & 0.8\\
   &  &  &  &  &  &  & 2005-08-05 & 53587.17 & 3600 & 43 & 0.8\\
   &  &  &  &  &  &  & 2005-08-05 & 53587.26 & 3600 & 36 & 0.8\\
   &  &  &  &  &  &  & 2005-08-05 & 53587.21 & 3600 & 42 & 0.8\\
   &  &  &  &  &  &  & 2005-08-10 & 53592.29 & 3600 & 45 & 0.8\\
   &  &  &  & RED564 & 42310 & 8.1 & 2007-11-03 & 54407.02 & 3000 & 57 & 1.0\\
   &  &  &  &  &  &  & 2008-01-25 & 54490.03 & 3000 & 75 & 1.0\\
  BPSCS30315-029 & CES2334-2642 & 23:34:26.70 & -26:42:14.0 & BLU390 & 40970 & 7.6 & 2007-11-06 & 54410.99 & 3600 & 36 & 1.0\\
   &  &  &  &  &  &  & 2007-11-08 & 54412.09 & 3600 & 44 & 1.0\\
   &  &  &  &  &  &  & 2007-11-08 & 54412.13 & 3600 & 40 & 1.0\\
\end{longtable}
\end{landscape}

\end{appendix}

\end{document}